\documentclass[numberedappendix,twocolappendix]{emulateapj}
\usepackage{apjfonts}
\usepackage{natbib}
\usepackage{graphics}
\usepackage{multirow}
\usepackage{amsbsy}
\usepackage{mathrsfs}
\usepackage[normalem]{ulem}
\usepackage{enumitem}

\def\bicep{{\sc Bicep}}

\def\bicepone{{\sc Bicep1}}

\def\spider{{\sc Spider}}

\def\keck{\textit{Keck Array}}
\def\synfast{\texttt{synfast}}
\def\anafast{\texttt{anafast}}

\def\anafast{{\tt anafast}}
\def\synfast{{\tt synfast}}

\def\deg{^\circ}
\def\emode{$E$-mode}
\def\bmode{$B$-mode}
\def\BB{$BB$}

\def\lcdm{$\Lambda$CDM}

\def\ttp{\textit{T}$\rightarrow$\textit{P}}
\def\etb{\textit{E}$\rightarrow$\textit{B}}

\begin{document}

\title{\bicep2 III: Instrumental Systematics}

\author{BICEP2 Collaboration - P.~A.~R.~Ade\altaffilmark{1}}
\author{R.~W.~Aikin\altaffilmark{2}}
\author{D.~Barkats\altaffilmark{3}}
\author{S.~J.~Benton\altaffilmark{4}}
\author{C.~A.~Bischoff\altaffilmark{5}}
\author{J.~J.~Bock\altaffilmark{2,6}}
\author{J.~A.~Brevik\altaffilmark{2}}
\author{I.~Buder\altaffilmark{5}}
\author{E.~Bullock\altaffilmark{7}}
\author{C.~D.~Dowell\altaffilmark{6}}
\author{L.~Duband\altaffilmark{8}}
\author{J.~P.~Filippini\altaffilmark{2,9}}
\author{S.~Fliescher\altaffilmark{10}}
\author{S.~R.~Golwala\altaffilmark{2}}
\author{M.~Halpern\altaffilmark{11}}
\author{M.~Hasselfield\altaffilmark{11}}
\author{S.~R.~Hildebrandt\altaffilmark{2,6}}
\author{G.~C.~Hilton\altaffilmark{12}}
\author{K.~D.~Irwin\altaffilmark{13,14,12}}
\author{K.~S.~Karkare\altaffilmark{5}}
\author{J.~P.~Kaufman\altaffilmark{15}}
\author{B.~G.~Keating\altaffilmark{15}}
\author{S.~A.~Kernasovskiy\altaffilmark{13}}
\author{J.~M.~Kovac\altaffilmark{5}}
\author{C.~L.~Kuo\altaffilmark{13,14}}
\author{E.~M.~Leitch\altaffilmark{16}}
\author{M.~Lueker\altaffilmark{2}}
\author{C.~B.~Netterfield\altaffilmark{4}}
\author{H.~T.~Nguyen\altaffilmark{6}}
\author{R.~O'Brient\altaffilmark{6}}
\author{R.~W.~Ogburn~IV\altaffilmark{13,14}}
\author{A.~Orlando\altaffilmark{15}}
\author{C.~Pryke\altaffilmark{10}}
\author{S.~Richter\altaffilmark{5}}
\author{R.~Schwarz\altaffilmark{10}}
\author{C.~D.~Sheehy\altaffilmark{16,10,XX}}
\author{Z.~K.~Staniszewski\altaffilmark{2}}
\author{R.~V.~Sudiwala\altaffilmark{1}}
\author{G.~P.~Teply\altaffilmark{2}}
\author{J.~E.~Tolan\altaffilmark{13}}
\author{A.~D.~Turner\altaffilmark{6}}
\author{A.~G.~Vieregg\altaffilmark{17,16}}
\author{C.~L.~Wong\altaffilmark{5}}
\author{K.~W.~Yoon\altaffilmark{13,14}}

\altaffiltext{1}{School of Physics and Astronomy, Cardiff University, Cardiff, CF24 3AA, UK}
\altaffiltext{2}{Department of Physics, California Institute of Technology, Pasadena, CA 91125, USA}
\altaffiltext{3}{Joint ALMA Observatory, ESO, Santiago, Chile}
\altaffiltext{4}{Department of Physics, University of Toronto, Toronto, ON, Canada}
\altaffiltext{5}{Harvard-Smithsonian Center for Astrophysics, 60 Garden Street MS 42, Cambridge, MA 02138, USA}
\altaffiltext{6}{Jet Propulsion Laboratory, Pasadena, CA 91109, USA}
\altaffiltext{7}{Minnesota Institute for Astrophysics, University of Minnesota, Minneapolis, MN 55455, USA}
\altaffiltext{8}{SBT, Commissariat \`a l'Energie Atomique, Grenoble, France}
\altaffiltext{9}{Department of Physics, University of Illinois at Urbana-Champaign, Urbana, IL 61820 USA}
\altaffiltext{10}{Department of Physics, University of Minnesota, Minneapolis, MN 55455, USA}
\altaffiltext{11}{Department of Physics and Astronomy, University of British Columbia, Vancouver, BC, Canada}
\altaffiltext{12}{National Institute of Standards and Technology, Boulder, CO 80305, USA}
\altaffiltext{13}{Department of Physics, Stanford University, Stanford, CA 94305, USA}
\altaffiltext{14}{Kavli Institute for Particle Astrophysics and Cosmology, SLAC National Accelerator Laboratory, 2575 Sand Hill Rd, Menlo Park, CA 94025, USA}
\altaffiltext{15}{Department of Physics, University of California at San Diego, La Jolla, CA 92093, USA}
\altaffiltext{16}{Kavli Institute for Cosmological Physics, University of Chicago, Chicago, IL 60637, USA}
\altaffiltext{17}{Department of Physics, Enrico Fermi Institute, University of Chicago, Chicago, IL 60637, USA}
\altaffiltext{XX}{Corresponding author: csheehy@uchicago.edu}

\begin{abstract}

In a companion paper, we have reported a $>5\sigma$ detection of degree scale $B
$-mode polarization at 150~GHz by the \bicep2\ experiment. Here we provide a
detailed study of potential instrumental systematic 
contamination to that measurement.
We focus extensively on spurious polarization 
that can potentially arise from beam imperfections. 
We present a heuristic classification of beam imperfections according to their
symmetries and uniformities, and discuss how resulting contamination
adds or cancels in maps that combine observations made at multiple orientations
of the telescope about its boresight axis.  
We introduce a technique, which we call ``deprojection,''
for filtering the leading order beam-induced
contamination from time-ordered data,
and show that it reduces power in \bicep2's actual and null-test $BB$ spectra
consistent with predictions using high signal-to-noise beam shape measurements.
We detail the simulation pipeline that we use to directly simulate instrumental
systematics and the calibration data used as input to that pipeline. Finally, we
present the constraints on $BB$ contamination from individual sources of
potential systematics. We find that systematics contribute $BB$ power that is a
factor $\sim10\times$ below \bicep2's three-year statistical uncertainty, and
negligible compared to the observed $BB$ signal. The contribution to the
best-fit tensor/scalar ratio is at a level equivalent to $r=(3-6)\times10^{-3}$.

\end{abstract}


\section{Introduction}

Since the the discovery of the $2.7$~K cosmic microwave background (CMB) by
\citet{penzias65}, rapid progress in instrumental sensitivity has permitted the
detection of progressively subtler effects. The $\sim100~\mu$K temperature
anisotropies, measured to high precision by the \textit{WMAP} and \textit{Planck} satellites
\citep{wmap9params,planckxvi} and by ground-based telescopes
\citep{sievers13,story13,das13,hou14}, are $\sim10^{-5}$ fluctuations in the
2.7~K background. The degree scale primary CMB temperature anisotropies are polarized at the
$\sim1\%$ level \citep{kovac02}, with fluctuations of the order of $1~\mu$K. This
polarization, which arises as a natural consequence of the same acoustic
oscillations that source the temperature anisotropies \citep{bond84}, is
curl-free (\emode) and its angular power spectrum is uniquely predicted given the temperature ($T$)
spectrum with the addition of no additional cosmological parameters. The
agreement of the \emode\ spectrum with the predictions given the best fitting $T$
spectrum is a striking, independent confirmation of \lcdm, modern cosmology's
basic paradigm \citep{pryke09,quiet12,barkats13,crites14,actpol14}.

Fainter still is the divergence-free (\bmode) polarization of the CMB that would
be caused by gravitational waves present in the Universe at the time of
recombination \citep{polnarev85,kamikosostebbins97,seljak97a,seljak97b}. Because the production of a
stochastic background of gravitational waves is a generic prediction of
inflationary models
\citep{grishchuk75,starobinskii79,rubakov82,fabbri83,abbott84}, the detection of the
cosmological \bmode\ polarization would constitute direct evidence for an era of
cosmic inflation. The amplitude of the cosmological \bmode\ spectrum is
parametrized by the tensor/scalar ratio $r$. An $r=0.1$ \bmode\ signal has
degree scale fluctuations of the order of $100~$nK, a factor $10$ smaller than the
\emode\ anisotropy, a factor $10^3$ smaller than the unpolarized anisotropy, and
a factor $10^8$ smaller than the CMB monopole.

Measuring CMB polarization anisotropy is made difficult by its weakness relative
to the unpolarized anisotropy and by the additional sources of systematic error
specific to polarization measurements. Effects that convert CMB temperature
anisotropy into a false polarization signal are of particular importance. This
is especially true for \bmode\ measurements because both the temperature and the
expected inflationary \bmode\ spectra peak at similar angular scales. Detecting
and characterizing a \bmode\ polarization signal of this magnitude requires controlling
systematics to a level to match the experiment's unprecedentedly low
instrumental noise.

In \citet{biceptwoI}, hereafter the \textit{Results Paper}, we present a
detection of \bmode\ power in $>5\sigma$ excess over
the lensed-\lcdm\ CMB expectation.
In this paper, we present extensive studies of possible systematic
contamination in this measurement using
detailed calibration data that allow us to
directly predict or place stringent upper limits on it.  We find
that systematics contribute power at a level subdominant to \bicep2's
statistical noise and negligible compared to the measured \bmode\ spectrum.

The structure of this paper is as follows. In Section~\ref{sec:instrument} we briefly
review the aspects of the \bicep2\ instrument that are most important for an
understanding of potential systematic contamination.
In Section~\ref{sec:noisemodel} we review the noise estimation procedure and show
that our debiased auto spectrum procedure is equivalent to a cross
spectral analysis.
In Section~\ref{sec:cancellation}
we review how \bicep2's specific observing strategy modulates the contamination
from beam systematics in the signal maps and in our internal consistency
checks.
In Section~\ref{sec:deprojection} we introduce the deprojection algorithm we use
to mitigate contamination from beam imperfections.
In Section~\ref{sec:beammeas} we review external beam shape measurements.
In Section~\ref{sec:simpipeline} we detail the simulation
pipeline used to predict the level of spurious polarization due to
imperfect beam shapes.
In Section~\ref{sec:jackknives} we review \bicep2's
``jackknife'' internal consistency null tests and discuss the classes of systematics to
which each is sensitive.
In Section~\ref{sec:deprojperformance} we check that deprojection of CMB
data does indeed recover the known beam non-idealities within
uncertainties, even in the presence of realistic template noise.
In Section~\ref{sec:syslevels} we present the constraints on
many potential sources of systematic contamination.
We conclude in Section~\ref{sec:conclusions}.
In a series of four appendices we provide the formal
definition of our elliptical Gaussian beam parametrization
(Appendix~\ref{sec:beamparam}), an expanded discussion of beam 
shape mismatch (Appendix~\ref{sec:beamheuristic}), the mathematical and
practical details of deprojection (Appendix~\ref{sec:deprojmath}), and a
discussion of the uncertainties in the beam mismatch simulations
(Appendix~\ref{sec:beammapsimappendix}).

\section{Instrument design and observational strategy}
\label{sec:instrument}

The \bicep2\ instrument is discussed in depth in \citet{biceptwoII}, hereafter
the \textit{Instrument Paper}. Here we highlight the details most relevant to
systematics, and in particular those that can cause false polarization. In this section we describe how
effects can arise in the antennas (beam shape and pointing), in the bolometers
(thermal mismatch), or in the readout (crosstalk).  We also describe several
aspects of the observing strategy that serve to suppress these systematics
and/or to aid in identifying them.

\subsection{Instrument Design}
\label{sec:instdesign}

Each camera ``pixel'' in \bicep2's focal plane consists of two orthogonally
polarized beam-forming antennas \citep{obrient12,biceptwoV} that couple
incoming radiation to two bolometric detectors (each antenna is coupled to its
own detector).  We label the members of an antenna/detector pair (which we refer
to simply as a ``detector pair'') ``A'' and ``B.''  The A and B antennas within
a pair are spatially coincident in the focal plane so they nominally observe the
same location on the sky. The time-ordered data, or ``timestreams,'' from the A
and B detectors are summed to measure the total intensity of the incoming
radiation and differenced to measure its polarized component.  Therefore, any
mechanism other than the intrinsic polarization of the sky signal that produces
a differential signal in the A and B detectors will produce spurious
polarization if not properly accounted for.

The response of an antenna to incoming radiation as a function of angle is called
its beam. One class of systematics that can cause a false polarization is a
difference in the beam shape or beam center (``centroid') of the A and B
detectors. Beam shape imperfections or centroid offsets that are common to A
and B do not cause a false polarization. We observe that \bicep2's beams
exhibit significant systematic centroid mismatch within a pair, which we call
``differential pointing,'' and which we have precisely characterized.

In the time-reversed sense, each antenna illuminates the telescope aperture with
a nearly Gaussian pattern \citep{kuo08}.  The illumination pattern (i.e. the
``near-field beam'') is truncated on a $26.4$~cm cold aperture stop. The
asymmetric truncation of the near-field beams will induce an expected far-field
beam asymmetry.  We observe an expected dependence of detectors' beam
ellipticity on the radial position in the focal plane. Because we treat beam shapes
and centroids fully empirically, a precise understanding of the mechanisms
governing them is not required for assessing systematic contamination.  A brief
review of the parametrization and measurements of \bicep2's beams is given
in Section~\ref{sec:ellipparam} and Section~\ref{sec:beammeas}, respectively.  A fully
detailed treatment is given in \citet{biceptwoIV}, hereafter the \textit{Beams
  Paper}.

We have designed the telescope shielding system and our observation strategy to
mitigate contamination from the ground and the Galaxy. A co-moving forebaffle
and fixed ground shield ensure that at the lowest observing elevation rays
originating from the ground must diffract twice before entering the telescope
aperture. The brightest parts of the Galaxy are always well outside of the angle
intersected by the co-moving forebaffle.  The lowest galactic latitude of the
observations is $b=-39\deg$, and we have measured that for a typical detector
$<0.1\%$ of the total integrated power is found outside of $25\deg$ from the
main beam with the co-moving forebaffle installed. Details are in the Beams
Paper.

\bicep2's bolometers are transition edge sensors (TESs). We measure the amount
of incident radiation by tracking, as a function of time, the
amount of electrical power (presumed to be in addition to the
radiative power) required to maintain the TES at a fixed point in the
superconducting/normal transition. Thermal drifts in the focal plane thus
produce spurious signals in the detector timestreams. A false
\textit{polarization} signal arises if the responses of the A and B bolometers to
these thermal fluctuations are different.  We mitigate thermal drift using
a combination of passive thermal filters and active thermal control
\citep{kaufmanthesis}. We then continuously measure any remaining thermal
fluctuations to high precision using neutron transmutation doped (NTD) germanium
thermistors located on the focal plane, allowing us to directly constrain
spurious signal from thermal drift (see Section~\ref{sec:thermalinstability}).

The bolometers are read out using multiplexed superconducting quantum
interference devices (SQUIDs) \citep{irwin02}. The use of SQUID readouts
introduces susceptibility to pickup from magnetic fields.  \bicep2\ employs a
combination of high magnetic permeability and superconducting shielding to block
external magnetic fields, and its scan strategy allows for nearly perfect
filtering (``ground subtraction'') of pickup that is constant in time and a
function of telescope pointing direction, as is expected of most magnetic
fields. The multiplexing of detector timestreams \citep{dekorte03} creates
crosstalk between channels in the cryogenic and room temperature readout
hardware. Crosstalk, which we have measured in a variety of ways, can also
produce false polarization.

Using calibration data we make detailed calculations of the impact of the above
effects in Section~\ref{sec:syslevels} below.

\subsection{Observational Strategy and Data Cuts}
\label{sec:obsstrat}

The \bicep2\ telescope was situated on an azimuth/elevation mount that performed
constant elevation scans at a fixed azimuth center. The scans spanned just over
$60\deg$ in azimuth and were re-centered on a new azimuth at approximately one
hour intervals, during which time the sky moved in azimuth by $15\deg$.  Because the sky
changed position with respect to the scan boundaries, we can differentiate
between signals that are scan synchronous (ground-fixed signal), and signals that
rotate with the sky.  By subtracting the mean of all scans from each scan we
exactly remove any contaminating signal that is a function of scan position and
is constant over hour-long timescales.  We refer to this filtering as ``ground
subtraction.'' This method was used successfully by \bicep1
\citep{chiang10,barkats13} and by the QUIET experiment \citep{quiet12}.

The \bicep2\ mount also allowed for a third axis of motion, the rotation of the
entire telescope about the boresight. \bicep2\ observed at four distinct
boresight orientations, or ``deck angles'': $68\deg$, $113\deg$, $248\deg$, and
$293\deg$.\footnotemark[1]\footnotetext[1]{The Instrument Paper notes that different deck angles were
  used early in the 2010 season. Given their low weights in the final data set,
  however, they are largely irrelevant for the present analysis.}  (At $0\deg$,
the rows of \bicep2's focal plane were roughly perpendicular to the horizon.)
Because \bicep2's detector polarization angles were all aligned in the focal
plane, reconstructing maps of Stokes $Q$ and $U$ requires a minimum of two deck
angles, optimally separated by $45\deg$, $135\deg$, or $225\deg$. A valid deck
angle pair cannot be separated by $180\deg$.
With \bicep2's
four deck angles, a map formed from one valid deck angle pair (e.g. $68\deg$
and $113\deg$) is complementary to the map made from the other deck angle pair
(e.g. $248\deg$ and $293\deg$). The deck angle pair that is complementary to any
of the four valid pairs is rotated $180\deg$ from it.

We guard against systematics arising from unusually functioning detectors by
removing them during map making.  The map making process uses data from only a
subset of the nominally functioning (i.e. optically responsive) detectors. We
implement a series of channel cuts that exclude detector pairs having certain
properties outside a pre-defined range. The details are discussed
in Section~ 13.7 of the Instrument Paper. When we have \textit{a
  priori} reason to believe that a systematic will contaminate a few detectors
much more strongly than others, we can also perform a detector pair exclusion
test in which we remake maps cutting the most contamination-prone pairs.
For the test to be considered passed, we
require that the change in the resulting maps and power spectra is consistent
with the corresponding changes in systematics-free simulations.

\subsection{Summary}
\label{sec:instrsummary}

We address systematics using a combination of five general strategies.  Three
strategies reduce contamination in the final maps.

\begin{enumerate}
\item \textit{Natural mitigation}: \bicep2's maps are built up from observations
  made with many detectors. A systematic that varies between detector pairs will
  thus statistically average down in the final map. \bicep2's maps are also built
  from observations at four deck angles.  Some systematics cancel with
  instrument rotation.  This is discussed further in Section~\ref{sec:cancellation}.
\item \textit{Time-domain filtering}: We remove atmospheric $1/f$ noise by applying
  a third-order polynomial filter to the timestreams. Atmospheric noise is not a
  systematic because it averages down over time and is accounted for in the noise model,
  but such a filter also removes any large angular scale contamination
  that might not average down. In addition, we also exactly remove any
  remaining signal that is fixed with respect to the ground or scan 
  (as opposed to the sky) by applying the ground subtraction filter discussed
  in Section~\ref{sec:obsstrat}.
\item \textit{Deprojection}: We also filter out the map modes most contaminated by
  beam imperfections.
  If they are ignored, differences in beam shape between the two detectors of
  a detector pair will transform bright temperature anisotropies into false
  polarization anisotropies.  We have developed a technique
  to explicitly filter the handful of map modes contaminated by several major
  types of beam mismatch, and to account for this removal in power spectrum
  estimation. This technique is described in Section~\ref{sec:deprojection} and in 
  Appendices A-D.
\end{enumerate}

\noindent Two strategies characterize the level of contamination remaining in the maps.

\begin{enumerate}
\item \textit{Jackknife maps}: Many classes of systematics produce different
  contamination in different subsets of data. As part of our internal
  consistency checks, we split \bicep2's data set into two halves, form $Q$ and
  $U$ maps from each of the halves, difference these maps, and test whether the
  resulting residuals are consistent with the difference of systematics-free,
  signal-plus-noise simulations.  We refer to these null tests as ``jackknives,'' and they
  are discussed in more detail in Section~\ref{sec:jackknives}. We refer to the
  un-differenced maps, made with the full data set, from which the science
  analysis in the Results Paper derives, as the ``signal'' maps. We refer to the
  angular power spectra of those maps as the signal spectra.
\item \textit{Time-domain simulations}: Our analysis pipeline generates simulated
  realizations of time-ordered data (signal and noise) for each detector, which
  is then processed in exactly the same manner as our real data.  We have
  extended our pipeline to optionally incorporate the effects of various
  instrumental systematics into these simulated data, which allows us to model
  their effects on the final power spectra and $r$ estimate.  This pipeline is
  described in Section~\ref{sec:simpipeline}, with particular regard given to
  simulating beam mismatch effects. Measurements of beam mismatch are presented
  in Section~\ref{sec:beammeas}. The results of these studies are presented
  in Section~\ref{sec:syslevels}.
\end{enumerate}

Generally speaking, time-domain simulations allow us to model the consequences
of known systematic effects.  Jackknife maps are useful for empirically
constraining contamination from both known and unknown systematics.

\section{Noise estimation}
\label{sec:noisemodel}

The Results Paper describes the construction of ``noise pseudosimulations'' that
we use to estimate the noise bias and uncertainty of our measured auto
spectrum. We construct these pseudosimulations by differencing the two maps made
from two halves of a random permutation of $17,000$ temporal subsets of the full
data set, which are long enough (approximately 1h each) to have minimal noise correlations.  
We impose a constraint that each half have the same total weight. Jackknife noise
pseudosimulations are similarly constructed by randomly permuting the subsets
within a jackknife half and differencing the two maps in each half separately.
As described in the Results Paper, this noise estimation procedure has been
checked against two alternative techniques and all are found to yield equivalent results.

More formally, the $j$th random permutation splits the full data set to
define two half maps $M_{1j},M_{2j}$, which can be 
recombined by summing or differencing:
\begin{eqnarray}
M & = & {\textstyle\frac{1}{2}}(M_{1j}+M_{2j}) \nonumber \\
N_{j} & = & {\textstyle\frac{1}{2}}(M_{1j}-M_{2j}) .
\end{eqnarray}

\noindent $M$ is our standard full map and is the same for any split,
while $N_j$ is the noise realization.  The auto spectra of
these two maps can be written
\begin{eqnarray}
M \!\! \times \!\! M & = & {\textstyle\frac{1}{4}}
[M_{1j} \!\! \times \!\! M_{1j} + 2(M_{1j} \!\! \times \!\! M_{2j}) 
+ M_{2j} \!\! \times \!\! M_{2j}] \\
N_j \!\! \times \!\! N_j & = & {\textstyle\frac{1}{4}}
[M_{1j} \!\! \times \!\! M_{1j} - 2(M_{1j} \!\! \times \!\! M_{2j}) 
+ M_{2j} \!\! \times \!\! M_{2j}] .
\end{eqnarray}

\noindent Subtracting these gives
\begin{equation}
M \!\! \times \!\! M - N_j \!\! \times \!\! N_j = M_{1j} \!\! \times \!\! M_{2j} .
\end{equation}

\noindent We see that subtracting the auto spectrum of a single noise pseudosimulation
 $N_j$ from that of the full map is identical to taking the cross-spectrum of the
two corresponding half maps.

Our actual noise bias and uncertainty estimation uses an ensemble of $N\sim500$
noise pseudosimulations.  We noise debias the auto spectrum of the full map by
subtracting the mean of the auto spectra of the noise realizations,
\begin{equation}
M \!\! \times \!\! M - \left< N_j \!\! \times \!\! N_j \right> =
\left< M \!\! \times \!\! M - N_j \!\! \times \!\! N_j \right> =
\left< M_{1j} \!\! \times \!\! M_{2j} \right> .
\end{equation}

\noindent where brackets represent mean over the $j=1...N$ realizations of the ensemble.
This shows that our debiasing procedure is equivalent to computing the 
mean of cross-spectra between data subsets for a large number of splits.
Similarly, 
the higher order statistics (variance, skewness, etc.) of the noise
pseudosimulations are mathematically equivalent to the higher order statistics
of the cross-spectra formed between the data subsets. 
One can go on to demonstrate that our procedure is also equivalent to taking
the mean of cross-spectra between many smaller data split
chunks~\citep{fowler10,lueker10,story13}.
As in any such cross-spectrum
analysis, in the limit of uncorrelated noise
between data subsets, there can be no residual noise bias from incorrect noise
modeling, as our ``noise model'' is in fact not a model, but rather a linear
combination of the data themselves.

The main effect that could possibly correlate noise among data subsets is
anisotropic turbulent structure in the atmosphere. The spatial structure of the
turbulence above the telescope averages down over time but persists on
timescales of the order or the height of the turbulent layer divided by the wind speed
at that altitude. (The timescale only becomes shorter if the turbulent structure
is not assumed to be ``frozen in'' in the frame of the moving atmosphere but
instead also evolves in time.)  For a height of $5$~km and a wind speed of
$5$~m~s$^{-1}$, the timescale is $\sim15$~minutes. The data subsets we use are
approximately $1$~hr in duration, so even in the unpolarized pair sum
timestreams, the noise properties of which are dominated by turbulent atmospheric
emission, we expect very little noise correlation between data
subsets. Furthermore, because the atmosphere is almost totally unpolarized,
pair-differencing of detector pairs almost completely eliminates the noise due
to 
atmospheric turbulence, leaving only the white noise of random photon arrival
times. The cancellation of unpolarized atmospheric turbulent emission is
apparent in Figure 22 of the Instrument Paper, which shows that the
instantaneous temporal power spectrum of the unfiltered pair-difference
timestreams is dominated by white noise, with a possible contribution
from atmospheric turbulence at most a few percent at the lowest frequencies.

Lastly, any remaining polarization noise correlations are further suppressed by
the time-domain filtering described in Section~\ref{sec:instrsummary}, which downweights
the lowest frequency Fourier modes along the scan direction. These are the modes
with the highest fractional contribution of atmospheric turbulence to the total
noise.

\section{Beam systematics in maps}
\label{sec:cancellation}

\begin{deluxetable*}{lccc}
\tablecolumns{4} \tablewidth{0pc} \tablecaption{Transformation of beam mismatch
  leakage under rotation\label{tab:cancellation}} \tablehead{\colhead{Rotation}
  & \colhead{\shortstack{Monopole \\ (e.g.\ Diff. Gain, Beamwidth)}} &
  \colhead{\shortstack{Dipole \\ (e.g.\ Centroid Offset)}} &
  \colhead{\shortstack{Quadrupole \\ (e.g.\ Diff. Ellipticity)}}} \startdata $45\deg$
& $E \rightarrow B$, $B \rightarrow E$ & $E \rightarrow (E+E')/\sqrt{2}$,
$B \rightarrow (B+B')/\sqrt{2}$ & $E \rightarrow E$, $B \rightarrow B$ \\ $90\deg$ &
$E \rightarrow -E$, $B \rightarrow -B$ & $E \rightarrow E'$, $B \rightarrow B'$ &
$E \rightarrow E$, $B \rightarrow B$ \\ $180\deg$ & $E \rightarrow E$, $B \rightarrow B$
& $E \rightarrow -E$, $B \rightarrow -B$ & $E \rightarrow E$, $B \rightarrow B$ \enddata
\tablecomments{In a map formed by a detector pair at one set of projected orientations
  on the sky, this table summarizes how the spurious signal from beam mismatch
  of the given symmetry is transformed in a second map made from the same
  detector pair at a second set of orientations rotated from the first by the given
  angle.}
\end{deluxetable*}

We refer to any differential response to incoming unpolarized radiation between
the A and B members of a detector pair as ``beam mismatch.'' In the presence of
beam mismatch, the pair-difference signal will, in general, be non-zero even
when observing an unpolarized source. This signal directly enters polarization
maps and so must be filtered out or otherwise accounted for.
One can think of such potential contamination as the
unpolarized temperature field ``leaking'' into the pair-difference signal of a
given detector pair. We refer to this as temperature-to-polarization (\ttp)
leakage.  At high galactic latitude at $150$~GHz, CMB $T$ is much brighter than
foregrounds and is the dominant unpolarized signal sourcing \ttp\ leakage.

The leaked signal, $d$, that enters the pair-difference data of a given
detector pair is the convolution of the unpolarized sky with the difference of
the pair's A and B beams,

\begin{eqnarray}\label{eq:diff}
d_{T \rightarrow P} & = & T(\mathbf{\hat{n}}) \ast \left[
  B_A(\mathbf{\hat{n}}) - B_B(\mathbf{\hat{n}}) \right] \nonumber \\ 
& \equiv & T(\mathbf{\hat{n}}) \ast B_{\delta}(\mathbf{\hat{n}})
\end{eqnarray}

\noindent where $T$ is the unpolarized temperature field, $B$ is the response of
a detector, and $\mathbf{\hat{n}}$ is the sky coordinate.  If the difference
beam, $B_{\delta}$, is non-axially symmetric, then $d_{T\rightarrow P}$ is a
function of both the pointing direction of the detector pair and the projected
orientation of $B_\delta$ on the sky.

Given measurements of $T(\mathbf{\hat{n}})$ and $B_{\delta}(\mathbf{\hat{n}})$,
Equation~\ref{eq:diff} is sufficient to predict the instantaneous \ttp\ leakage 
in a detector pair's pair-difference timestream as a function of that pair's
pointing direction. Predicting how this timestream level contamination manifests
in polarization maps requires
knowledge of the observing strategy. In principle, timestream level simulations of beam
mismatch that go all the way to final maps capture the map
level contamination without the need for any heuristic understanding.
Nonetheless, to gain confidence that these simulations accurately reflect
reality, it is helpful to build intuition about the way
in which different classes of beam mismatch interact with the observing strategy
to produce the map level contamination. The remainder of this section attempts
to develop this intuition.

We treat each detector pair's difference beam as the linear combination of different
components, or modes,

\begin{equation}\label{eq:modesum}
B_{\delta}(\mathbf{\hat{n}}) = \sum_{k} a_k B_{\delta k}(\mathbf{\hat{n}})
\end{equation}

\noindent Our map making procedure is a linear process. Thus,
the contamination in the final maps is a linear combination of the contamination
produced by each of these modes individually.  How each mode contaminates the
final map depends upon its {\it amplitude} $a_k$, its {\it coherence} across detector
pairs in the focal plane, and its {\it symmetry} under rotation of the
instrument with respect to the sky.  Amplitude sets the magnitude of the
systematic in time-ordered data, while coherence and symmetry determine the
degree of cancellation in maps made from multiple detectors and at multiple deck
angles.

\subsection{Incoherence Across the Focal Plane}
\label{sec:incor}

When combining data from multiple detector pairs to form a map, the
\ttp\ leakage from a difference beam mode that randomly varies among detector
pairs will average down if $\left< a_k \right> = 0$,
where the expectation value of the $k$th mode is over detector pairs. Since any
map pixel is only sampled by a finite number of detectors, the averaging
down is only partial. Nonetheless, because the contamination in maps made from
different subsets of detector pairs will be different, the jackknife tests
described in Section~\ref{sec:jackknives} that check for consistency between detector
pairs will fail. In general, jackknife maps have the same noise level as the
signal maps. Because a randomly varying beam systematic will contaminate the
signal map as much as a pair selection jackknife,
we expect pair selection jackknives to fail when the contamination in the signal map
is comparable to \bicep2's statistical uncertainty.

More worrisome are beam systematics that are correlated between detector pairs,
the leakage from which does not necessarily average down and can potentially
evade jackknives. \bicep2's many pair selection jackknives test for consistency
between subsets of detectors whose beam mismatch is expected to be different for
various mechanisms, e.g.\ varying by position in the focal plane or by multiplex
column.

\subsection{Symmetry}

A difference beam mode that is common to all
detector pairs (i.e. fully coherent across the focal plane) will not produce any
contamination of pair selection jackknives and will not average down when
combining data from detector pairs. However, under an azimuthal rotation of the
beam about its center, the leakage from modes of certain symmetries will change
sign.  When combining data from detectors at different projected orientations on
the sky, the leakage from even fully coherent mismatch will sometimes nearly
exactly cancel in the signal maps \citep{odea07,shimon08,quiet11}. Whether or
not this occurs depends on the azimuthal symmetry of the mode.  \bicep2 heavily
exploits this cancellation effect by performing deck angle rotation.  When this
cancellation occurs in the signal maps, the contaminating signals in both halves
of the corresponding deck angle jackknife map are equal to each other but
opposite in sign, so that the jackknife experiences no such
cancellation. In this case, the deck angle jackknife will fail for levels of
contamination that are negligible in the signal map. Appropriate deck angle jackknifes are
thus highly sensitive probes of \ttp\ leakage from these beam systematics.

In analogy with the azimuthal symmetry of pure monopoles, dipoles, and
quadrupoles, we classify difference beam modes as having monopolar symmetry
(i.e. invariant under rotation, i.e. azimuthally symmetric), dipolar symmetry
(reversing sign under $180\deg$ rotation), or quadrupolar symmetry (reversing
sign under $90\deg$ rotation); other symmetries are possible for complex beam
shapes, but are not modeled here.  Table~\ref{tab:cancellation} summarizes how
$d_{T \rightarrow P}$ from these modes is reconstructed as a false polarization
signal in a polarization map depending on the mode's projected orientation on the sky.
The reconstructed leakage from a monopole symmetric mode changes sign under a
$90\deg$ rotation. (Thus, leakage to $+E$ and $+B$ at one orientation leaks to
$-E$ and $-B$ at the second; adding these two maps results in cancellation of
the leakage, and subtracting them to form a jackknife multiplies the
contamination by two.)
The reconstructed leakage from a dipole symmetric mode changes sign under a
$180\deg$ rotation.  The reconstructed leakage from a quadrupole symmetric mode
is invariant under rotation.

In a given map pixel, the cancellation of \ttp\ leakage from a monopole or
dipole symmetric mode will occur if that pixel is sampled at appropriate
orientations by the same detector pair. If the pixel is sampled by different
detector pairs, then it is only the leakage from the common component that
cancels due to the rotation.  Full cancellation of the \ttp\ leakage from the monopole or dipole
symmetric modes thus requires that one of two corresponding criteria
be met: either (1) the sky coverage of any given detector pair is the same at
all deck angles, or (2) the contribution to any final map pixel is from detector
pairs with identical $a_k$.

\begin{figure}[t]
  \begin{center}
    \begin{tabular}{c}
      \includegraphics[width=1\columnwidth]{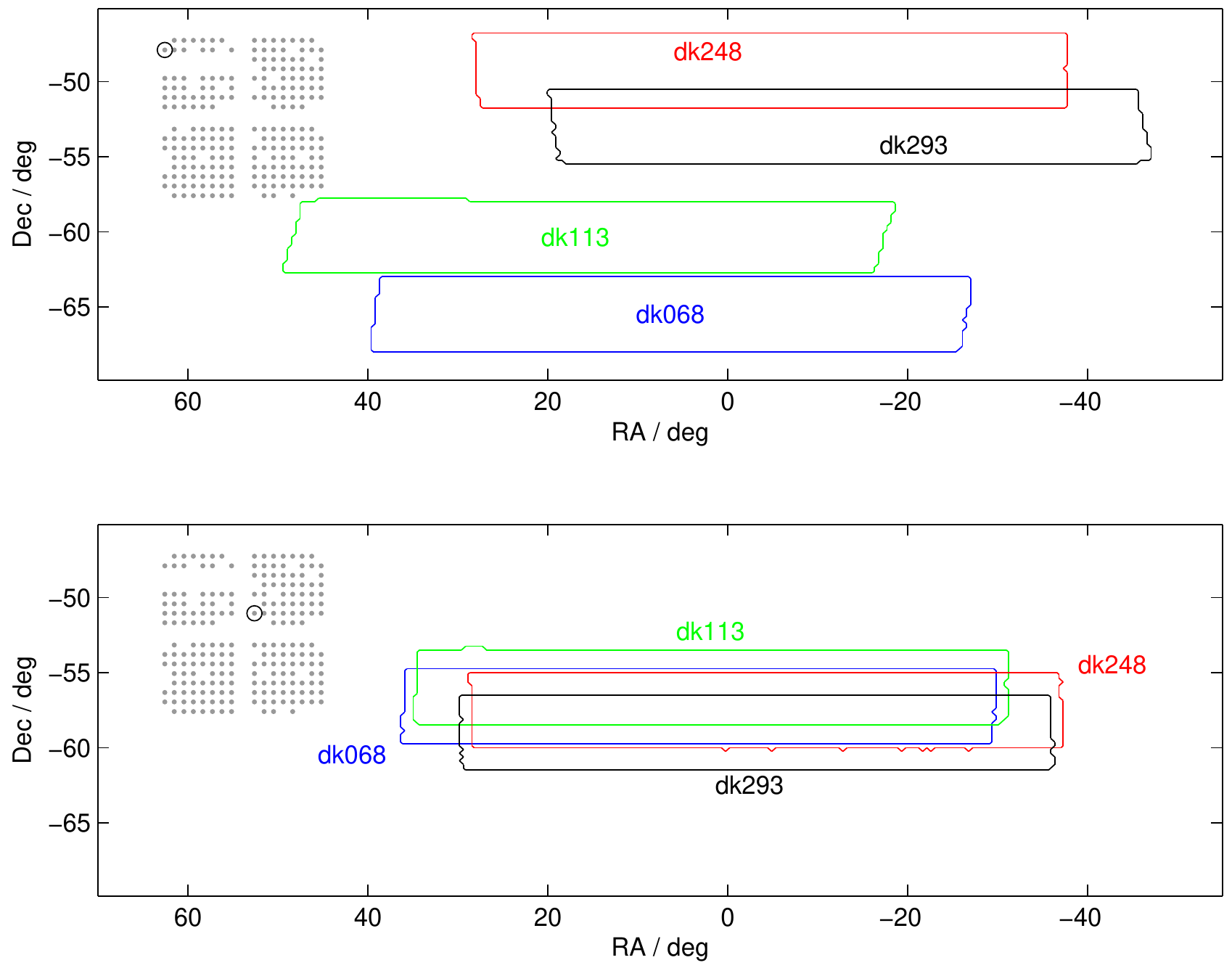}
    \end{tabular}
  \end{center}
  \caption[example] { \label{fig:perchmap} Map coverage of a single \bicep2 detector
    pair located (top panel) near the edge of the focal plane and (bottom
    panel) near the center of the focal plane. The coverage at different deck
    angles overlaps significantly for central detectors but not at all for edge
    detectors. The inset (not drawn to scale) indicates the location of the
    detector pair in the focal plane.}
\end{figure}

Boresight rotation, in addition to rotating a detector pair's beam, also changes
its pointing direction.  Because the instantaneous field of view of \bicep2's
focal plane is large compared to the overall map boundaries, the area of
sky mapped by a given detector is different at different deck angles.  This is
illustrated in Figure~\ref{fig:perchmap}, which shows the map regions sampled by
two detector pairs, one located near the center of the focal plane and one
located near the edge. The coverage of the detector pair located near the center
of the focal plane is largely the same at different deck angles; the coverage of
the detector pair located near the edge of the focal plane is very different at
different deck angles. As a consequence, detector pairs near the center of the
focal plane (and thus the central regions of the signal maps) satisfy criterion
(1) and experience highly efficient cancellation. Detector pairs near the edge
of the focal plane still experience cancellation, but only in so far as they
satisfy criterion (2).

The remainder of this section considers in more detail the cancellation of
leakage from difference beams of different symmetries.

\begin{deluxetable*}{lccc}[h]
\tablecolumns{4} \tablewidth{0pt}
\tablecaption{Summary of Beam Mismatch Leakage Effects\label{tab:cancelsummary}}
\tablehead{\colhead{Symmetry:} & \colhead{Monopole} & \colhead{Dipole} & \colhead{Quadrupole}}
\startdata
\textit{Incoherent across focal plane} &  &  &  \\
In signal map:          & Averages down & Averages down & Averages down \\
In pair selection jackknife: & Potentially contaminates & Potentially contaminates & Potentially contaminates\\
\\
\textit{Coherent across focal plane} &  &  & \\
In signal map:          & Cancels under $90\deg$ rot.\ & Cancels under $180\deg$ rot.\ & Does not cancel \\
In deck angle jackknife: & Contaminates in $90\deg$ jackknife & Contaminates in $180\deg$ jackknife & Does not contaminate \\
\enddata
\tablecomments{In a map formed by many detector pairs at multiple projected
focal plane orientations on the sky this table summarizes the
behavior of \ttp\ beam systematics having various symmetries.}
\end{deluxetable*}

\subsubsection{Monopole Symmetric Difference Beam}
\label{sec:monopole}

Examples of monopole symmetric difference beams are the difference of two
circular Gaussians with different peak heights or widths, as illustrated in the
upper and lower left panels of Figure~\ref{fig:differencebeams}. We focus on
these particular modes because the calibration measurements presented
in Section~\ref{sec:beammeas} indicate that they describe the majority of \bicep2's
monopole symmetric beam mismatch. However, we note
that the discussion here is generally applicable to \emph{any} monopole
symmetric difference beam.

If $d_{T \rightarrow P}$ for a detector pair pointed at some location on the sky
is from a monopole symmetric difference beam, it remains constant under rotation
of the difference beam. However, because the polarization sensitivity of the pair
(i.e. the interpretation of that signal under the assumption that it is not a
systematic and ``on the sky'') rotates as well, how $d_{T \rightarrow P}$ is reconstructed in the final
map does change.  If the leakage is reconstructed as a false polarization with
some magnitude and direction at one orientation, rotating the detector pair
$90\deg$ causes it to be reconstructed as false polarization with equal
magnitude but rotated $90\deg$ from the first. Rotating a polarization vector by
$90\deg$ simply transforms $+Q\rightarrow-Q$ and $+U\rightarrow-U$, so combining
the measurements cancels the \ttp\ leakage.

\bicep2's scan strategy did not cancel leakage from monopole symmetric
difference beams in this way. \bicep2's observation strategy included only
$180\deg$ deck angle pair complements and no $90\deg$ complements. In maps made
from deck angle pairs separated by $180\deg$, the \ttp\ leakage from monopole
symmetric difference beams is reconstructed as $Q$ and $U$ identically. This
leakage adds in the signal map and cancels in the deck jackknife, making the
deck jackknives \bicep2\ forms insensitive to this type of leakage.

\bicep2's successor experiment, the \keck\ \citep{sheehy10,kernasovskiy12},
consists of five \bicep2-like receivers with common boresight pointing and
oriented at $72\deg$ increments to one another.  This leads to an effective
fivefold increase in the number of deck angles and thus a certain degree of
cancellation of monopole symmetric beam mismatch in the final coadded map.
Monopole symmetric mismatch that is common between the focal planes of the two
experiments will thus be suppressed in cross-spectra taken between them.
Beginning in 2013, the \keck\ added the additional four $90\deg$ complementary
deck angles necessary to fully cancel leakage from coherent monopole symmetric
difference beams and to form deck jackknives that can test for it. Recently,
\citet{k15} demonstrated consistency between \bicep2 and \keck's auto
and cross spectra.

The predecessor experiment to \bicep2\ was \bicep1\ \citep{yoon06}. \bicep1 also
observed at the same deck angle intervals as \bicep2, but because the
polarization angles of \bicep1's detector pairs were not uniformly oriented in
the focal plane like \bicep2\ and \keck's, a monopole symmetric difference beam
common to \bicep1\ and \bicep2\ will also be suppressed in a cross-spectrum.

In summary, even though monopole symmetric beam mismatch does not contaminate
\bicep2's deck jackknives, it will (1) contaminate the \keck's $90\deg$ deck angle
jackknife, (2) contaminate \bicep1's pair selection jackknives, and (3) not
produce fully correlated power in cross-spectra formed between any of these
experiments. Lastly, we expect the deprojection technique described
in Section~\ref{sec:deprojection} to fully remove \ttp\ leakage from gain and
beamwidth mismatch, both of which have monopole symmetric difference beams. We
empirically test this last proposition via the beam map simulations described
in Section~\ref{sec:simpipeline}.

\subsubsection{Dipole Symmetric Difference Beam}
\label{sec:dipolecancel}

An example of a difference beam having dipolar symmetry is the difference of two
identical circular Gaussians with offset centroids, as illustrated in the top middle
and right panels of Figure~\ref{fig:differencebeams}. As discussed in
 Section~\ref{sec:beammeas}, this ``differential pointing'' is also \bicep2's dominant
source of \ttp\ leakage.

Dipole symmetric difference beam $d_{T\rightarrow P}$ changes sign under
a $180\deg$ rotation. Because the rotation of the detector polarization angles
is also $180\deg$, the reconstructed spurious polarization is equal in magnitude
and opposite in sign. Again, averaging the maps cancels the leakage; subtracting
the maps to form \bicep2's deck jackknife boosts the contamination by a factor
of two. \bicep2's set of deck angles does include $180\deg$ complements. The
high degree of cancellation in the signal map relative to the deck jackknife
makes the deck jackknife a powerful probe of dipole symmetric contamination. This is discussed in
more detail in Section~\ref{sec:deprojperformance}.

\subsubsection{Quadrupole Symmetric Difference Beam}
\label{sec:quadrupolecancel}

An example of a quadrupole symmetric difference beam is the difference of two
elliptical Gaussians with mismatched magnitudes and/or directions of
their elongations, and
is illustrated in the bottom middle and right panels of
Figure~\ref{fig:differencebeams}.
In this case it is the difference between the pair polarization sensitivity
angle and the orientation angle of the quadrupolar pattern
which determines the nature of the leakage --- $0\deg$ and $90\deg$ leak
$T\rightarrow \pm E$ while $\pm 45\deg$ leak $T\rightarrow \pm B$ \citep{shimon08}.

A quadrupole symmetric difference beam $d_{T \rightarrow P}$ changes sign under a
$90\deg$ rotation. This is the same periodicity as a real polarized
sky signal, so no amount of boresight rotation can distinguish it from real
polarization for a single pair.
As explained in Section~\ref{sec:incor}, leakage from incoherent
beam mismatch with any symmetry averages down over pairs in the signal map
and potentially contaminates pair selection jackknives.
Coherent quadrupolar mismatch
produces leakage that is indistinguishable from real sky polarization. No
possible jackknife can test for this. For this reason, coherent quadrupole
symmetric beam mismatch is especially pernicious and must be carefully
controlled. In Section~\ref{sec:beamsim}, we accurately simulate the real
beam mismatch and correctly predict the effects of
ellipticity mismatch in our data (this being the dominant quadrupole symmetric
component).

\subsection{Summary}

Table~\ref{tab:cancelsummary} summarizes the situation.  Any component of
$B_{\delta}(\mathbf{\hat{n}})$ that varies randomly across the focal plane(s)
averages down to at least some degree --- even for quadrupolar effects so long
as the orientations are random --- and in general we expect residual
contamination to be as strong in the jackknife maps as in the signal map.  For a
component of $B_{\delta}(\mathbf{\hat{n}})$ that is coherent across the focal
plane(s), whether or not there is cancellation in the signal map under instrument
rotation depends on the symmetry of the component, as does the jackknife split
required to expose the systematic.  A subtlety is the issue of whether each pair
self-cancels under instrument rotation.  This will be true in the limit that the
focal plane field of view is small compared to the size of the map, and becomes
less true as the field of view approaches the size of the map
(as is the case for \bicep2).

\section{Deprojection technique}
\label{sec:deprojection}

As introduced in Section~{IV.F} of the Results Paper, we have developed an analysis
technique, which we call ``deprojection,'' to filter out \ttp\ leakage from beam
mismatch (and potentially other effects). Such a filter renders our analysis
immune to contamination from leading order beam imperfections.
In this section, we describe the
technique as we have implemented it for the \bicep2 analysis.  Testing of the
performance of the algorithm in our case is deferred
to Section~\ref{sec:deprojperformance}.

\subsection{Beam Parametrization}
\label{sec:ellipparam}

We model $B_{\delta}(\mathbf{\hat{n}})$ as the difference of two elliptical
Gaussians.  In principle, we are free to choose any model with which to
parametrize and mitigate \ttp\ leakage, but the elliptical Gaussian
parametrization is convenient.

Six parameters define an elliptical Gaussian: one for peak height, two for the
center of the ellipse (centroid), one for width, and two specifying
ellipticity. The two parameters for ellipticity are often taken as a magnitude
and orientation. We choose an alternate but equivalent basis --- plus- and
cross-ellipticity, denoted $p$ and $c$ --- that describes an ellipse oriented
either vertically/horizontally or at $\pm45\deg$ to the horizontal axis. The
mathematical details of the parametrization are given in
Appendix~\ref{sec:beamparam}.

We model intra-pair gain mismatch (differential gain) as a difference in
Gaussian peak height; the difference beam mode for differential gain,
$B_{\delta g}(\mathbf{\hat{n}})$, is therefore just a circular
Gaussian.  We model differential pointing as a centroid offset
in an $x/y$ coordinate system fixed with respect to the focal plane and centered
on the nominal beam center; the corresponding difference beam modes,
$B_{\delta x}(\mathbf{\hat{n}})$ and $B_{\delta y}(\mathbf{\hat{n}})$, are the
differences of circular Gaussians offset in either the $x$ or $y$
direction. (\bicep2's beams are $\sim0.5\deg$~FWHM, so making
the flat sky approximation and parametrizing the ellipse on a Cartesian
coordinate system centered on each beam center is an adequate approximation.)
Beamwidth mismatch is parametrized by a difference in Gaussian width
$\sigma$. Differential plus- and cross-ellipticity are defined as the
differences of purely plus-elliptical or purely cross-elliptical Gaussians whose
orientations are defined with respect to the same focal plane fixed coordinate
system in which differential pointing is described. 

\begin{figure}
  \begin{center}
    \begin{tabular}{c}
      \includegraphics[width=1\columnwidth]{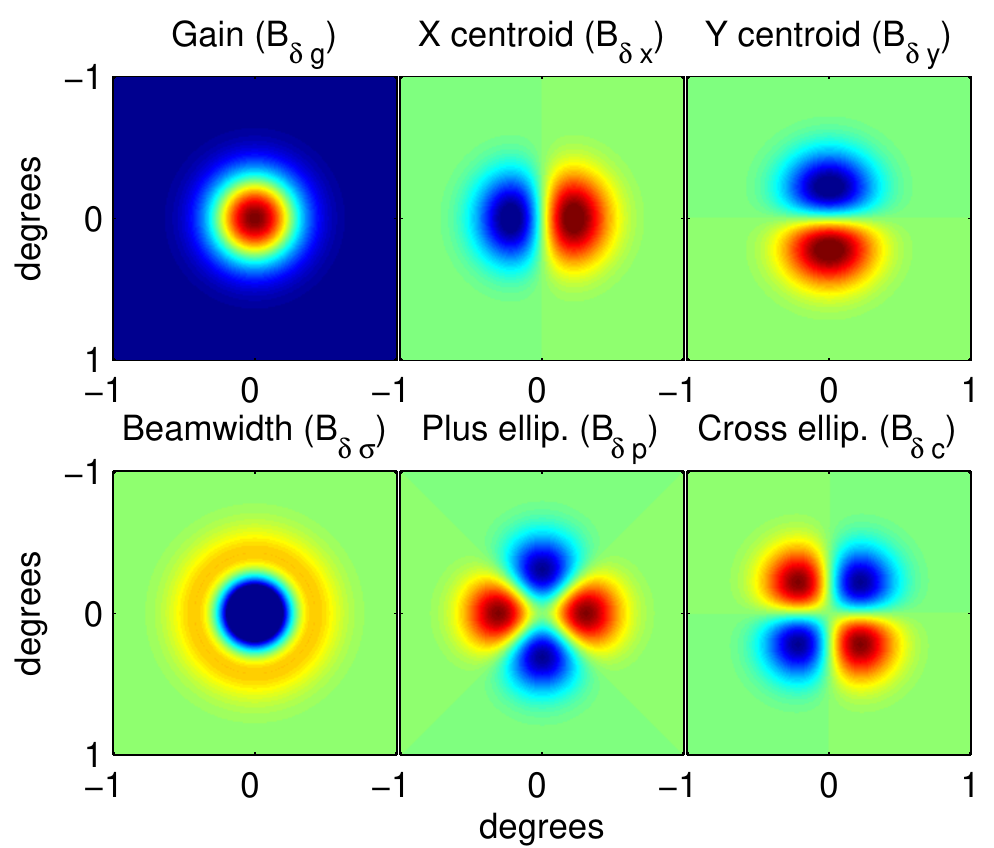}
    \end{tabular}
  \end{center}
  \caption[example] { \label{fig:differencebeams} Differences of elliptical
    Gaussian beams, which we choose for $B_{\delta k}(\mathbf{\hat{n}})$. The
    total difference beam, $B_{\delta}(\mathbf{\hat{n}})$, is a linear
    combination of these modes.  Differential gain and beamwidth produce
    monopole symmetric difference beams, differential pointing a dipole
    symmetric difference beam, and differential ellipticity a quadrupole
    symmetric difference beam.  These difference beams couple to different
    derivatives of the underlying CMB temperature field.  }
\end{figure}

Figure~\ref{fig:differencebeams} shows the differential elliptical Gaussian
modes.  We consider the total difference beam to be a linear combination of
these modes in isolation, so that the sum in Equation~\ref{eq:modesum} is over
$k=\{g,x,y,\sigma,p,c\}$.

\subsection{Algorithm}
\label{sec:deprojalgorithm}

Because the \ttp\ leakage from beam mismatch is deterministic and beam shapes
are constant in time, we can filter some of it out by
constructing leakage templates corresponding to the
differential modes of elliptical Gaussians, fitting them to our data, and subtracting
them. Such a method prevents contamination arising from the component of \bicep2's beams
described by elliptical Gaussians from entering the maps. It
requires no \textit{a priori}
knowledge of the actual magnitude of the mismatch
\citep{aikinthesis,sheehythesis}.

To second order, the individual modes of a differential elliptical Gaussian
couple to distinct linear combinations of $T(\mathbf{\hat{n}})$ and its first
and second derivatives \citep{hu03}. Appendix~\ref{sec:beamheuristic} provides a
heuristic description of this coupling.  Given maps of $T(\mathbf{\hat{n}})$ and
its spatial derivatives (which we refer to as the
``template maps'') and knowledge of the pointing of each of \bicep2's detector
pairs as a function of time (as required for map making), we sample the template
maps along each detector pair's pointing trajectory to create derivative
timestreams. We use the chain
rule for derivatives to express the derivatives with respect to the
\bicep2\ focal plane coordinate system as projected on the sky at each step in
the time series. The derivative timestreams are given by

\begin{equation}
d_{i,j}(t)=\nabla^i_j \tilde{T}(t)
\end{equation}
\noindent where the $i$th spatial derivative is defined with respect to the
focal plane coordinate $j=\{x,y\}$, 

\begin{equation}
\nabla^i_j  \equiv \frac{\partial^i}{\partial j^i},
\end{equation}
\noindent and the tilde denotes that the template map
has been pre-convolved by a circular Gaussian beam of nominal width, $\sigma$.

We then form the linear combinations of $ d_{i,j}(t)$ that correspond to leakage
from differential elliptical Gaussian modes. We call these linear combinations
the ``leakage templates'' and denote them $d_{\delta k}(t)$ for the $k$th mode.

The net leakage corresponding to mismatched elliptical Gaussians is then a
linear combination of the leakage templates,

\begin{equation}
d_{\delta}(t) = \sum_{k=g,x,y,\sigma,p,c} a_k d_{\delta k}(t).
\end{equation}

We fit the leakage templates to a detector pair's timestreams to obtain
$a_k$ and subtract the fitted templates to filter out the leakage. We also have the
option to directly measure differential beam parameters from external
calibration data, in which case we can fix $a_k$ at its measured value and
subtract scaled leakage templates to remove leakage.

Table~\ref{tab:deprojection} summarizes the proportionality between the fit
coefficients, $a_k$, and the differential beam parameters, $\delta k$, for the six modes of
our elliptical Gaussian beam parametrization.  Table~\ref{tab:deprojection} also
summarizes the linear combinations of $d_{i,j}(t)$ that
comprise the leakage templates, $d_{\delta k}(t)$.  The derivation of the leakage templates and a
discussion of the practical implementation of deprojection is given in
Appendix~\ref{sec:deprojmath}.

\begin{deluxetable}{lcccc}
\tablecolumns{5} \tablewidth{0pc} \tablecaption{Deprojection templates and fit
  coefficients \label{tab:deprojection}} \tablehead{\colhead{Differential Mode}
  & \colhead{Symbol} & \colhead{Definition} & \colhead{Fit Coefficient} &
  \colhead{Template}} \startdata Gain & $\delta g$ & $g_A-g_B$ & $\delta g$ &
$\tilde{T}$ \\ Pointing, x & $\delta x$ & $x_A-x_B$ & $\delta x $ &
$\nabla_x\tilde{T}$ \\ Pointing, y & $\delta y$ & $y_A-y_B$ & $\delta y $ &
$\nabla_y\tilde{T}$ \\ Beamwidth & $\delta\sigma$ & $\sigma_A-\sigma_B$ &
$\sigma\delta\sigma$ & $(\nabla^2_x+\nabla^2_y)\tilde{T}$ \\ Ellipticity, +
& $\delta p$ & $p_A-p_B$ & $(\sigma^2/2)\delta p$ &
$(\nabla^2_x-\nabla^2_y)\tilde{T}$ \\ Ellipticity, $\times$ & $\delta c$ &
$c_A-c_B$ & $(\sigma^2/2)\delta c$ & $2\nabla_x\nabla_y\tilde{T}$ \enddata
\tablecomments{A qualitative description of the coupling of elliptical Gaussian
  beam mismatch to the first and second spatial derivatives of the nominal beam
  convolved temperature field, $\tilde{T}$, is given in
  Appendix~\ref{sec:beamheuristic}. The formal derivations of the templates are
  given in Appendix~\ref{sec:deprojmath}}.
\end{deluxetable}

Like any filtering, deprojection removes non-leakage signal modes from the final
map, and thus affects the inferred power spectra.
In practice, only a tiny fraction of the $Q$ and $U$ maps are removed.
However, along with timestream filtering and sky cut effects,
deprojection does cause relevant mixing of $E$ into $B$.
This can be corrected for in the mean using simulations, but instead
we remove the contaminated spatial modes from the map using the
``matrix purification'' method described in Section~{VI.B} of the Results Paper.

\section{``External'' beam measurements}
\label{sec:beammeas}

\begin{figure}[t]
\begin{center}
\includegraphics[width=1\columnwidth]{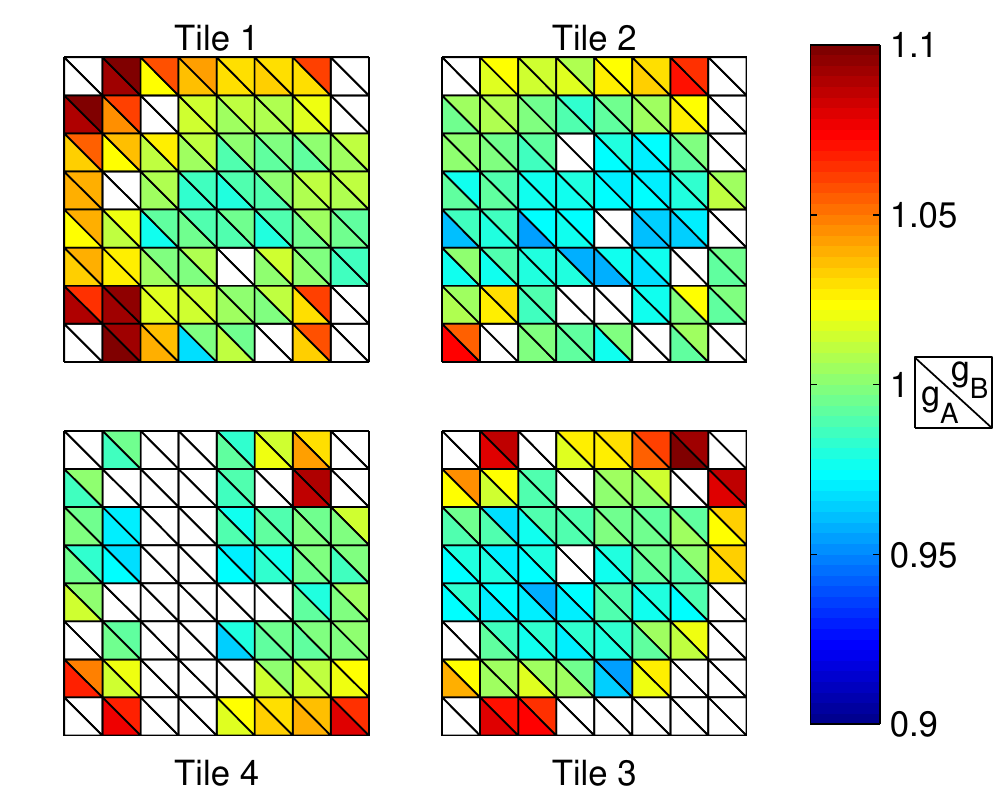}
\end{center}
\caption
    {Measured absolute gain for each detector included in \bicep2's maps.
       The gains
      are normalized such that the median gain is one. The distribution within
      the focal plane is represented schematically. Each detector pair is
      depicted as a small square. The A (B) member of a detector pair is
      depicted as the lower (upper) triangle of the square.}
\label{fig:abscal}
\end{figure}

\begin{figure}[]
\begin{center}
\includegraphics[width=1\columnwidth]{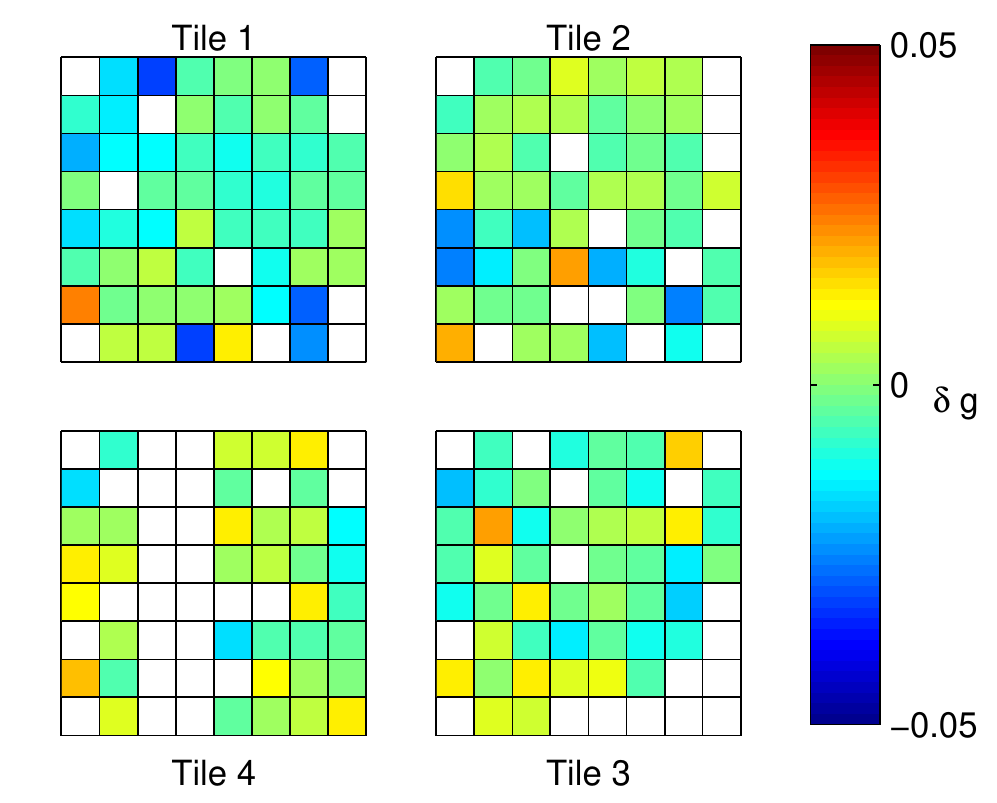}
\end{center}
\caption
    {Measured fractional differential gain, $(g_A-g_B)/[g_A+g_B)/2]$, for each
      detector pair included in \bicep2's maps.}
\label{fig:relgain}
\end{figure}

We emphasize that the deprojection algorithm described above does not
require any external measurements of beam imperfections --- the necessary
coefficients, $a_k$, are fit for (marginalized over) from the CMB data itself.
However, checking the operation of the technique and determining the residual
contamination remaining
after deprojection of any given set of modes requires external measurements of
the actual instrument beams.

As summarized in Section~11.2 of the Instrument Paper, we have made high
signal-to-noise beam maps of each detector by rastering the telescope over
a chopped thermal source located $195$~m from the telescope's aperture --- for
full details, see the Beams Paper.  In
this paper, we use these beam maps in two ways: (1) we fit elliptical Gaussians to
them and cross check the fit parameters against those derived from the
deprojection algorithm (Section~\ref{sec:beammapconsist}), and (2) we use them as direct inputs to simulations to
predict the \ttp\ leakage in the real data signal and jackknife maps while varying the set of modes
deprojected (Section~\ref{sec:beamsim}). Both offer highly robust checks that the beam
maps correspond to reality.

During beam mapping, the instrument is put in a rather different state than that
used for routine CMB observing, and the frequency spectrum of the source is not
the same as that of the CMB. Beam shapes (especially differential beam shapes)
and centroids are relatively
insensitive to changes in the source spectrum, but differential gain --- which
typically arises from the coupling of intra-pair bandpass mismatch to
the difference between the frequency spectrum of the atmosphere and the CMB ---
 is not. Therefore, the differential gain measured in beam
maps is not a reliable estimate of the CMB value.  Instead we estimate it by
cross-correlating single detector $T$ maps coadded over the full data set against the
\textit{Planck} 143~GHz map in a per-detector analog of the absolute gain calibration
described in Section~13.3 of the Instrument Paper. Figure~\ref{fig:abscal} shows the
results, the measured absolute gain, $g$ for each of \bicep2's
detectors. Figure~\ref{fig:relgain} shows the measured fractional differential gain for
each of \bicep2's detector pairs, $(g_A-g_B)/[(g_A+g_B)/2]$.

Differential pointing can be measured either from the beam maps or from the 
per-detector cross-correlation against the Planck 143~GHz maps described in Section~11.9 of
the Instrument Paper.  The results are very similar.

\begin{figure}
\begin{center}
\includegraphics[width=1\columnwidth]{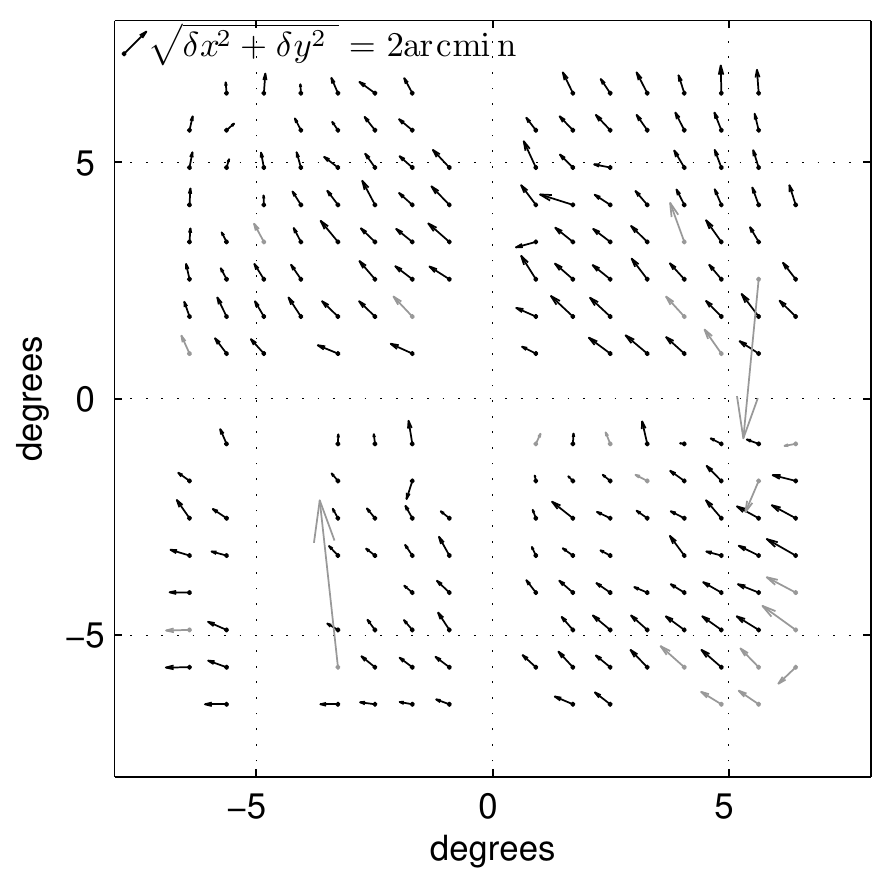}
\end{center}
\caption{Differential pointing in the \bicep2 focal plane as projected onto the
  sky at deck$=90\deg$. As drawn, the vectors originate at the nominal beam
  center and point from detector B to detector A. Their magnitudes are
  drawn $\times20$ for display purposes. All functioning pairs are
  plotted, but grayed out vectors indicate detector pairs that are excluded from
  the final maps. }
\label{fig:dipolequiver}
\end{figure}

Figure~\ref{fig:dipolequiver} shows \bicep2's differential pointing measured
from per-detector cross-correlation,
which shows a strong coherent component across the focal plane.  The coherent
part of the pattern will cancel in the final signal map and be enhanced in a
$180\deg$ split jackknife as described in Section~\ref{sec:dipolecancel}.  The
incoherent part will average down in the signal map and also potentially cause
jackknife failure.

\begin{figure}[!t]
\begin{center}
\includegraphics[width=1\columnwidth]{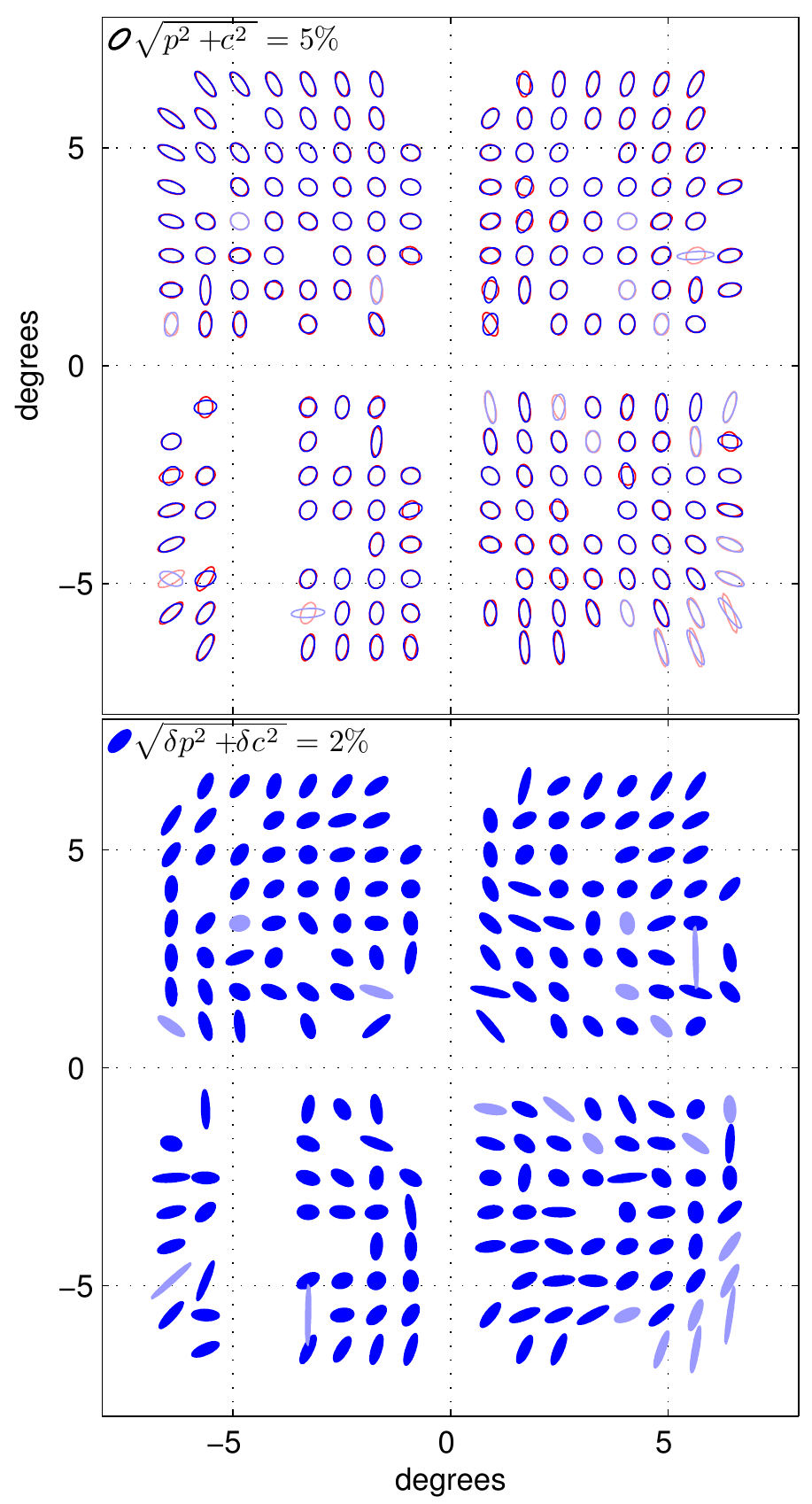}
\end{center}
\caption
{Top: per-detector beam ellipticity in the \bicep2 focal plane as
  projected onto the sky at deck$=90\deg$. Ellipticity is exaggerated for
  clarity.  Red and blue denote A and B members of a detector pair,
  respectively. All functioning pairs are plotted, but light colors indicate
  detectors that are excluded from the final map.  Bottom: per-pair
  differential ellipticity, defined as $\sqrt{(\delta p)^2+(\delta c)^2}$. The
  orientation of the ellipse indicates the orientation of the difference beam
  quadrupole. Detector polarization angles are aligned with the horizontal and
  vertical axes.}
\label{fig:ellipquiver}
\end{figure}

Figure~\ref{fig:ellipquiver} shows \bicep2's measured beam
ellipticity and differential ellipticity. The differential ellipticity shows
strong pair to pair variation in angle, so we expect some averaging down of
leakage in the signal maps as described in Section~\ref{sec:quadrupolecancel}. We also
expect that jackknife tests that split the data according to pair will be
sensitive to it.

\section{Simulation pipeline}
\label{sec:simpipeline}

\bicep2's power spectrum analysis is Monte-Carlo-based, requiring simulations of
maps ``as seen'' by the experiment \citep{hivon02}. We simulate both noiseless
(signal-only) and noise-only maps.  The standard simulations introduced in
Section~{V.A} of the Results Paper include only differential pointing at the
measured values shown in Figure~\ref{fig:dipolequiver}.  Here we extend the
signal-only simulations to include many different types of systematics.

\subsection{Input Maps and Interpolation}
\label{sec:standardsim}

The simulation pipeline produces signal-only timestreams by
sampling an input Healpix map along individual detectors' trajectories. The
simulated timestream data is then passed through the same map maker as the real
data to produce simulated $T$, $Q$, and $U$ maps that are filtered identically to
the data. Our pipeline extensions optionally introduce many different
systematics at the timestream generation stage, allowing us to model their
effects on the final maps.

Both the main simulations and our dedicated systematics simulations use input
maps of Nside=2048.
We perform the simulations of systematic \ttp\ using the \textit{Planck} HFI 143~GHz $T$
map --- pre-smoothed by \bicep2's nominal, circular Gaussian beam as described in
Appendix~\ref{sec:practical} --- as input. We use the downgraded resolution, 
Nside=512 version of the same map as the deprojection template. To predict \ttp,
we set the input $Q$ and $U$ maps to zero so that any
non-zero signal in the resulting polarization maps and spectra are due entirely to
leakage. To simulate systematics that primarily leak \etb\ we use as input maps
\texttt{synfast} generated realizations of \lcdm\ and do not set the $Q$ and $U$
maps to zero. We difference the spectra simulated with and without the
systematic included and average over 10 realizations to estimate the
\etb\ leakage.

All the simulations except the beam map simulations described
in Section~\ref{sec:beamsimpipeline} interpolate the input map to
timestreams using a second order Taylor expansion around the $T$, $Q$, and $U$ pixel
centers using the derivative maps that are a standard output of
\texttt{synfast}.  Assuming a polarization angle and efficiency, we combine a
single detector's $T$, $Q$, and $U$ timestream into a single timestream. Using
simulated input maps of progressively higher resolution allows us to simulate
timestreams to arbitrary accuracy. Doing this, we find that using an Nside=2048
map produces negligible fractional differences from a still higher resolution
input map.

\subsection{Elliptical Gaussian Beam Convolution}
\label{sec:ellipconv}

Leakage from differential pointing is naturally handled in all the simulations
discussed above because each detector is allowed to
have its own pointing trajectory on the pre-smoothed input maps.

In studies of systematics where we wish to vary the simulated elliptical Gaussian
beam shape, we use multiple input maps which have each been pre-smoothed with
circular Gaussians of different widths. Convolution on the sphere is fast and
exact for any beam that is circularly symmetric \citep{wandelt01}.

To simulate beam widths that vary from detector
to detector, we use a perturbative method in which two or more Healpix maps of
bracketing widths are simultaneously read in and interpolated between at each
time step to approximate the timestream from a beam of intermediate width. Using
bracketing maps of closer and closer spacing allows simulation of differential
beamwidth to arbitrarily high accuracy, which we use to verify that our choice
of bracketing widths simulates leakage from beamwidth mismatch to sufficient
accuracy.

Elliptical beam convolution is handled by approximating elliptical beams as the
superposition of three or more circular sub-Gaussians of different widths, centers,
and amplitudes, the choice of which is a function of $p$, $c$ and $\sigma$ and is
predetermined from 2-d fits to elliptical Gaussians. Input maps pre-smoothed to
different circular Gaussian widths are read in and each is interpolated along
the sub-Gaussians' trajectories. The individual timestreams are then combined to
approximate the timestream from a detector with an elliptical Gaussian beam. The
amplitudes, widths, and relative centers of the sub-Gaussians are fixed, but the
orientation of the ellipse can vary along a scan trajectory according to the
beam's projected orientation.  We have verified the accuracy of this approach
with special simulations using intrinsically flat input maps and explicit 2D
convolution.  As with beamwidth, we can simulate elliptical beams to
arbitrarily high accuracy using superpositions of greater numbers of circular
Gaussians.
Defining ellipticity $e=(\sigma_{maj}^2-\sigma_{min}^2)/(\sigma_{maj}^2+\sigma_{min}^2)$
we find that our procedure, which uses three Gaussians, produces
timestreams from elliptical beams that are accurate for $e<0.15$.

\subsection{Arbitrary Beam Shape Convolution}
\label{sec:beamsimpipeline}

The preceding methods allow for nearly exact simulation of
elliptical Gaussian beams. We also allow for arbitrary beam shape convolution. 

We perform arbitrary beam shape convolution
by forming a flat map projection of the input Healpix map, convolving this
projection directly with a 2D kernel, and interpolating off the flat map to
form simulated timestreams. We call these ``beam map simulations.'' Ordinarily,
such a brute force algorithm would be very computationally expensive when
simulating a large number of detectors observing over a long time period.  For
\bicep2 we have considerably reduced the expense by exploiting the fact that
(1) the telescope's deck angle remains fixed during CMB scans, 
(2) there is no sky rotation at the South Pole, and (3) \bicep2's scan pattern
is highly redundant. Thus, for a fixed deck angle,
each detector observes a given location on the sky with only one orientation,
and the convolution of the kernel with a flat sky map need only be performed
once per detector per each of the four deck angles.

This method suffers from distortion away from the center of the
projection. However, because the distortion is common to both members of a
detector pair, the difference signal is still predicted with high accuracy. We
test this by comparing the \ttp\ leakage simulated using the multiple Gaussian
approach described in Section~\ref{sec:ellipconv} (which, again, does not suffer from any flat sky
distortion effects and which we perform to high accuracy) to beam map
simulations that use as the convolution kernels elliptical Gaussians constructed
to reflect identical beam parameters. Any difference in the \ttp\ leakage from
the two methods is attributed to algorithmic limitations of the beam map
simulation procedure. We have verified that the method of flat sky beam
convolution is sufficient to accurately predict the level of leakage from all
modes of an elliptical Gaussian, both before and after deprojection. These
simulations make no assumptions of elliptical Gaussian beam structure, so this
test verifies that beam map simulations will accurately predict \ttp\ leakage
from arbitrary beam shape mismatch.

Deprojection is performed on these beam map simulations in the same way as in the
standard simulations. Therefore, the leakage templates suffer from no
corresponding distortion effects, and the main impact of projection distortion
in beam map simulations is to slightly degrade the ability of deprojection to
filter leaked power from the timestreams. This results in an artificial
``floor'' at $\simeq 10^{-5}~\mu$K$^2$ below which power
due to the mismatch of elliptical Gaussians will not deproject in a beam map
simulation. Beam map simulations thus always predict at least as much residual
contamination as is present in the real data.

We have developed the beam map simulation procedure so that we can use measured
beam maps as inputs. Because these empirical beam maps make no assumption of
elliptical Gaussian structure, their ability to reproduce the behavior of real
data spectra, both signal and jackknife, under
different deprojection options is powerful evidence against residual, unmodeled,
and undeprojected contamination from beam mismatch (see  Section~\ref{sec:beamsim}).

\section{Jackknife tests}
\label{sec:jackknives}

\bicep2's most basic guard against systematics is jackknife tests
\citep{pryke09,chiang10}. As already described in Section~\ref{sec:instrsummary}, we
split the data into two subsets, form $T$, $Q$, and $U$ maps from each subset,
and difference the maps. Under the hypothesis that the observed signal is real
and ``on the sky,'' the difference map should be consistent with the
distribution of systematics-free signal-plus-noise simulations.  If some or all of the observed signal is
from an instrumental systematic, then, depending on the type of hypothesized
systematic, the different halves of a split will contain either different amplitudes
or different spatial patterns of contamination.  The \bicep2 jackknife tests
were discussed in Section~{VII.C} of the Results Paper.  Here we review and give some
fuller details.

\begin{figure}
\begin{center}
\includegraphics[width=1\columnwidth]{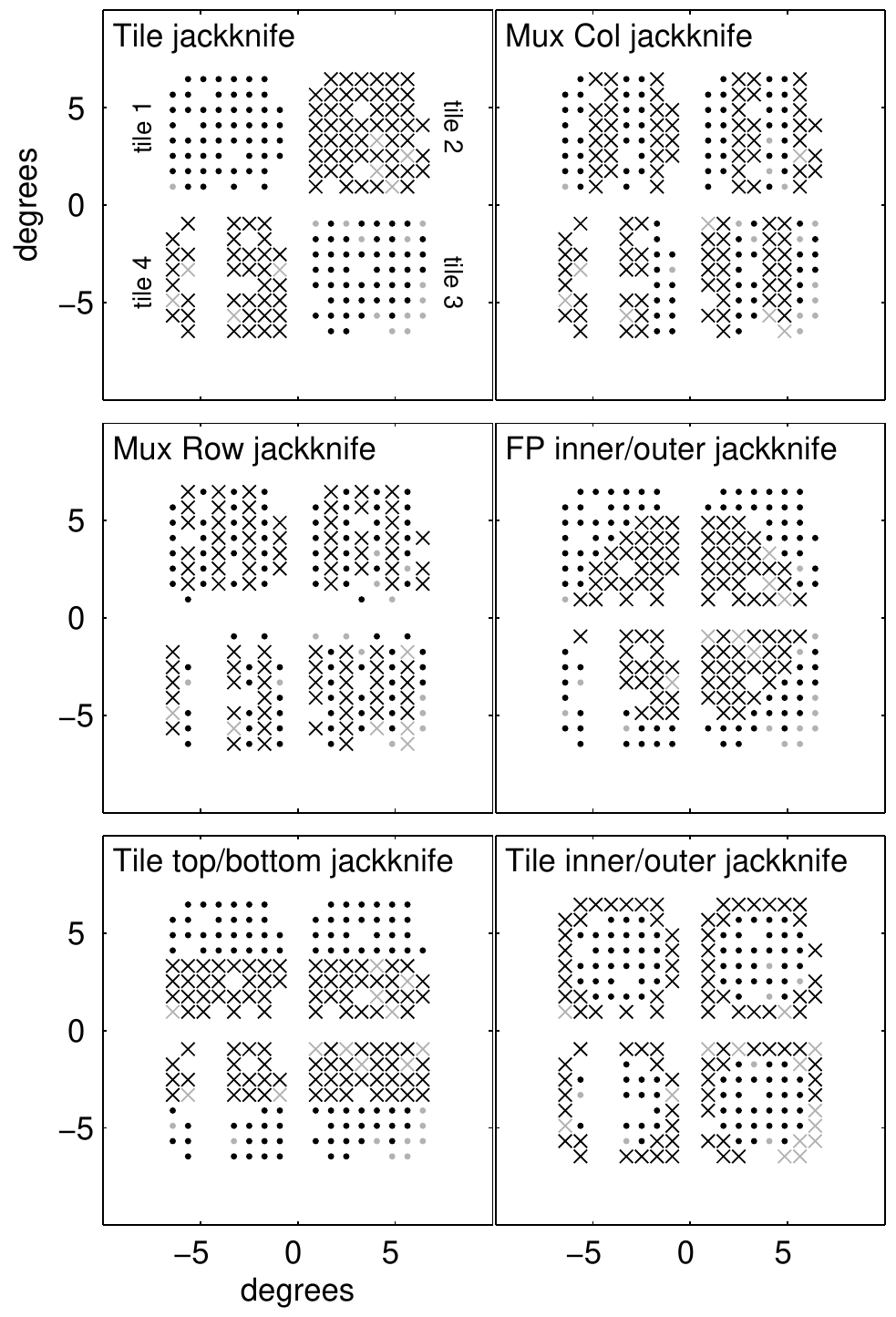}
\end{center}
\caption{ Map of the \bicep2\ focal plane projected onto the sky at
  deck$=90\deg$ illustrating detector pair selection jackknife splits. Dots
  denote detector pairs that are coadded to form one half of the jackknife
  split; X's denote detector pairs coadded to form the other half. All
  functioning pairs are shown. Light gray symbols indicate pairs that are
  excluded from the final map.}
\label{fig:chjackfpmap}
\end{figure}

Different jackknives probe for different classes of systematics. Some jackknives
split the data according to the observing cycle, some according to detector pair
selection, and one according to a combination of both. Detector pair selection
jackknives are illustrated in Figure~\ref{fig:chjackfpmap}.  Most systematics
will produce different contamination in the two halves for at least one of the
jackknife splits we form.  The following is a description of \bicep2's
jackknives and the types of systematics that are expected to cause each to fail.

\begin{figure}[t]
  \begin{center}
    \begin{tabular}{c}
      \includegraphics[width=1\columnwidth]{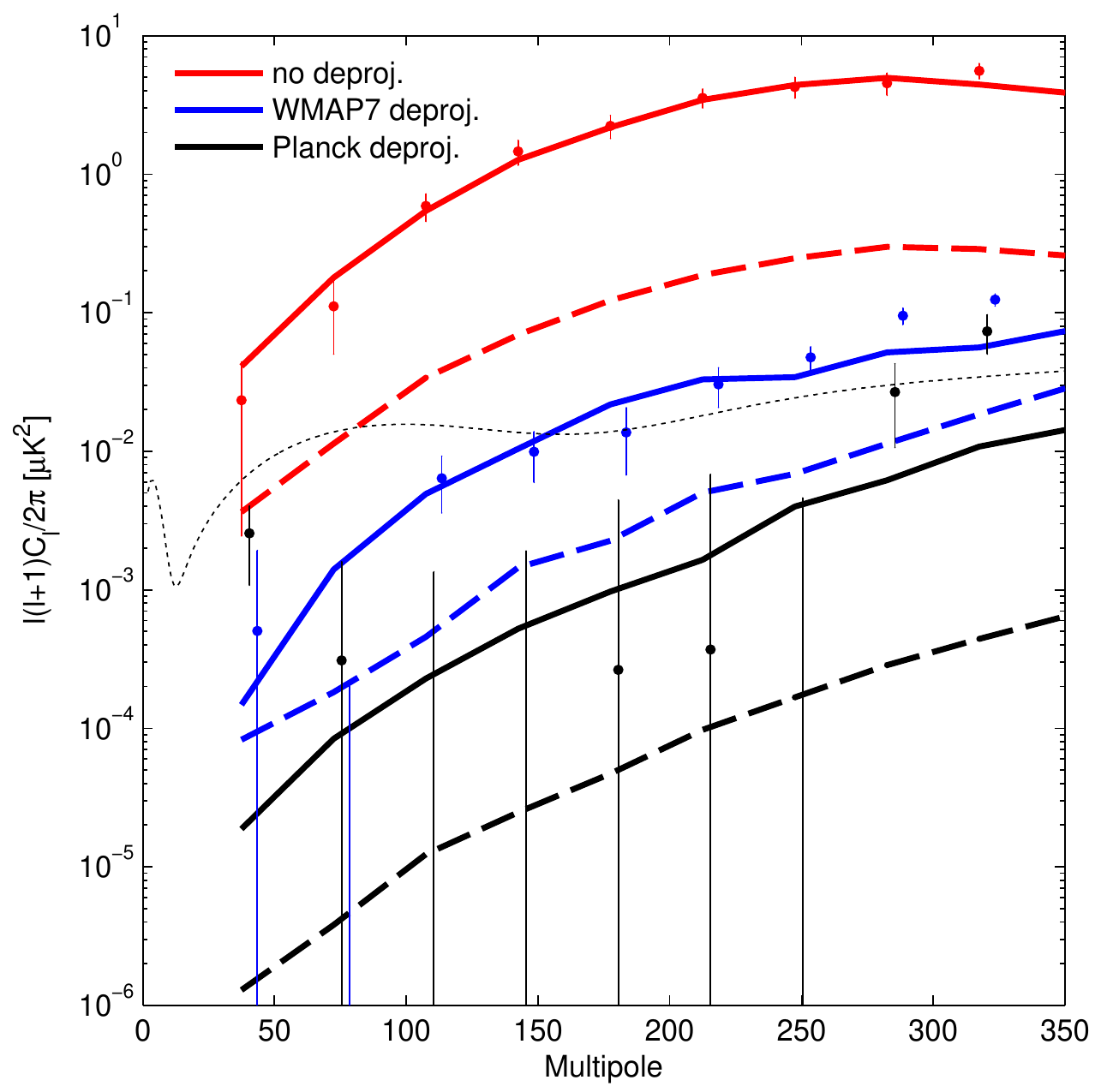}
    \end{tabular}
  \end{center}
  \caption[example] { \label{fig:deprojperf} Points with error bars are
    \bicep2's deck jackknife bandpowers with (red) no deprojection, (blue)
    differential pointing deprojected with a \textit{WMAP7} V-band template, and (black)
    differential pointing deprojected with a \textit{Planck} 143 GHz template, with error
    bars taken as the standard deviation of \lcdm\ plus instrumental noise
    simulations that include \bicep2's measured differential pointing.  The
    solid lines are the corresponding simulated deck jackknife spectra, computed
    as the mean of 50 noiseless simulations of \lcdm\ $T$ and \bicep2's measured
    differential pointing, deprojected with templates containing simulated
    template noise.  The dashed lines show the corresponding simulated non-jackknife, signal
    $BB$ leakage.  The dotted line shows a lensed \lcdm\ + $r=0.2$ spectrum for
    reference.}
\end{figure}

\vspace{1em}
\begin{description}[nolistsep,itemindent=0cm,leftmargin=.5cm,before={\renewcommand\makelabel[1]{\normalfont ##1}}]
\item[\textnormal{\textit{Deck angle}}:] Splits data according to boresight orientation,
  $68\deg+113\deg$ vs.\ $248\deg+293\deg$; highly sensitive to systematics that
  change sign under a rotation of the instrument, such as beam mismatch with
  dipolar symmetry (see Section~\ref{sec:dipolecancel}). Because of this, \bicep2's
  differential pointing contaminates the deck jackknife more strongly than the
  signal map (see Figure~\ref{fig:deprojperf}).
\item[\textnormal{\textit{Alternative deck}}:] Same as deck, but $68\deg + 293\deg$ vs.\ $113\deg$ +
  $248\deg$; similar to the deck jack, probes contamination that varies with
  boresight orientation.
\item[\textnormal{\textit{Temporal split}}:] Splits data into equal weight halves by date; sensitive to
  any long-term drifting of instrument properties.
\item[\textnormal{\textit{Scan direction}}:] Splits data according to the telescope scanning direction,
  left-going vs.\ right-going; sensitive to detector transfer function
  mismatch. \ttp\ leakage from transfer function mismatch contaminates the scan
  direction jackknife more strongly than the coadded map. Because it is the
  jackknife with the lowest predicted residuals, it is also the jackknife most
  sensitive to noise model errors.
\item[\textnormal{\textit{Azimuth}}:] Splits data according to interleaved 10~hr blocks of time
  (phases) within the three-day observing cycle (phases B+E+H vs.\ C+F+I; see
  Section~12.3 or Table 6 of the Instrument Paper for details).  Because these phase
  groups are offset from each other in azimuth, this jackknife probes azimuth
  fixed contamination, such as would be expected from polarized ground pickup.
\item[\textnormal{\textit{Moon up/down}}:] Splits according to times when the moon is above vs.\ below
  the horizon; sensitive to contamination due to the moon.
\item[\textnormal{\textit{Tile}}:] Splits data by detectors, tiles 1+3 vs.\ tiles 2+4; sensitive to
  differences in detector properties, e.g. bandpass.
\item[\textnormal{\textit{Tile/deck}}:] Tiles 1/2 at deck $68\deg$/$113\deg$ + tiles 3/4 at deck
  $248\deg$/$293\deg$ vs.\ tiles 1/2 at deck $248\deg$/$293\deg$ + tiles 3/4 at
  deck $68\deg$/$113\deg$; sensitive to effects that are common between tiles.
  (Rotating the receiver by $180\deg$ places new tiles at a given projected
  location on the sky. However, the physical orientations of tiles 1 and 2 as
  installed in the focal plane are rotated $180\deg$ from tiles 3 and 4, so that
  the new tiles have the same projected orientation after rotation. Thus, the
  regular deck jackknife does not directly probe for tile fixed effects that are
  common among tiles.) Because \bicep2's instantaneous field of view is large
  compared to the map area, this jackknife map has smaller useful coverage than
  the other jackknives.
\item[\textnormal{\textit{Focal plane inner/outer}}:] Splits according to the inner 50\% of detectors
  vs.\ the outer 50\% of detectors in the focal plane; sensitive to beam shape
  mismatch that varies with distance from the center of the focal plane, as
  would be expected of ellipticity induced by variable beam truncation in the
  aperture plane.
\item[\textnormal{\textit{Tile top/bottom}}:] Splits according to top of each tile vs.\ bottom of each
  tile, where the sense of top and bottom is defined with respect to the tile as
  fabricated, not globally within the focal plane; sensitive to effects that
  vary within an individual tile.
\item[\textnormal{\textit{Tile inner/outer}}:] Splits according to the inner 50\% vs.\ the outer 50\%
  of detectors within a tile; sensitive to effects that vary within an
  individual tile.
\item[\textnormal{\textit{Mux column}}:] Splits according to detector multiplexing column, even vs.\ odd;
  sensitive to crosstalk contamination.
\item[\textnormal{\textit{Mux row}}:] Splits according to detector multiplexing row.
\item[\textnormal{\textit{Differential pointing best/worst}}:] Splits according to the 50\% of detector
  pairs with the smallest differential pointing and the 50\% of detector pairs
  with the greatest differential pointing. Like the deck and alt deck
  jackknives, it is more sensitive to differential pointing contamination than
  the signal maps.
\end{description}

\vspace{1em}
Table 1 of the Results Paper lists probability to exceed (PTE) values for four statistics,
computed separately for the $EE$, $BB$, and $EB$ spectra, for each of the above
14 jackknife spectra.
There are thus 168 PTE statistics but some of these are partially correlated.
There is one $BB$ or $EB$ PTE with a value $\leq 0.01$, the mux row
$BB$.
Of the 499 \lcdm\ signal + noise simulations used in the main analysis
(which should reproduce the correlations), 306 realizations have one or more
$BB$ or $EB$ PTE $\leq 0.01$ so this is unsurprising.
The real data contain six $EE$ PTEs $\leq
0.01$.
Of the 499 simulations, 2 have 6 or more $EE$ PTEs $\leq 0.01$.

The Results Paper offers an explanation for the apparently anomalous number of
low $EE$ PTEs: because of the high signal-to-noise of \bicep2's $EE$
measurements, variation in the mean gain from detector pair to detector pair
results in failures of the detector selection jackknives shown in
Figure~\ref{fig:chjackfpmap}. The $BB$ detection is, of course, highly
significant as well, but the $EE$ signal-to-noise ratio, which is $\sim 500$ at
$\ell=100$, makes even the smallest absolute calibration difference between jackknife
halves impact the PTE, even though such absolute calibration errors do not imply
systematic contamination of the signal map.
We include this effect in 10 of the signal
simulations by multiplying each detector pair's data by the mean of its measured
absolute gain, $(g_A+g_B)/2$, shown in Figure~\ref{fig:abscal}
(see Section~\ref{sec:gainvar}).  The difference of the $EE$ spectra with and without
gain variation is an estimate of the contaminating power, and is $\simeq
1\times10^{-3}~\mu$K$^2$ at $\ell=100$.  Including this contaminating power
results in 9 of the 499 realizations having six or more $EE$ PTEs $\leq 0.01$. 
Gain variation is not important for jackknives of the comparatively low signal-to-noise
$BB$ data.

\section{Deprojection Performance}
\label{sec:deprojperformance}

We characterize the performance of deprojection by specifying the residual
spurious power remaining in \bicep2's polarization power spectra after
deprojection. We split this characterization into two parts. First, we
approximate the beams as elliptical Gaussians and determine the residual
contamination from various mismatch modes using the simulations introduced
in Section~\ref{sec:ellipconv}.
This serves as a test of deprojection's fundamental limit.
Second, we use the beam map simulations described in Section~\ref{sec:beamsimpipeline}
to determine the actual residual contamination after deprojection, including
that from the portion of \bicep2's beams not described by elliptical Gaussians.
In this section, we deal only with the first characterization. The second is
described in Section~\ref{sec:beamsim}.

\subsection{Template Map Non-idealities}

We first consider how non-idealities in the deprojection template map limit the efficacy
of deprojection. By far, the most important non-ideality is simply statistical
noise in the deprojection template. We have deprojected \bicep2\ data with two different
templates --- a \textit{WMAP}7 V-band \citep{wmap10} and a \textit{Planck} HFI 143~GHz $T$ map
\citep{planckviii} --- which have different bandpasses and different noise
properties. We have also performed timestream simulations using the
measured elliptical Gaussian parameters discussed in Section~\ref{sec:beammeas}
and deprojected them with
templates containing simulated \textit{Planck} and \textit{WMAP} noise. (We describe the construction of simulated
template maps in Appendix~\ref{sec:practical}.) 

These simulations predict that \bicep2's
differential pointing is by far the dominant source of contamination in the deck
jackknife, and the dominant source of contamination in the signal spectra prior
to deprojection. Furthermore,
as expected given the discussion
in Section~\ref{sec:dipolecancel} and the substantially coherent measured differential
pointing shown in Figure~\ref{fig:dipolequiver}, 
the deck jackknife spectrum is far more
contaminated by differential pointing than the signal spectrum. 

We isolate the effect of differential pointing by simulating it separately from
other difference beam modes. Because we want to investigate the impact of
template map noise, we simulate \ttp\ leakage using noiseless realizations
of \lcdm\ $T$ as input and noise added versions of those same maps,
downgraded to Nside=512, as the deprojection templates.
Figure~\ref{fig:deprojperf} shows the results as well as
real data for \bicep2's deck jackknife.
In these simulations, the efficacy of deprojection is
entirely determined by the level of noise in the template map. 
The predicted
contamination in the signal spectrum after deprojection with either the \textit{WMAP}7 or
\textit{Planck} template (dashed blue and black lines)
is small compared to an $r=0.2$ IGW
spectrum at $\ell<150$. However, when deprojecting with the noisier \textit{WMAP}7
template, the \ttp\ leakage in the deck jackknife (solid blue line) is measurable and well
predicted by simulation.
Because the deck jackknife has much greater contamination than the
signal spectrum, it is a highly stringent test of contamination.
Our accurate prediction of residual contamination in the deck jackknife
is strong evidence against significant unmodeled leakage in the signal maps. In
\bicep2's main results, deprojection is performed with a \textit{Planck} 143 GHz
template, and \ttp\ leakage from differential pointing is negligible and
unmeasurable in even the deck jackknife.

We note that bandpass differences between \bicep2\ and the deprojection template
are not important.  The \textit{WMAP} V-band template is centered at 60~GHz while the
\textit{Planck} template is centered at 143~GHz, much closer to \bicep2's central
frequency. In principle, the \ttp\ leakage at different frequencies is not the
same because of unpolarized foregrounds with non-CMB-like spectral dependencies.
The agreement of the data points and the solid lines in
Figure~\ref{fig:deprojperf} indicates that for even significant bandpass
differences, undeprojected leakage from foregrounds not present in the
deprojection templates is negligible. Foregrounds present in the deprojection
template that are fainter in \bicep2's band would be a source of unmodeled
template noise, which Figure~\ref{fig:deprojperf} indicates is also not an issue. We
have also simulated adding point sources to the template map that are not
present in simulated \bicep2\ maps, and this also has a negligible effect on
deprojection.

\subsection{Consistency With Beam Maps}
\label{sec:beammapconsist}

\begin{figure}[t]
  \begin{center}
    \begin{tabular}{c}
      \includegraphics[width=1\columnwidth]{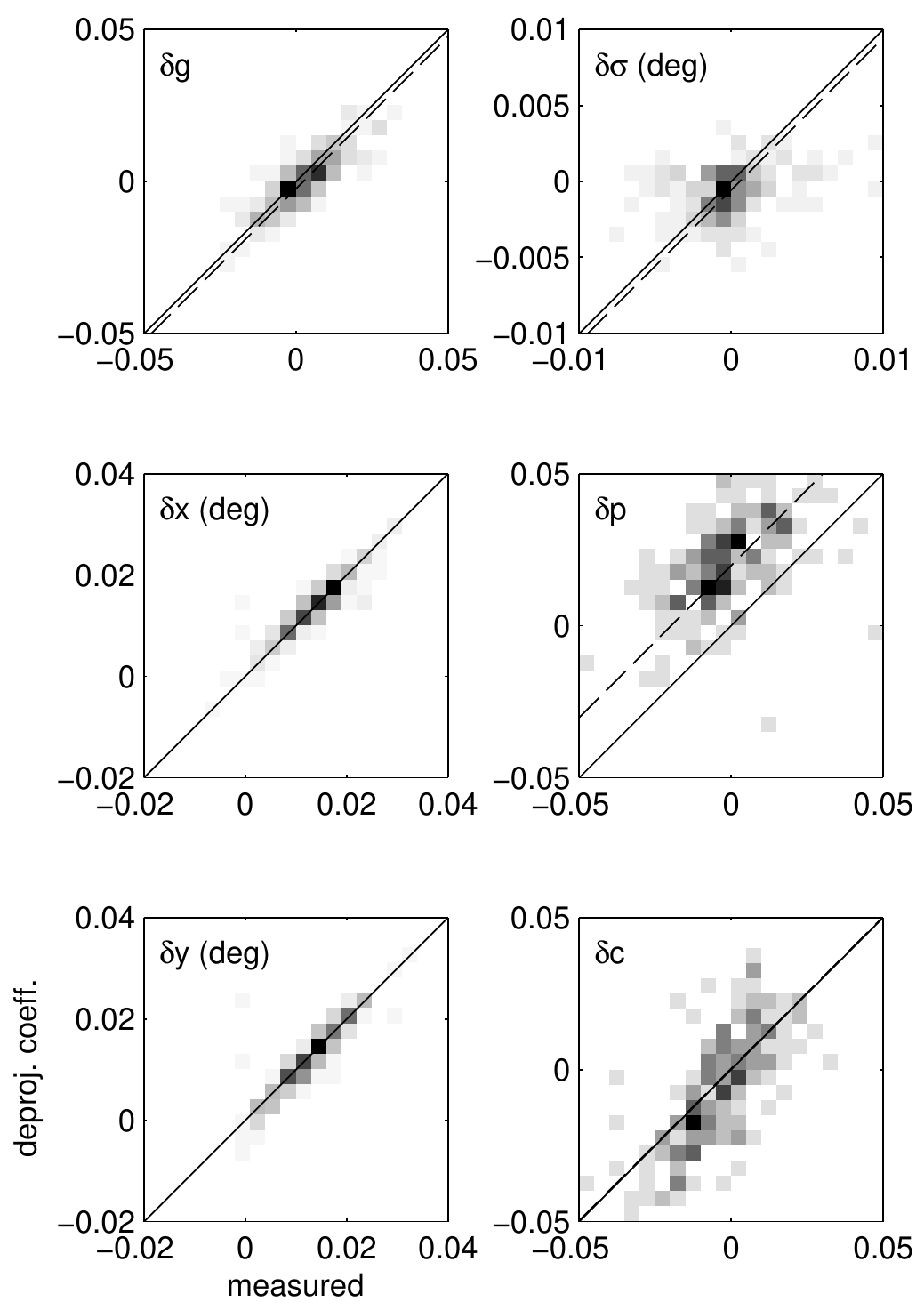}
    \end{tabular}
  \end{center}
  \caption[example] { \label{fig:dpvsmeas} Differential beam parameters measured
    from far-field beam maps (horizontal axis) and from template regression as
    used in deprojection (vertical axis), shown as 2D histograms over detector
    pairs.  (Differential gains are determined from cross-correlation of
    individual detector T maps with \textit{Planck}.) The solid line has a slope of 1 and
    a $y$-intercept of 0. The dashed line has slope of 1 but has been offset
    vertically by the bias in the recovered deprojection coefficients predicted
    from simulation. The scatter and bias in the observed relation is broadly
    consistent with that predicted from signal-plus-noise
    simulations. }
\end{figure}

We confirm that deprojection filters contamination consistent with our measured
difference beams by comparing the differential beam parameters implied by the
deprojection fit coefficients of \bicep2's real data (calculated according to
Table~\ref{tab:deprojection}) to the independent measurements of the same
parameters described in Section~\ref{sec:beammeas}. Figure~\ref{fig:dpvsmeas} shows
the correlation of the deprojection derived differential beam parameters with the
beam-map-derived differential beam parameters. (Note that $\delta x$ and $\delta y$
are measured from beam maps, not from correlation of per-detector $T$ maps with
\textit{Planck} maps, as in Figure~\ref{fig:dipolequiver}, and are thus fully independent
of the deprojection coefficients, if somewhat lower signal-to-noise.)
The uncertainties of the beam-map-derived parameters are somewhat difficult to accurately estimate. However, the
scatter in the observed relation is consistent with the scatter on the
deprojection coefficients predicted from signal-plus-noise simulations,
indicating that noise in the CMB data dominates the scatter in
Figure~\ref{fig:dpvsmeas}.

The significant bias visible in the plus-ellipticity deprojection coefficient
results from the inherent $TE$ correlation in \lcdm\ cosmology, which ensures
some correlation between the true CMB polarization signal and the deprojection
templates. This bias does not impair the filtering of \ttp\ leakage, but it does
cause additional filtering of cosmological \emode s (the effect on \bmode s is
negligible). The effect on both \emode s and \bmode s is automatically accounted
for in the filter/beam suppression factors derived from simulations that apply
the same choice of deprojection (see Section~ VI.C of the Results Paper).  We have
verified that the bias arises from 
\lcdm\ $TE$ correlation by observing that the bias disappears in simulations
with no $TE$ correlation.

Given good agreement between measured differential beam parameters and those
inferred from deprojection, we can choose to either deproject a given
differential mode or to subtract the contamination expected given our direct
measurements.  Differential gain can, in principle, have a significant time
variable component, so we choose to deproject it. (We perform the deprojection
regression on approximately 9~hr chunks of data; see
Appendix~\ref{sec:practical} for details.)
 Differential pointing is
measured with high signal-to-noise in beam maps and is expected to be constant
in time, but because it is \bicep2's largest source of \ttp\ leakage we
conservatively choose to deproject it to avoid any residual leakage arising from
noise in the calibration measurements.  Since differential ellipticity
deprojection preferentially filters our $TE$ and $EE$ spectra, we choose to fix the
deprojection coefficients to the beam-map-derived values and subtract the
scaled deprojection templates from the data, rather than fitting the templates.
In the results of the beam map simulations described in Section~\ref{sec:beamsim}, 
we find this to be empirically equivalent to deprojecting ellipticity

The simulation of \bicep2's best-fit elliptical Gaussian beam shapes that include
all six differential modes demonstrates that \ttp\ leakage from pure
elliptical Gaussian mismatch can be cleaned to the $r\sim1\times10^{-4}$ level
with deprojection using a template with \textit{Planck} 143~GHz noise levels. At this level,
the component of \bicep2's beam mismatch not fit by the difference of elliptical
Gaussians is the dominant source of \ttp\ leakage.

\section{Systematics error budget}
\label{sec:syslevels}

\begin{figure*}[t]
  \begin{center}
    \begin{tabular}{c}
      \includegraphics[width=2\columnwidth]{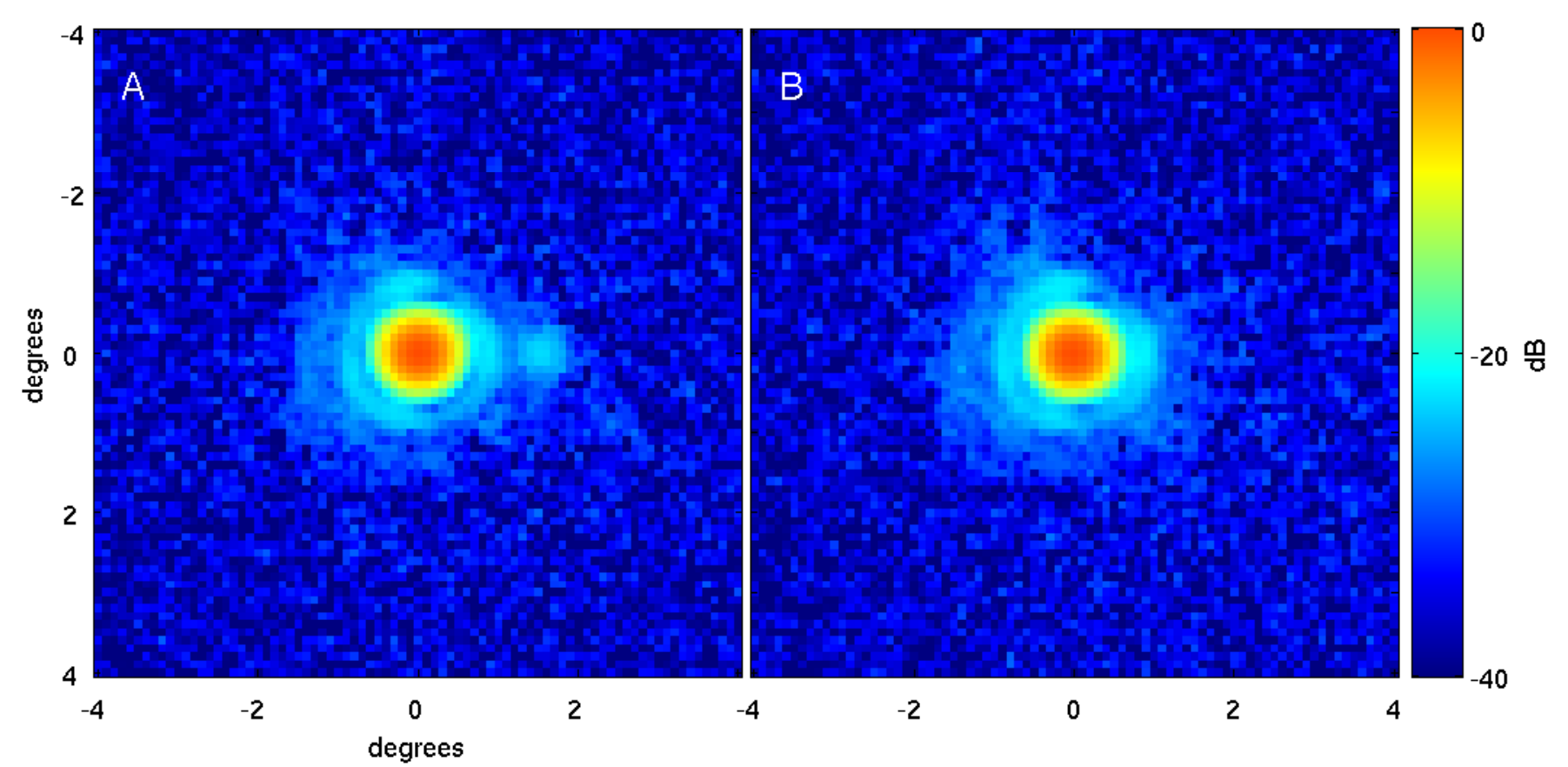}
    \end{tabular}
  \end{center}
  \caption[example] {\label{fig:beammap} Composite beam map for a representative
    detector pair, showing the A and B beams. ($1~\mbox{dB}=10\log_{10}$.) The
    coordinate system is centered on the mean pair centroid. The expected
    crosstalk feature with $\sim-25$~dB amplitude is visible in the A detector on
    the horizontal axis at a distance of $\sim+1.7\deg$ from the beam
    center. The first Airy ring is visible at a radius of $\sim1\deg$.  The
    difference beam (not shown) is dominated by a dipole structure.}
\end{figure*}

Jackknife tests fail when the magnitude of contamination exceeds
the noise in the jackknife maps, which, in general, is comparable
to the noise in the signal maps. If the
contamination is uncorrelated in the two halves of the jackknife split, then
jackknife tests can place upper limits on possible contamination only as low as
the level of \bicep2's statistical uncertainty. We therefore rely on the
jackknife tests described in Section~\ref{sec:jackknives} primarily as a safeguard
against unknown and unmodeled systematics.  Using special calibration data, we
constrain known possible systematics to much lower levels.

In this section, we use a few approaches to either directly determine or
place upper limits on the contamination from a given systematic. First, where a
systematic is strong enough relative to the sensitivity of calibration data, we
directly determine the $BB$ spectrum of the expected spurious signal using
simulations of the effect. Many of the calibration measurements are described in
the Instrument Paper, and are similar to those described in
\citet{takahashi10}. Second, where calibration data exist but the systematic
effect in question is not large enough to directly measure, we place upper
limits on the contamination given the sensitivity of the calibration
data. Third, in the absence of robust calibration data, we can determine the
level of a hypothesized systematic that would show an observable
effect in \bicep2's signal and jackknife spectra and set an upper limit this
way.

We quote the level of contamination from individual sources of systematics by
assigning a characteristic tensor/scalar ratio to the spurious \BB\ power they
generate. We compute this characteristic $r$-value using the ``direct
likelihood'' method developed in \citet{barkats13} and used in Section~ XI.A of the
Results Paper.  We first compute a weighted sum of bandpowers of the predicted
spurious signal. We use signal/variance weighting, with a signal equal to an
$r=0.1$ IGW spectrum and variance equal to the variance of bandpowers from
simulations of lensed-\lcdm\ signal + instrument noise. The ratio of this
weighted sum (multiplied by $0.1$) to the identically weighted sum of a pure
$r=0.1$ IGW spectrum is the characteristic $r$-value of the contamination. (In
practice, the choice of fiducial $r$ makes no difference.)

Because this procedure strongly de-weights bandpowers above $\ell\simeq120$,
contamination at these multipoles will not be reflected in the quoted
$r$-values. Nonetheless, we plot systematics spectra
at $\ell<350$ and can therefore verify that systematics are small at all
scales presented in the main analysis.

\subsection{Undeprojected Residual Beam Mismatch}
\label{sec:beamsim}

In Section~\ref{sec:deprojection}, we described the deprojection algorithm that allows
us to filter out \ttp\ leakage from mismatched beams and in Section~\ref{sec:deprojperformance}
demonstrated that for idealized elliptical beams the residual
\ttp\ contamination after deprojection using the \textit{Planck} 143~GHz template map is
well below \bicep2's noise. Because deprojection, as parametrized, filters only
power corresponding to the modes of the difference of elliptical Gaussians,
the portion of any detector pair's difference beam not described by this model
creates residual, undeprojected contamination. 

As described in Section~\ref{sec:beammeas} and in the Beams Paper, we have obtained high
signal-to-noise beam maps of every \bicep2\ detector.  The source was observed
3 times each at 4 deck angles to produce a total of 12 individual
$8\deg\times8\deg$ beam maps for each detector. The central region of each
detector's beam map, at radius $r\leq1.2\deg$, is covered by all 12
observations. This area contains $97\%$ of the total integrated beam power.  The
regions of the beams at $r>1.2\deg$ are not fully covered by observations
at a single deck angle.  Beam map pixels at $r\leq3\deg$ from the beam
center are observed at a minimum of two deck angles. Regions of the beam map at
$r>3\deg$ are generally observed at a single deck angle.

We combine the available observations to form one composite beam map for
each detector. We do this in two ways: (1) we median filter the full beam maps
to produce $8\deg\times8\deg$ maps, and (2) we set to zero the portion of the
beam maps at $r>1.2\deg$ and mean filter the observations.  We refer to these
two composite maps as the (1) extended and (2) main beams.  The median filter is
necessary for the outer regions of the beam maps because of
artifacts in some of the observations.  The extended composite beam map for a
representative detector pair is shown in Figure~\ref{fig:beammap}.

We apply a gain mismatch by normalizing each detector's beam map to reflect the
differential gain measurements shown in Figure~\ref{fig:relgain}. (We normalize
each detector pair's two beam maps such that the mean gain is one and the
intra-pair ratio of the mean of the square root of the azimuthally averaged beam
window functions, $B_{\ell}$, in the multipole range $100<\ell<300$ equals the
ratio of the measured absolute gains.  This procedure ensures we apply the
differential gain in simulation to the same multipole range as in which it was
measured.)

\subsubsection{Undeprojected Residual in Signal Maps}

We use these beam maps as inputs to the beam map simulation algorithm described
in Section~\ref{sec:beamsimpipeline} and compare the resulting \ttp\ leakage to the
real data.  

\begin{figure*}[t]
  \begin{center}
    \begin{tabular}{c}
      \includegraphics[width=2\columnwidth]{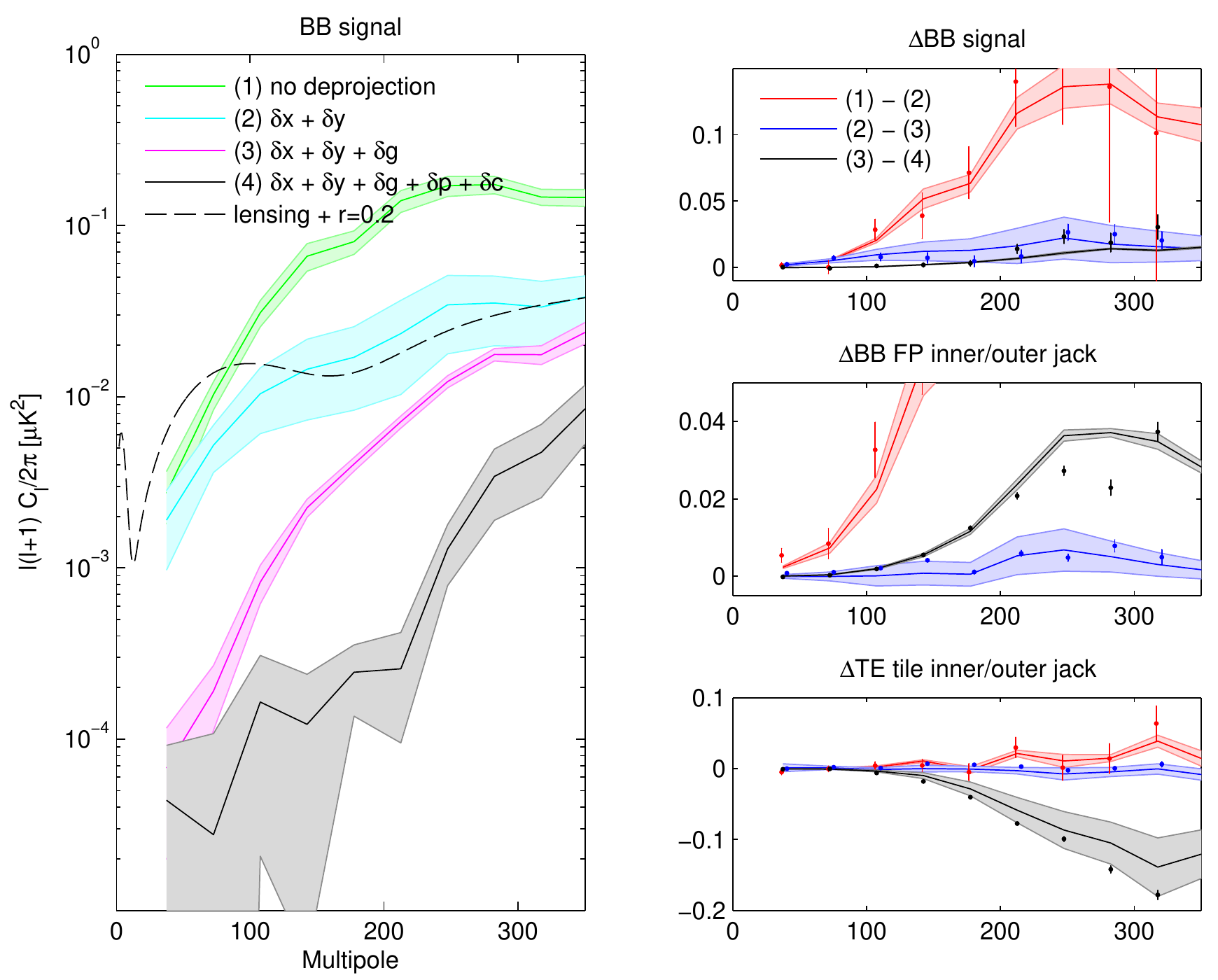}
    \end{tabular}
  \end{center}
  \caption[example] {\label{fig:beammapjacks} Left panel: $BB$ contamination
    predicted from beam map simulations of \bicep2's measured main beams
    (temperature only simulations using the Planck HFI 143 GHz $T$ map convolved
    with measured, per-detector beam maps).  The shaded bands indicate the
    $1\sigma$ uncertainty of the contamination given the sensitivity of the beam
    maps and gain mismatch measurements. The colors correspond to different choices of deprojection: (1) no
    deprojection; (2) deprojection of differential pointing ($\delta x + \delta y$); (3)
    deprojection of differential pointing and differential gain ($\delta x +
    \delta y + \delta g$); and (4)
    deprojection of differential pointing, differential gain, and differential
    ellipticity ($\delta x + \delta y + \delta g + \delta p + \delta
    c$). Right panels: Changes in bandpowers with 
    different deprojection choices for: (top) the $BB$ signal
    spectrum, (middle) the $BB$ focal plane inner/outer jackknife, and
    (bottom) the $TE$ tile
    inner/outer jackknife. 
    The solid lines and shaded bands again indicate the mean and $1\sigma$
    uncertainty of the predicted leakage given the sensitivity of the beam
    maps and gain mismatch measurements. The points with error bars are the real data bandpower differences,
    with error bars computed as the rms of \bicep2's standard,
    lensed-\lcdm\ signal plus instrumental noise simulation set. }
\end{figure*}

The left panel of Figure~\ref{fig:beammapjacks} shows the predicted
$BB$ contamination from the main beam map simulations using different
deprojection options. The colored bands indicate the $\pm1\sigma$ uncertainty of
the predicted leakage, which is set by noise in the beam maps and the absolute
gain measurement uncertainty. The top right
panel of Figure~\ref{fig:beammapjacks} shows the change in simulated leakage
when applying deprojection as colored bands, as well as the observed change in
\bicep2's bandpowers under different choices of deprojection as points. Again,
the shaded bands indicate the $\pm1\sigma$ uncertainty of the beam map
simulations. The error bars on the points are the standard deviation of
bandpower differences from simulations that include lensed-\lcdm\ signal and
instrumental noise. The details of the estimation of the leakage uncertainty
are given in Appendix~\ref{sec:beammapsimappendix}.

Figure~\ref{fig:beammapjacks} shows that deprojection of differential
pointing is absolutely necessary and differential gain is also
important.
Differential ellipticity is a smaller effect.
Once these deprojections are in effect the residual contamination
is seen to be very small.

\subsubsection{Undeprojected Residual in Jackknife Maps}

We can further compare the action of deprojection on real data and simulations
for each of the jackknives described in Section~\ref{sec:jackknives}.  As described
in Section~\ref{sec:cancellation}, some beam systematics undergo considerably less averaging
down due to incoherence across the focal plane and cancellation
due to instrument rotation in certain jackknifes.
In these cases, we can
investigate the behavior of deprojection in circumstances where it has to
``work harder'' than in the full signal map.

Examples of this are the focal plane inner/outer and tile inner/outer splits
illustrated in Figure~\ref{fig:chjackfpmap}.  As seen in
Figure~\ref{fig:ellipquiver} \bicep2's beam ellipticities exhibit a dependence
on distance from the focal plane center
while the differential ellipticity is
strongest around the edges of individual tiles.  The right center panel of
Figure~\ref{fig:beammapjacks} shows that the focal plane inner/outer jackknife
has a
much stronger response in $BB$ to differential ellipticity deprojection than the full signal
map, and that the degree of this response matches between real data and
simulations.  Likewise, the bottom right panel shows that the tile inner/outer
jackknife responds as predicted in the $TE$ spectrum.

In general the simulated jackknife residuals match the real data for all the
jackknives, under all deprojection combinations.  Even \textit{without}
differential ellipticity deprojection, the contamination of the $BB$ spectrum is
negligible, yet we still detect it in the jackknives that ought to be most
sensitive to it.  These many additional tests build confidence that we
understand \ttp\ leakage from beam shape mismatch to an accuracy and precision
surpassing that required by our error budget.

\subsubsection{Undeprojected Residual Correction}

The simulated main beam leakage with differential gain, pointing and ellipticity
deprojection is robustly measured and is shown as the black line in the left
panel of Figure~\ref{fig:beammapjacks}. This leakage corresponds to
$r=1.1\times10^{-3}$ and is subtracted from the $BB$ bandpowers prior to fitting
$r$ in Section~ VIII.A of the Results Paper. The expected main beam contamination in the final
results is therefore zero.

The extended beam simulations are noisier than the main beam simulations and
the median filter makes statistics derived from them less robust. The predicted
extended beam leakage is consistent with zero and we adopt its $1\sigma$
uncertainty as the upper limit of possible remaining \ttp\ leakage from beam
shape mismatch after the main beam residual correction.  Because
the extended beam maps include the main beam, this upper limit includes the
uncertainty of the residual leakage correction. Moreover, because the extended
beam maps include crosstalk beams, it includes \ttp\ leakage from multiplexer
crosstalk. The upper limit is shown in Figure~\ref{fig:sysfigs} and indicates
that beam shape mismatch contributes \ttp\ leakage corresponding to $r<3.0\times10^{-3}$.

\subsection{Further Consideration of Gain Mismatch}
\label{sec:gain}

Deprojection filters \ttp\ leakage from gain mismatch with such effectiveness
that the subtle choices of multipole ranges and normalization constants
described in Section~\ref{sec:beamsim} make
virtually no difference. We have simulated up to three times the level of
measured relative gain mismatch and found no change in the predicted residual
contamination after deprojection. The ``extended beam'' upper limit in
Figure~\ref{fig:sysfigs} includes contamination from gain mismatch.

Other lines of evidence against systematic contamination by gain mismatch are
the cross-spectrum of \bicep2 and \bicep1, and the passing of
jackknives. Because we calibrate the relative response of our detectors hourly
by executing elevation dips (the ``el nods'' described in Section~ 12.4 of the Instrument Paper)
and observing the large response from changing atmospheric loading, we expect
that a gain miscalibration will primarily be the result of intra-pair bandpass
mismatch coupling to differences between the color spectrum of the CMB and the
atmosphere at the South Pole \citep{bierman11}.  As discussed
in Section~\ref{sec:monopole}, it is only coherent gain mismatch that will evade
\bicep2's jackknife tests. Because \bicep1 and \bicep2's bandpasses are
physically defined in very different ways (horn and mesh filter vs. antenna and
lumped-element filter), we expect that a coherent mismatch will not correlate
between the two experiments.  

Also as discussed in Section~\ref{sec:monopole},
while coherent differential gain will not
contaminate jackknives in \bicep2, it will (1) contaminate jackknives in
\bicep1, and (2) contaminate the signal map \emph{differently} in \bicep1
because of the different layout of polarization angles. Thus, while the power
spectrum of contamination from uniform gain mismatch could be similar in \bicep1
and \bicep2\, correlation would not be expected. Thus, the
consistency of the \bicep1$\times$\bicep2\ cross-spectrum with the \bicep2\ auto
spectrum as presented in Figure~7 of the Results Paper is evidence against residual uniform
gain mismatch.  Incoherent gain mismatch is still expected to contaminate pair
selection jackknives.
Additionally, a coherent gain mismatch
common to \bicep2\ and to the \keck\ will not produce correlated power. In
\keck\ maps from 2013 and after, a coherent gain mismatch will fully cancel
in signal maps as
well as contaminate $90\deg$ split jackknives.

\subsection{Gain Variation}
\label{sec:gainvar}

\bicep2 applies a single absolute calibration to the final coadded
maps. Because the map coverage region is not the same for all detector pairs,
a variation of mean gain from pair to pair will cause
\etb\ leakage, even if the intra-pair differential gain is zero.

The matrix-based map purification discussed in Section~ VI.B of the Results Paper
ensures that the \etb\ leakage from timestream filtering and map apodization is
at a level corresponding to $r<10^{-4}$.  We simulate \etb\ leakage from gain
variation within the focal plane by applying the per-pair mean of the
absolute gains shown in Figure~\ref{fig:abscal} to
signal-only simulations containing unlensed \lcdm\ $E$-mode power. The accuracy
of this procedure is limited by the 
matrix purification, and so is an upper limit.  We find that gain
variation within the focal plane contributes \etb\ leakage
corresponding to $r < 5.3\times 10^{-5}$.

A separate issue is temporal gain variation. Temporal gain variation per
se is not a systematic (the full season coadded maps are calibrated against
\textit{Planck}), nor is static differential gain or temporal variation of
differential gain on timescales longer than $\sim9$~hr, the timescale over
which we perform the fit of the deprojection templates to the data (see
 Appendix~\ref{sec:practical}).  However, temporal variation of the differential gain on
timescales shorter than the 9~hr deprojection timescale will produce \ttp\ leakage that does not
fully deproject.  We therefore reject $\sim1$~hr blocks of data (``scansets'')
from channels
whose el-nod-derived gains change by more than $30\%$ as measured at the
beginning and end of the scanset, and reject pairs whose ratio of
gains changes by more than $10\%$ (see Section~13.7 of the Instrument Paper). As
discussed in Appendix~\ref{sec:practical} and in Section~ IV.F of the Results Paper, the
deprojection timescale was chosen as a compromise between the desire for
robustness against temporal variation of sytematics (favoring shorter
timescales) and the desire to minimize unnecessary filtering of signal (favoring
longer timescales). We note that before this timescale was settled upon, the
power spectrum results using deprojection performed on hour-long timescales were
consistent with those using deprojection performed on 9~hr long timscales,
modulo the additional \etb\ variance resulting from the more aggressive
filtering. We regard this as empirical evidence against the existence of leakage
from unknown temporal differential gain variation at relevant levels.

\subsection{Crosstalk}
\label{sec:crosstalk}

The leakage from the forms of crosstalk we expect in \bicep2
\citep{brevikthesis} is easily incorporated into our simulation pipeline. The
simulated timestreams from a detector's multiplexer neighbors are simply multiplied by
constants reflecting the level of crosstalk and added to the detector's
timestream.  We have measured levels of crosstalk between channels in a variety
of ways: first, we use cosmic ray hits on the focal plane that induce large
changes in the signal on a given detector to map out the relative pickup on
other detectors, yielding a non-symmetric $N_{\mathrm{detector}}\times N_{\mathrm{detector}}$
matrix of crosstalk coefficients. Second, we determine crosstalk coefficients
for nearby detectors by cross-correlating individual detector CMB $T$ maps offset
by the known angular distance to the channel next to it in the multiplexing
ordering scheme. Third, we determine the crosstalk from nearby detectors by
fitting a 2D Gaussian to the secondary beams that are seen with high
signal-to-noise in individual detector beam maps, such as in
Figure~\ref{fig:beammap}. Fourth, we have extended the deprojection algorithm to
remove crosstalk leakage. (We do this by fitting the differential gain leakage
template of a detector pair's two multiplex neighbors to that detector pair;
averaging the coefficients over three years is a measure of the crosstalk
coefficient.)

Crosstalk \ttp\ leakage will partially cancel when coadding detectors within a
multiplexing column into maps \citep{sheehythesis}.  Detector pairs that are
nearest neighbors in the multiplexing ordering scheme are second nearest
neighbors in the physical layout of the focal plane. Incrementing in multiplex
samples, every other detector along a physical row is first sampled, then the
interleaved detectors are sampled in the reverse physical direction. As a
consequence, the crosstalk induced \ttp\ leakage on two physically adjacent
detector pairs is equal in magnitude but opposite in sign when they are pointed
at the same location on the sky.
If the crosstalk coefficient is the same for all
detectors, the leakage almost fully cancels when adding the data from adjacent
pairs to form maps.  Timestream simulations confirm that the cancellation
mechanism is highly effective as long as the average crosstalk is similar
between channels upstream in the multiplexing order and channels
downstream in the multiplexing order, which our various measurements indicate 
is the case.

Direct simulations of crosstalk coefficients derived using all of the methods
described above predict similarly small levels of \ttp\ leakage. The least noisy
and most easily interpretable of these methods is the fitting to beam
maps. Simulation of the measured per-pair crosstalk, which has a median of
$\simeq0.3\%$, predicts leakage corresponding to $r \simeq 3.2\times10^{-3}$,
which we adopt as the predicted systematic contamination.

\subsection{Ghost Beams}
\label{sec:ghostbeams}

In addition to the $8\deg\times8\deg$ beam maps described
in Section~\ref{sec:beamsim}, we map the beam response out to radius $\sim20\deg$ using a
bright non-thermal source.  We observe a
small-amplitude ``ghost beam'' for each detector located at the position of that
detector's beam reflected across the boresight axis.  These likely result from
reflections in the optics chain. The peak amplitude of these ghost beams is
small, $\simeq4\times10^{-4}$ relative to the main beams. We can detect them
because in the large beam maps we use a brighter microwave source than that used
for the main and extended beam maps. We fit and measure
the differential elliptical Gaussian parameters of these ghost beams, which are
generally different from those of the corresponding main beam.  We directly
simulate the \ttp\ leakage from mismatched ghost beams by using the elliptical
Gaussian convolution approach described in Section~\ref{sec:ellipconv}. The predicted
leakage is small, corresponding to $r\simeq7.2\times10^{-6}$.

\subsection{Polarization Angles}

We divide the residual \etb\ leakage from polarization angle miscalibration into
a systematic (fully coherent) and a random component. 

\subsubsection{Systematic Polarization Angle Error}
\label{sec:syspol}

Section~VIII.B of the Results Paper describes \bicep2's
procedure for self-calibrating the overall polarization angle orientation of the
detectors, which removes the systematic component.  Summarizing this procedure, we
find that, prior to calibration, the high-$\ell$ $TB$ and $EB$ spectra are
consistent with a coherent $-1.1\deg$ polarization angle error and apply an equal and
opposite rotation to the polarization maps prior to computing power
spectra. Doing so filters the \etb\ leakage from a systematic polarization angle
error.  Given 
the analytic expression for \etb\ leakage found in Equation~5 of \citet{keating13} and
assuming a \lcdm\ $EE$ spectrum, we then calculate the maximum possible residual
miscalibration by determining the coherent rotation at which \bicep2's $TB$ and $EB$
spectra would show significant non-zero power. For coherent angle errors $\ll 1$~rad,
the contamination of $BB$ scales quadratically with the angle error, while 
contamination of $TB$ and $EB$ scales linearly. The $TB$ and $EB$ spectra are
therefore contaminated more strongly 
than $BB$ for a given angle error, and the resulting 
upper limit on $BB$ contamination is negligible. Given the sensitivity of
\bicep2's $TB$ and $EB$ spectra, a systematic
polarization angle rotation of $0.20\deg$ produces a failure of the $EB$
$\chi$ statistic, which tests for coherently positive or negative residuals
(and is defined in Equation~8 of the Results paper) in $95\%$ of
\bicep2's signal-plus-noise simulations, which 
limits the possible \etb\ leakage to $r<4.0\times10^{-4}$.

\subsubsection{Random Polarization Angle Error}

Self-calibration removes the leakage from a systematic error in polarization
angle, but errors in relative polarization angles between detectors still
produce additional \etb\ leakage. We measure detector polarization angles with a
dielectric sheet calibrator~\citep{takahashi10}. The measurements are described
in detail in Section~ 11.4 of the Instrument Paper. After accounting for the $-1.1\deg$
systematic rotation, the difference between the measured and nominal
polarization angles is small. The distribution is approximately Gaussian, with
an rms of $0.14\deg$. We have estimated that the precision of these measurements is
$\sim0.2\deg$~\citep{aikinthesis}, so the relative misalignment of individual
detector polarization angles is not measured with high significance.

Leakage from random polarization angle errors
is easily simulated. We simply assume one set of per-detector polarization
angles in the simulation stage and use another in the map making stage. The
resulting $Q$ and $U$ maps contain \lcdm\ $E$-mode power that has been rotated into
\bmode\ power. The difference between spectra estimated from maps made with the
``wrong'' polarization angles and the known, ``correct'' polarization angles is
the \etb\ leakage from polarization angle error.  Simulation of the
\etb\ leakage from a random polarization angle error of $0.2\deg$ rms predicts
contamination corresponding to $r\lesssim5.0\times10^{-5}$.

\subsection{Cross-polar Response}

In addition to the miscalibration of the polarization angle, there can be
higher order cross-polar response terms in the beam that give rise to
\etb\ leakage.  If a pair that is analyzed assuming it responds only to $Q$
polarization actually has some response to $U$, there will be polarization
rotation leading to \etb\ leakage.  Any $U$ response that is uniform across the
beam (i.e. a monopole) will be included in the polarization angle calibration.
However, non-uniform $U$ response cannot be fully removed by polarization angle
calibration.  We have measured the beam patterns of response to $Q$ and $U$ for
\bicep2\ with a rotating polarized source in the far field. 
After pair-differencing, the response to $U$ at any location is
$\lesssim0.8\%$ of the response to $Q$ at the peak.
The corresponding \etb\ leakage is
at the level of $r\lesssim10^{-3}$.  Boresight rotation and variation among
detectors would further reduce this effect, so this level is a conservative
upper limit.  See the Beams Paper for further details.

\subsection{Thermal Instability}
\label{sec:thermalinstability}

As mentioned in Section~\ref{sec:instdesign} fluctuations
in the focal plane temperature will produce spurious polarization if the
response of detectors to a change in focal plane temperature differs within a
detector pair. Two NTD thermistors are located on the \bicep2\ focal plane and
are read out at the same rate as the detectors. Using the heaters on the focal
plane normally used for active thermal control, we directly measured
individual detector's response to a change in focal plane temperature by varying
the focal plane temperature over a range $\sim10$~mK.

We estimate the leakage from thermal fluctuations by replacing each detector's
timestream with the measured focal plane temperature multiplied by that
detector's thermal response. We then make maps exactly as for the real data, but
using these timestreams substituted for the real ones. (We co-add focal plane
temperature data from only the 2011 and 2012 seasons because the NTD thermistor
biases rendered focal plane temperature data from 2010 noisy.) This procedure
naturally includes the mitigation of leakage from ground subtraction
and averaging down across detectors. The polarization maps produced in
this manner are consistent with the readout noise of the NTD thermistors. The
$BB$ spectrum of these maps is a directly measured upper limit on leakage from
thermal drift in the focal plane and corresponds to $r<1.2\times10^{-5}$.

\subsection{Detector Transfer Functions}

The temporal response of \bicep2's detectors is very fast. Typical detector time
constants are $\tau\sim1$~ms, with a few detectors having $\tau=5-8$~ms. We
therefore do not deconvolve the detector response function from the time ordered
data. In principle, a mismatch of detector response results in \ttp\ leakage.

We have measured each detector's transfer function (the Fourier transform of the
temporal response) with high signal-to-noise --- details are in Section~10.6 of the
Instrument Paper.  We simulate a conservative upper limit of the \ttp\ leakage
from transfer function mismatch by convolving simulated detector timestreams
with exponential response functions having $10\times$ the measured time
constants. This simulation predicts \ttp\ leakage at a level corresponding to
$r\simeq5.7\times10^{-4}$. We also verify from simulation that the scan
direction jackknife is contaminated by transfer function mismatch more strongly
than the signal spectra, making it a robust additional check against
leakage.

\subsection{Magnetic Pickup}

We do not attempt to directly simulate magnetic pickup in the SQUIDs. We
nonetheless have multiple lines of evidence disfavoring significant magnetic
contamination. First, and most importantly, the magnetic shielding employed by
\bicep2\ was found to suppress magnetic pickup from external sources by a factor
$\sim10^6$ (see Section~5.3 of the Instrument Paper for details).

Second, ground subtraction filtering exactly removes any signal that is constant
over hour-long timescales and fixed with respect to the telescope scan or to the
ground. Ground subtraction is performed on individual channels separately, so
detector to detector differences in magnetic response are accounted for.  Such a
scan- or ground-fixed signal includes the Earth's magnetic field or any other
magnetic field that is fixed with respect to the telescope superstructure. Only
the slight misalignment in azimuth of the time-ordered points of corresponding
telescope scans would cause imperfect subtraction. Simulation of this effect
shows that ground subtraction filters scan synchronous signals to below
$r\lesssim1\times10^{-8}$.

Third, magnetic pickup varies from channel to channel (especially across
multiplexing columns) due to differences in shielding environment.
Investigation of channels with deliberately severed TES-SQUID links (dark
SQUIDS) and special calibration runs with detectors in the normal state do show
column-to-column differences in magnetic pickup. We therefore expect the Mux
column jackknife to be a moderately sensitive probe of magnetic
contamination. (Within a given column, multiplex neighbors show highly
correlated magnetic sensitivity, so that the Mux row jackknife, which splits the
data within a column by interleaved channels, is a very weak test of magnetic
pickup.)

Lastly, we note that \bicep1\ did not use SQUID readouts and thus had no
sensitivity to magnetic fields, so that a \bicep1$\times$\bicep2
cross-spectrum will show no contamination from magnetic pickup.

\subsection{Electromagnetic Interference}

\begin{figure}
\begin{center}
\includegraphics[width=1\columnwidth]{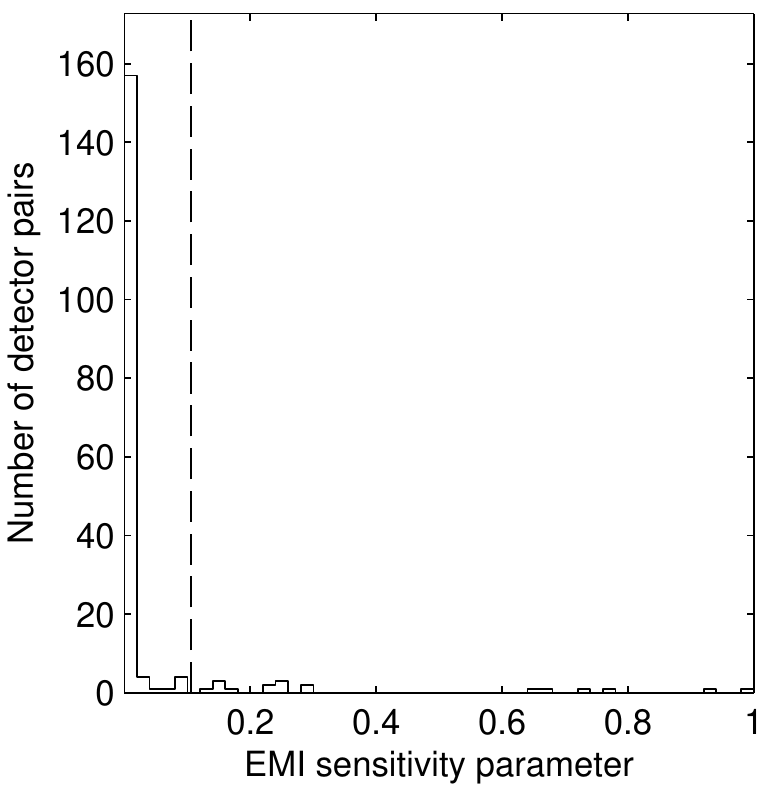}
\end{center}
\caption{EMI sensitivity parameter for all detector pairs included in \bicep2's
  maps. The parameter is proportional to the contribution of possible EMI from a
  given detector pair to \bicep2's polarization power spectra. The dashed line
  indicates the cut threshold used in constructing the EMI sensitivity pair
  exclusion test.}
\label{fig:satcomcut}
\end{figure}

After the completion of \bicep2\ observations, analysis of non-ground-subtracted
galactic maps revealed clear contamination during specific temporal periods
resulting from a satellite transmitter operating at the Amundsen-Scott South
Pole research station. The transmitter uplink operates in the S-band (2~GHz) for
approximately 7~hr per sidereal day. Details are given in Section~11.8 of the
Instrument 
Paper.

A few factors limit the impact of electromagnetic interference (EMI) in CMB
observations. One, because the satellite uplink schedule is locked to sidereal
time, it so happened that the CMB field mapped by \bicep2\ was always in the opposite direction from
the transmitter when it was on. The opposite was true for the galactic field mapped
by \bicep2. Second, a small subset of detector pairs shows much stronger
differential sensitivity to the EMI than others, indicating that pair selection
jackknives should fail if EMI were contributing significant power. Third, the
EMI in these few pairs is visible in raw pair-difference timestreams prior to
ground subtraction so that we can study its phenomenology. We find that the EMI
is fixed in azimuth and constant in time, and that ground subtraction filters it
nearly perfectly. The only contamination that occurs is when the transmitter
turns on or off during a scanset, causing imperfect ground subtraction.

We have performed a pair exclusion test to test for EMI.
Figure~\ref{fig:satcomcut} shows an EMI sensitivity parameter for all
\bicep2\ detector pairs used in the final maps. The parameter is proportional to
the square of the level of EMI pickup seen in non-ground-subtracted
pair-difference maps of the Galactic field. The
contribution of a given pair's contamination to power spectra should thus
scale with this parameter.

Because a few pairs dominate possible contamination from EMI, a pair exclusion
test is more sensitive than a jackknife that splits based on the EMI
statistic. We performed the test by re-coadding the real data maps and 50
signal-plus-noise simulations, excluding the 18 most sensitive pairs. The change
in the resulting $BB$ bandpowers, $\Delta C_{\ell}^{BB} = C_{\ell}^{BB,\mathrm{cut}} -
C_{\ell}^{BB}$, is consistent with the slightly altered noise and weighting of
the new map and is statistically insignificant.  In fact the first five bandpowers
shift slightly up when making the cut.
The $\chi$ statistic (also used in Section~\ref{sec:syspol})
 has PTE~$\simeq0.05$. The ratio of the mean EMI sensitivity parameter
with and without the pair cut implies that the cut reduces any EMI contamination
that is present by $\simeq90\%$ in the polarization power spectra. Taking this
into account, we place an upper limit on contamination from EMI at
$r\lesssim1.7\times10^{-3}$ with 95\% confidence. At contamination greater than
this, the pair exclusion test we performed would have a $95\%$ likelihood of
producing statistically significant negative $\Delta C_{\ell}^{BB}$'s.

We also note that while the coupling of EMI is not fully known, it did not
appear to involve the detector antennas, most likely coupling directly to the TES
islands.  The mechanism should therefore manifest differently or not at all in
\bicep1, which did not use TES technology.  The contaminating power will
therefore not be present in a \bicep1$\times$\bicep2\ cross-spectrum.

\begin{deluxetable}{lc}[t]
\tablecolumns{2} \tablewidth{0pc} \tablecaption{Instrumental systematics
\label{tab:sysfinal}} \tablehead{\colhead{Systematic}
  & \colhead{Characteristic $r$}} 
  \startdata 
  Crosstalk & $\simeq3.2\times10^{-3}$ \\
  Beams (including gain mismatch) & $<3.0\times10^{-3}$ \\
  EMI & $\lesssim1.7\times10^{-3}$ \\
  Cross polar response & $\lesssim 10^{-3}$ \\
  Eetector transfer functions & $<5.7\times10^{-4}$ \\
  Systematic polarization angle error & $<4.0\times10^{-4}$ \\
  Gain variation \etb & $<5.3\times10^{-5}$ \\
  Random polarization angle error & $\lesssim5.0\times10^{-5}$ \\
  Thermal fluctuations & $<1.2\times10^{-5}$ \\
  Ghost beams & $\simeq7.2\times10^{-6}$ \\
  Scan synchronous contamination & $\lesssim1\times10^{-8}$ \\
  \hline \\
  Total & $\simeq(3.2-6.5)\times10^{-3}$
  \enddata
  \tablecomments{The comparable characteristic $r$ of \bicep2's statistical
    uncertainty is $r=3.1\times10^{-2}$.  }
\end{deluxetable}

\subsection{Overall Achieved Systematics Level}

\begin{figure}
\begin{center}
\includegraphics[width=1\columnwidth]{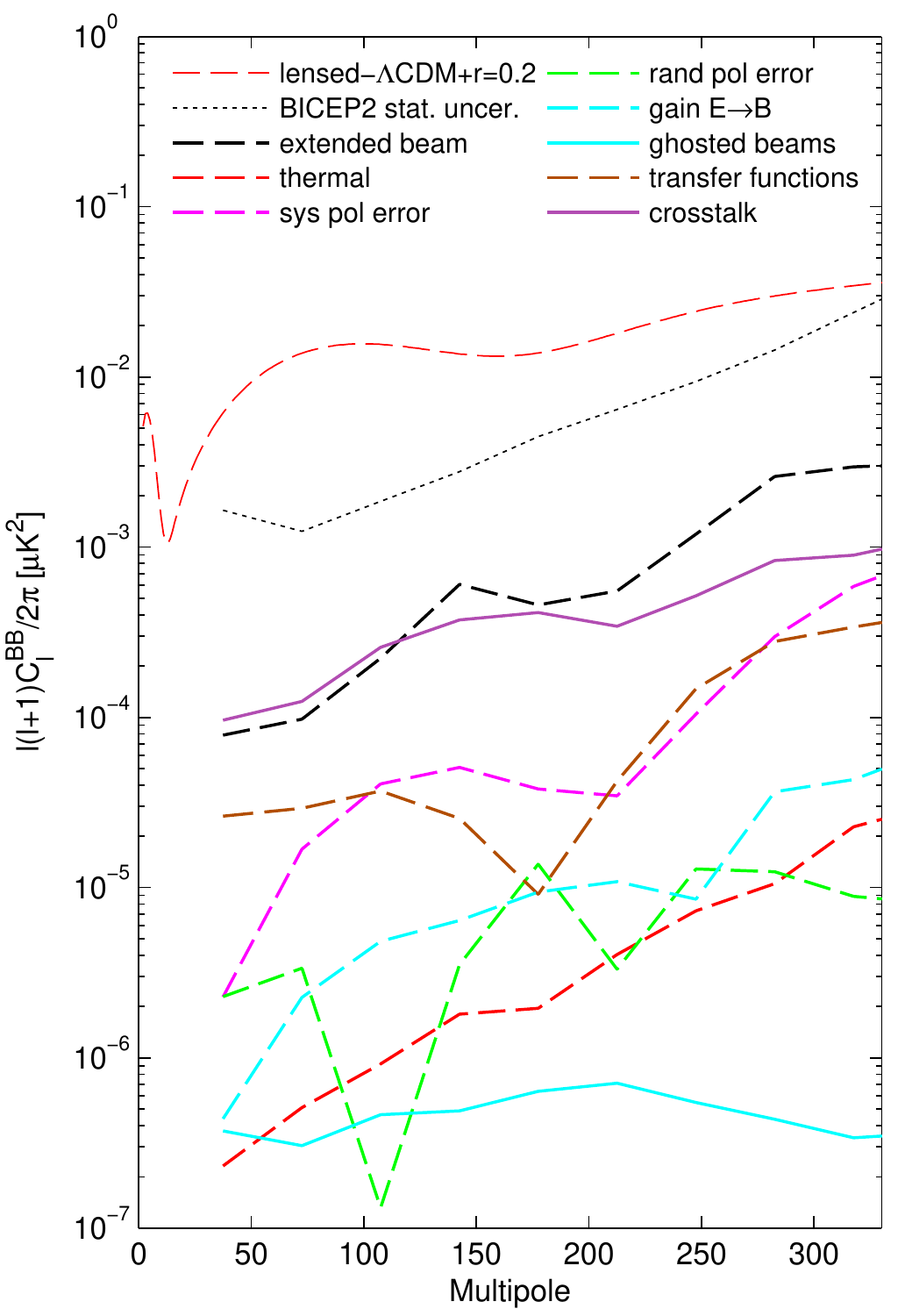}
\end{center}
\caption{Estimated levels of systematics as compared to a
  lensed-\lcdm+$r=0.2$ spectrum.  Solid lines indicate expected contamination.
  Dashed lines indicate upper limits. All systematics are comparable to or
  smaller than the extended beam mismatch upper limit, which is smaller than
  \bicep2's statistical uncertainty. }
\label{fig:sysfigs}
\end{figure}

Figure~\ref{fig:sysfigs} shows the expected $BB$ contamination or upper limits
on contamination from the individual sources of systematics considered in this
section.  Table~\ref{tab:sysfinal} summarizes the $r$-values quoted for them
above. To obtain a final constraint on instrumental systematics, we add the 
values for systematics quoted as predicted values and add the upper limits in
quadrature --- or the upper limit divided by two for those that are $95\%$
confidence upper limits (EMI and systematic polarization angle error).
We note
that magnetic pickup, which is expected to be negligible, has not been included.
The total contamination and its uncertainty is almost completely dominated by
the expected \ttp\ leakage from crosstalk and the uncertainty on the residual
\ttp\ leakage from beam shape mismatch.

\section{Conclusions}
\label{sec:conclusions}

\bicep2's systematic control demonstrates the validity of our
experimental approach for high signal-to-noise CMB polarimetry.  Instrumental
systematics are a negligible contributor to \bicep2's $BB$ auto spectrum. They
are also small compared to \bicep2's instrumental noise. Deprojection mitigates
\ttp\ leakage from beam mismatch to a level at least sufficient to detect
$r\simeq0.003$. Other calibration measurements allow us to limit additional
systematics to $r\lesssim0.006$. For comparison, \citet{bischoff13} claimed
a limit on instrumental systematics of $r<0.01$.

Cosmic variance limited measurements of CMB polarization promise to constrain
\lcdm\ cosmology with greater precision than temperature data alone
\citep{snowmass,galli14}. They will require control of systematics similar to
\bicep2's. Leakage from constant fractional beam mismatch (not including
differential gain) scales with beam size, with leakage peaking near
the beam scale. Telescopes with larger apertures than \bicep2\ but similar
fractional mismatch will have \ttp\ leakage that peaks at correspondingly
smaller angular scales. We expect deprojection to be equally effective at
filtering \ttp\ leakage at higher multipoles, as the \textit{Planck} temperature maps
should be sufficiently low-noise.  Even if they were not, it is possible to
deproject using an experiment's own temperature map, which should always have
sufficient sensitivity at the angular scales required. We do not take this
approach out of simplicity to avoid complications involved with map filtering.

In summary, \bicep2's proven systematics control demonstrates the power of
scanning, small aperture, pair differencing bolometric polarimeters that do not
use rotating half-wave plates or other polarization modulators.  Our
experimental approach will maintain its usefulness as we continue characterizing
the detected \bmode\ signal.

\acknowledgements

\bicep2 was supported by the U.S. National Science Foundation under
grants ANT-0742818 and ANT-1044978 (Caltech/Harvard) and ANT-0742592
and ANT-1110087 (Chicago/Minnesota).  The development of antenna-coupled
detector technology was supported by the JPL Research and Technology
Development Fund and grants 06-ARPA206-0040 and 10-SAT10-0017
from the NASA APRA and SAT programs.  The development and testing of
focal planes were supported by the Gordon and Betty Moore Foundation
at Caltech.  Readout electronics were supported by a Canada Foundation
for Innovation grant to UBC.  The receiver development was supported
in part by a grant from the W. M. Keck Foundation. Partial support for 
C. Sheehy was also provided by the 
Kavli Institute for Cosmological Physics at the University of Chicago through
grant NSF PHY-1125897 and an endowment from the Kavli Foundation and its founder
Fred Kavli. 
The computations in this paper were run on the Odyssey cluster
supported by the FAS Science Division Research Computing Group at
Harvard University.  Tireless administrative support was provided by
Irene Coyle and Kathy Deniston.

We thank the staff of the U.S. Antarctic Program and in particular
the South Pole Station without whose help this research would not have been possible.
We thank all those who have contributed past efforts to the \bicep /\keck\
series of experiments, including the \bicepone\ and \keck\ teams,
as well as our colleagues on the \spider\ team with whom
we coordinated receiver and detector development efforts at Caltech.
We dedicate this paper to the memory of Andrew Lange, whom we sorely miss.

\appendix

In Appendix~\ref{sec:beamparam}, we define the elliptical Gaussian
parametrization that we use to characterize \bicep2's beams. In
Appendix~\ref{sec:beamheuristic}, we offer a heuristic explanation of the
coupling of mismatched elliptical Gaussians to various linear
combinations of spatial derivatives of $T(\mathbf{\hat{n}})$. In
Appendix~\ref{sec:deprojmath} we formally derive these linear combinations,
which are the leakage templates summarized in Table~\ref{tab:deprojection}; we
also discuss the practical issues involved in implementing deprojection for
\bicep2. In Appendix~\ref{sec:beammapsimappendix}, we discuss how we estimate
the uncertainty of the \ttp\ predicted from beam map simulations of \bicep2's
measured beams, which is illustrated as shaded bands in
Figure~\ref{fig:beammapjacks} and sets the beam contamination upper limit in
Figure~\ref{fig:sysfigs} and Table~\ref{tab:sysfinal}.  \\

\section{Parametrization of beam shapes}
\label{sec:beamparam}

We define an elliptical Gaussian beam with respect to a coordinate
system, $(x,y)$, that is fixed with respect to the focal plane as projected onto
the sky. The parametrization of the beams is thus independent of the telescope
orientation.  The axes of the coordinate system are orthogonal great circles
intersecting at each detector's beam center.  A detailed treatment is given
in the Beams Paper. However, for the discussion below, we note that, to
a very good approximation over the extent of \bicep2's focal plane, the axes
point along the rows and columns of \bicep2's pixels.

In general, all parametrizations of elliptical Gaussians in Cartesian
coordinates are of the form
\begin{equation}
B(\mathbf{x}) = \frac{1}{\Omega}\exp\left[-(\vec{\mathbf{x}}-\vec{\boldsymbol{\mu}})^T
  \boldsymbol{\Sigma}^{-1} (\vec{\mathbf{x}}-\vec{\boldsymbol{\mu}})/2 \right]
\end{equation}

\noindent where $\vec{\mathbf{x}}$ is a 2D position vector,
$\vec{\boldsymbol{\mu}}$ is coordinate of the peak, $\Omega$ is a normalization
constant, and $\boldsymbol{\Sigma}$ is a covariance matrix.

Differences in parametrization arise in the specification of
$\boldsymbol{\Sigma}$. One common way of parametrizing an elliptical Gaussian is by
specifying its major and minor widths, $\sigma_{maj}$ and $\sigma_{min}$, and
the rotation $\theta$ of its major axis. In this parametrization,
\begin{equation}
\boldsymbol{\Sigma} = \mathbf{R}^{-1}\mathbf{C}\mathbf{R}
\end{equation}

\noindent with a covariance matrix $\mathbf{C}$ given by
\begin{equation}
\mathbf{C} = \left(\begin{array}{cc} \sigma_{maj}^2 & 0 \\ 0 & \sigma_{min}^2
\end{array}\right)
\end{equation}

\noindent and a rotation matrix given by
\begin{equation}
\mathbf{R} = \left(\begin{array}{cc} \cos\theta & \sin\theta \\ -\sin\theta & \cos\theta
\end{array}\right).
\end{equation}

We choose an alternate specification of the covariance matrix, given by
\begin{equation}
\boldsymbol{\Sigma} = \left(\begin{array}{cc} \sigma^2(1+p) & c\sigma^2 \\ c\sigma^2 &
  \sigma^2(1-p)
\end{array}\right),
\end{equation}

\noindent where $|p|<1$, $|c|<1$, and $(p^2+c^2)<1$. An elliptical Gaussian with
a horizontally oriented major axis has $+p$, and an elliptical Gaussian with a
vertically oriented major axis has $-p$. We refer to both as having
``plus-ellipticity,'' which we denote with $+$. An elliptical Gaussian with a
major axis oriented $\pm45\deg$ with respect to the x-axis has $\pm c$, which we
refer to as ``cross-ellipticity'' and denote with $\times$.

Expressed in the more familiar terms of $\sigma_{maj}$, $\sigma_{min}$ and
$\theta$, we can write
\begin{eqnarray}
\sigma^2 & = & \left(\sigma_{maj}^2+\sigma_{min}^2\right)/2 \label{eq:pc1} \\ p
& = & e \cos 2\theta \\ c & = & e \sin 2\theta
\end{eqnarray}

\noindent where we have defined the total ellipticity to be
\begin{equation}\label{eq:pc2}
e=\sqrt{p^2+c^2}=\left(\frac{\sigma_{maj}^2-\sigma_{min}^2}
{\sigma_{maj}^2+\sigma_{min}^2}\right).
\end{equation}

\section{Heuristic description of leakage from beam mismatch}
\label{sec:beamheuristic}

In this appendix, we discuss qualitatively how the differences of elliptical
Gaussians, illustrated in Figure~\ref{fig:differencebeams}, couple to the
spatial derivatives of the temperature field, $T(\mathbf{\hat{n}})$.

\subsection{Gain}
\label{sec:gainmismatch}

We model detector gain mismatch as a simple difference in normalization of
circular Gaussians of nominal width given by $\delta g = g_A - g_B$ and
$(g_A+g_B)/2=1$.  Because the difference beam, $B_{\delta g}(\mathbf{\hat{n}})$,
is simply a scaled version of the nominal beam,
the resulting spurious signal is
a scaled version of the nominal beam-smoothed temperature,
$\tilde{T}(\mathbf{\hat{n}})$ (where the tilde denotes convolution by the
nominal beam). Put another way and dropping the explicit dependence on
$\mathbf{\hat{n}}$, the zeroth derivative of $\tilde{T}$ multiplied by $\delta g$
is added (``leaks'') to the pair-difference timestream, $d_{T\rightarrow P}$.

\subsection{Differential Pointing}
\label{sec:pointingmismatch}

\begin{figure}[t]
  \begin{center}
    \begin{tabular}{c}
      \includegraphics[width=2.7in]{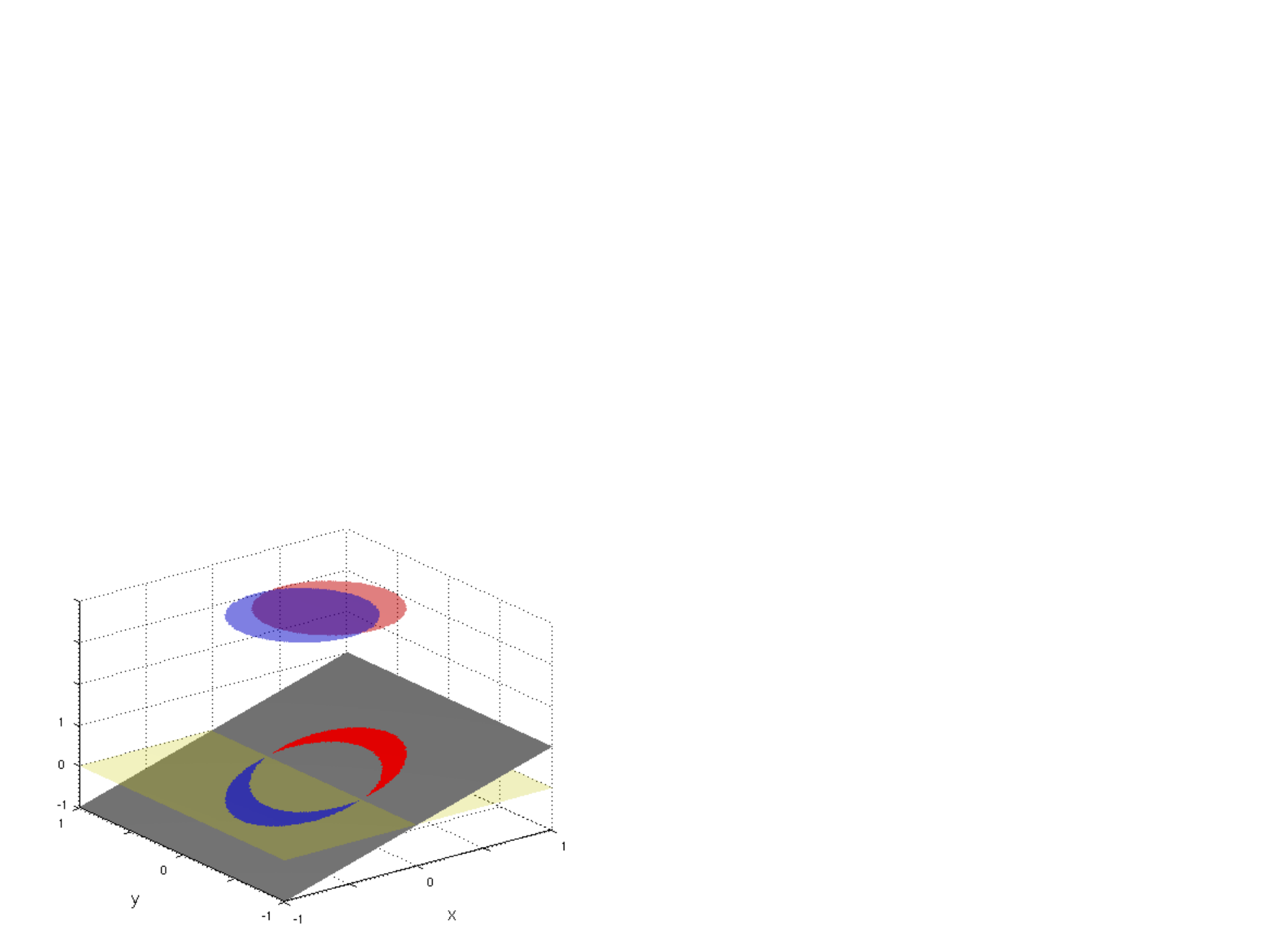} \\ (a)
      \\ \includegraphics[width=2.7in]{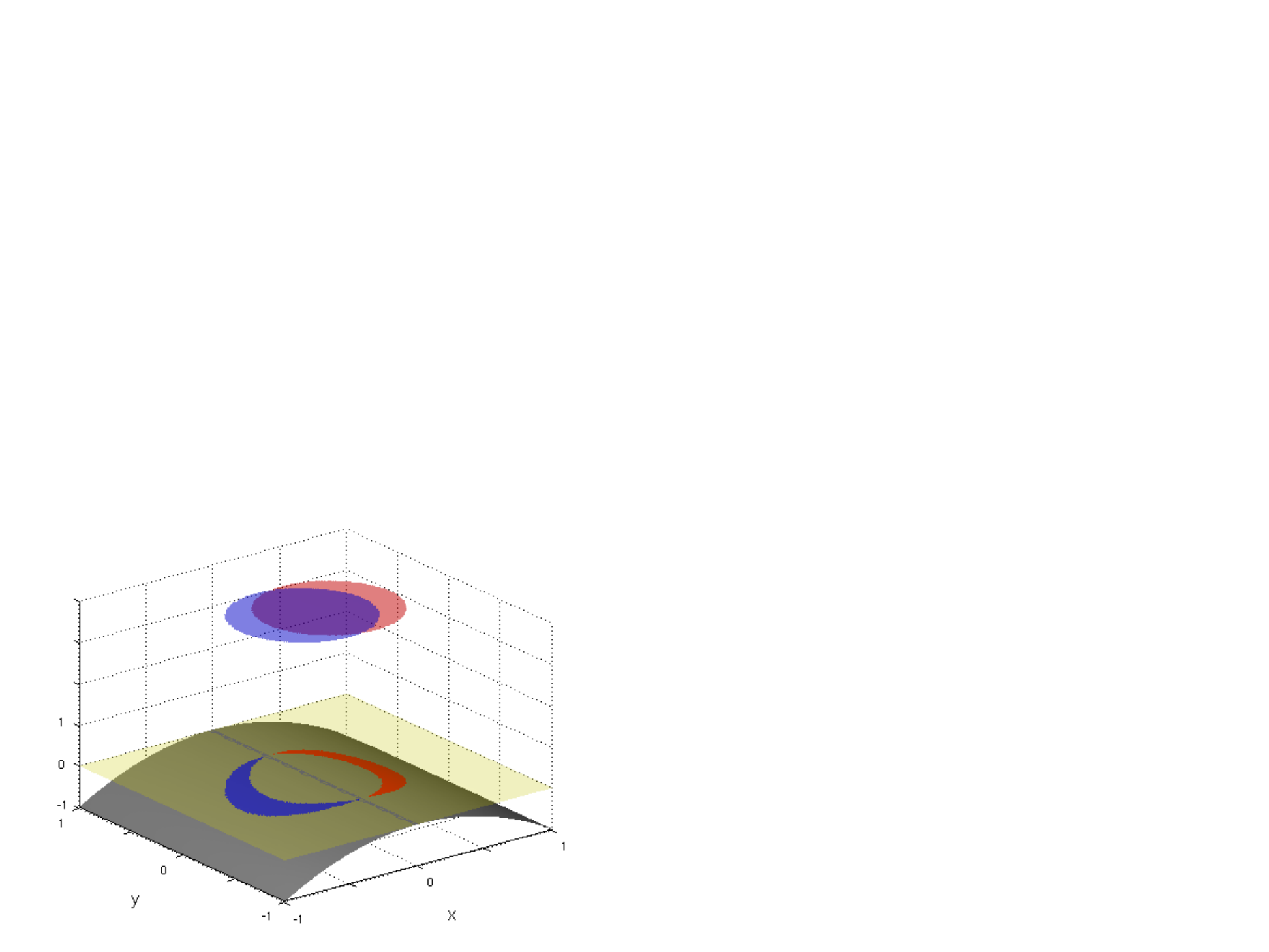} \\ (b) \\
    \end{tabular}
  \end{center}
  \caption[example] { \label{fig:manifolddiffpoint} Illustration of
    \ttp\ leakage resulting from differential pointing.  The gray plane
    represents a $T$ sky with (a) $\nabla_xT>0$, $\nabla^2_xT=0$,
    and (b) $\nabla_xT=0$, $\nabla^2_xT<0$.  The red and blue circles
    represent contour slices through the circular Gaussian beams of the A and B
    members of a detector pair, respectively.  The non-overlapping area is
    projected onto the $T$ plane. The transparent green plane at $z=0$ is simply
    for reference. The scenario in (a) leaks \ttp\ while (b) does not. }
\end{figure}

\begin{figure}[t]
  \begin{center}
    \begin{tabular}{c}
      \includegraphics[width=2.7in]{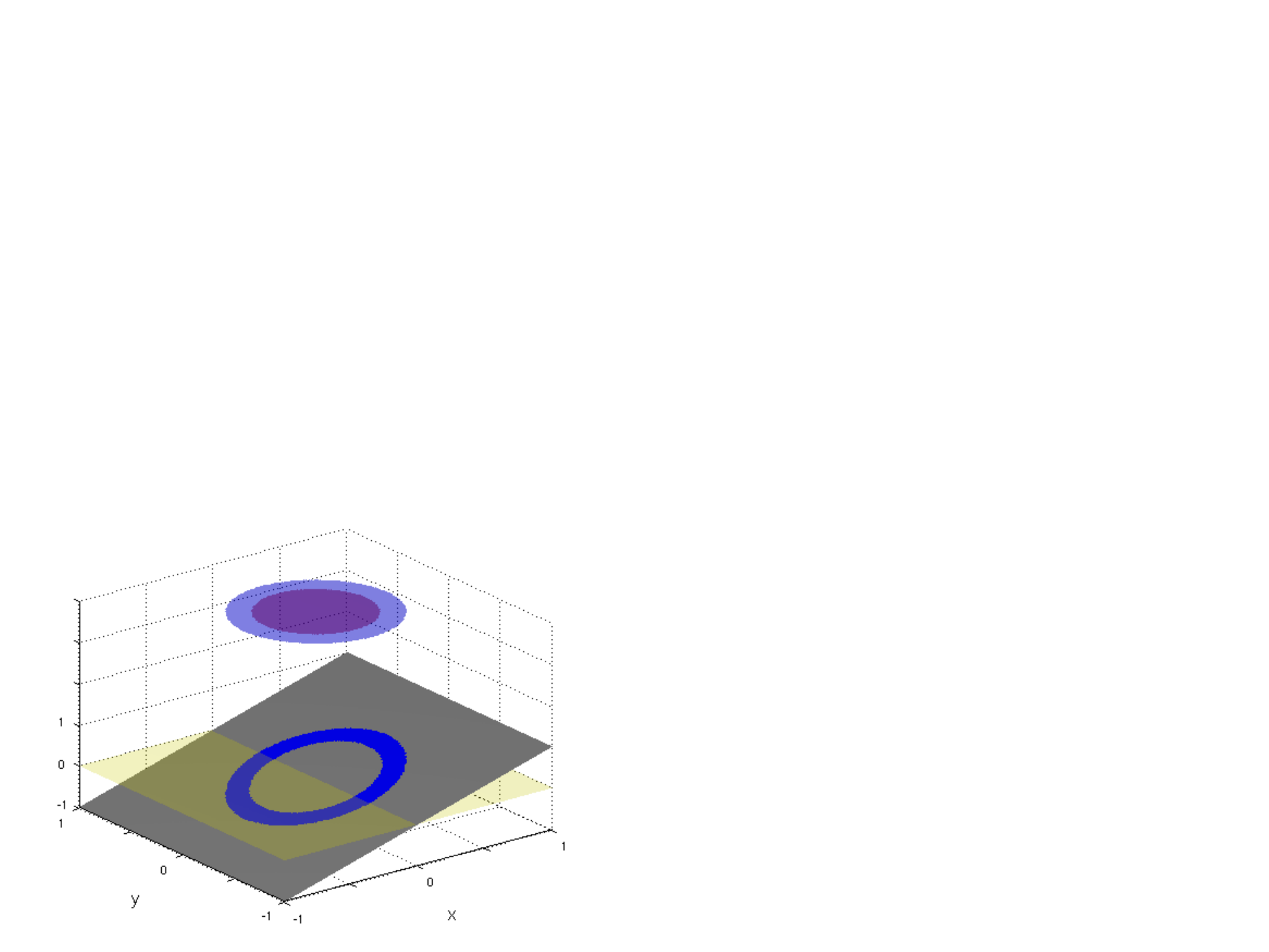} \\ (a)
      \\ \includegraphics[width=2.7in]{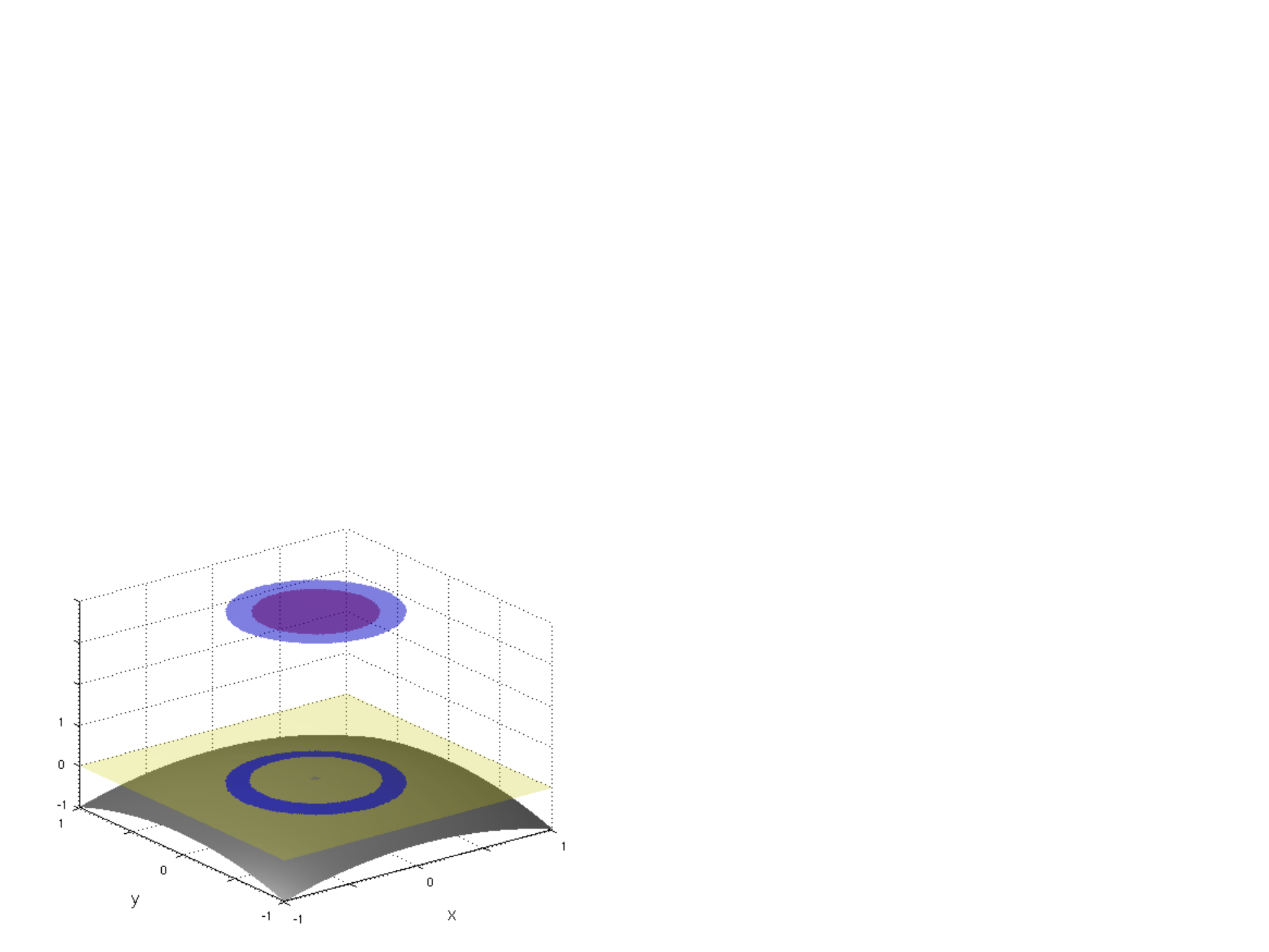} \\ (b)
    \end{tabular}
  \end{center}
  \caption[Differential beamwidth leakage]
          { \label{fig:manifoldbeamwidth}Illustration of \ttp\ resulting from
            differential beamwidth. The gray plane represents a $T$ sky with
            (a) $\nabla_xT>0$, $\nabla^2_xT=\nabla^2_yT=0$, and
            (b) $\nabla_xT=0$, $\nabla^2_xT=\nabla^2_yT<0$.  The
            scenario in (b) leaks \ttp\ while (a) does not.  }
\end{figure}

We model differential pointing as the difference of two circular Gaussians of
nominal width offset from each other in either the $x$-direction by an angular distance
$\delta x = x_A - x_B$ or the $y$-direction by a distance $\delta y = y_A -
y_B$.  The leakage from beams offset along arbitrary directions is a linear
combination of the leakage from these two orthogonal modes.

Differential pointing couples to the first derivative of $\tilde{T}$ in the
direction of the pointing offset. If the first derivative of $\tilde{T}$ is
zero, then regardless of where the A and B detectors are pointed, they both
observe the same temperature. (If $T$ were unpolarized, the resulting
pair-difference signal would be zero.) It is only if the first derivative of
$\tilde{T}$ in the direction of the centroid offset is non-zero that the
pair-difference timestream has contribution from $T$.

Figure~\ref{fig:manifolddiffpoint} illustrates the coupling of differential
pointing to the first derivative of $\tilde{T}$.  Because, in this scenario,
$\partial\tilde{T}/\partial x\equiv\nabla_x\tilde{T}>0$ and the
centroid offset is in the $x$-direction, the A detector measures a value for 
temperature that is slightly larger than the value at the mean A/B beam center
(pair centroid). The B detector measures a value that is slightly smaller. The
pair-difference signal, A$-$B, is therefore positive. In
Figure~\ref{fig:manifolddiffpoint}b, the first derivative at the pair centroid
is zero while the second derivative is non-zero.  In this case, both A and B
measure equally negatively offset signals and the pair-difference signal is
zero. Neither a non-zero first derivative in the direction perpendicular to the
centroid offset nor a non-zero second derivative in any direction produces
\ttp\ leakage.

An equivalent way of thinking about \ttp\ leakage resulting from beam mismatch
is to describe the leaked signal as that which results from the convolution of
$T$ with the difference beam, $B_{\delta}$ (Equation~\ref{eq:diff}). For differential pointing
the difference beam, illustrated in Figure~\ref{fig:differencebeams}, is a
dipole.  (Alternatively, in Figure~\ref{fig:manifolddiffpoint}, we could have
shown the dipole difference beam from Figure~\ref{fig:differencebeams} projected
onto the $T$ plane.)  Convolving $T$ with a dipole naturally produces a
beam-smoothed map of its first derivative. (The first derivative approximation
breaks down when $\delta x$ or $\delta y$ is much greater than the beamwidth.)

\subsection{Beamwidth}
\label{sec:beamwidthmismatch}

We model differential beamwidth as the difference of two circular Gaussians with
common beam centers but differing width given by
$\delta\sigma=\sigma_A-\sigma_B$. Differential beamwidth couples only to the
second derivatives of $T$, as illustrated in
Figure~\ref{fig:manifoldbeamwidth}. In Figure~\ref{fig:manifoldbeamwidth}a,
where the first derivative of $T$ is non-zero and the second derivative
zero, neither the A nor B detector measures a signal that is offset from the
value of $T$ at the pair centroid. There is no resulting \ttp\ leakage. In
Figure~\ref{fig:manifoldbeamwidth}b, where the first derivative of $T$ at
the pair centroid is zero but the second derivatives are non-zero, both
detectors measure a signal that is offset negatively.  However, the B detector,
which has a larger width than the A detector, measures a signal that is
\emph{more} negatively offset. The pair-difference signal is thus non-zero.
Also apparent from Figure~\ref{fig:manifoldbeamwidth}b is that a non-zero second
derivative in \emph{either} the $x$ or $y$ direction will produce \ttp\ leakage,
which explains why the net leakage couples to the sum of the orthogonal
derivatives.

\subsection{Ellipticity}
\label{sec:ellipticitymismatch}

We model differential ellipticity as the difference of two elliptical Gaussians
with either differing plus-ellipticity, given by $\delta p = p_A - p_B$, or
differing cross-ellipticity, given by $\delta c = c_A - c_B$.

\begin{figure*}[t]
  \begin{center}
    \begin{tabular}{cc}
      \includegraphics[width=2.7in]{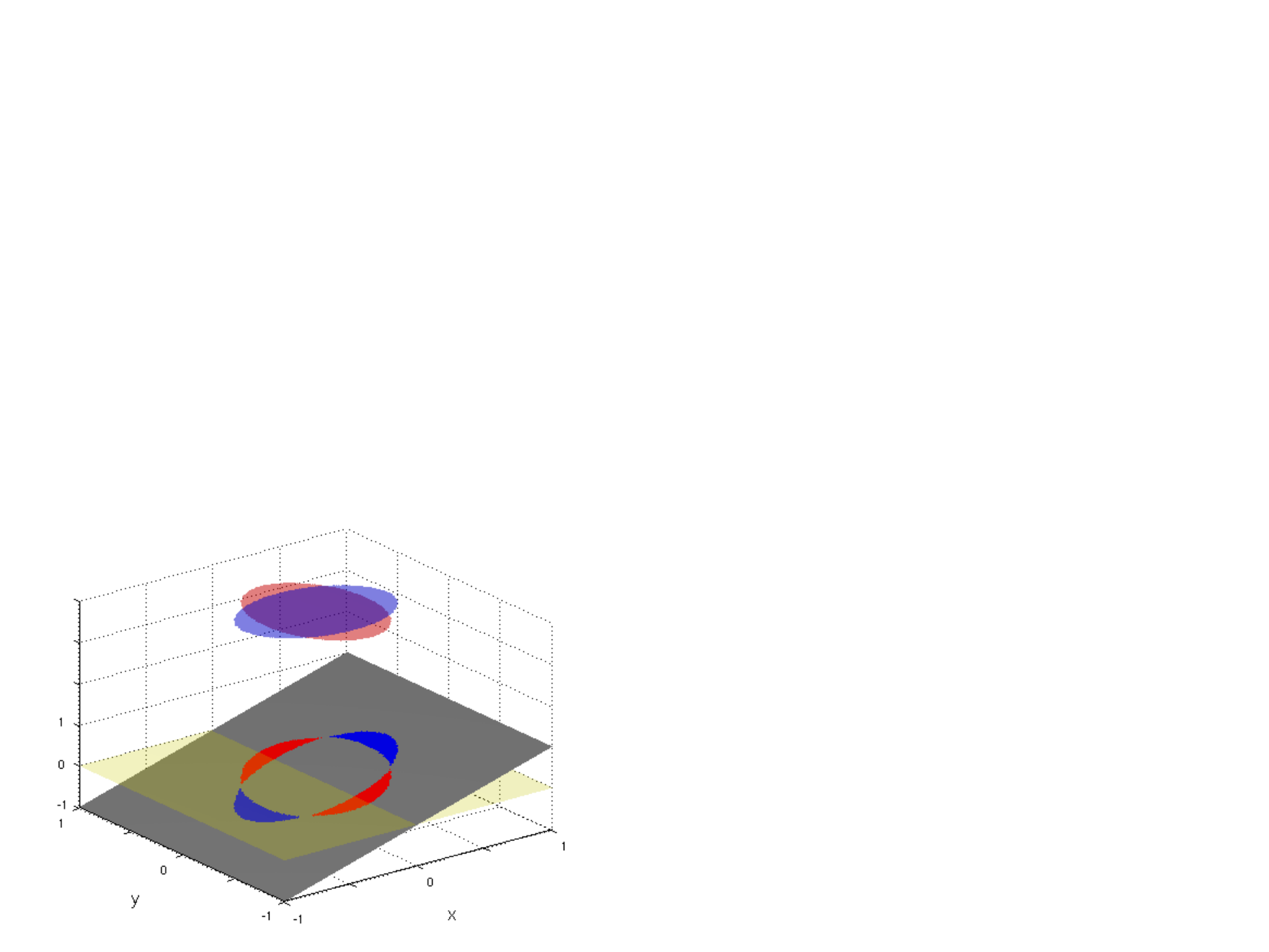} &
      \includegraphics[width=2.7in]{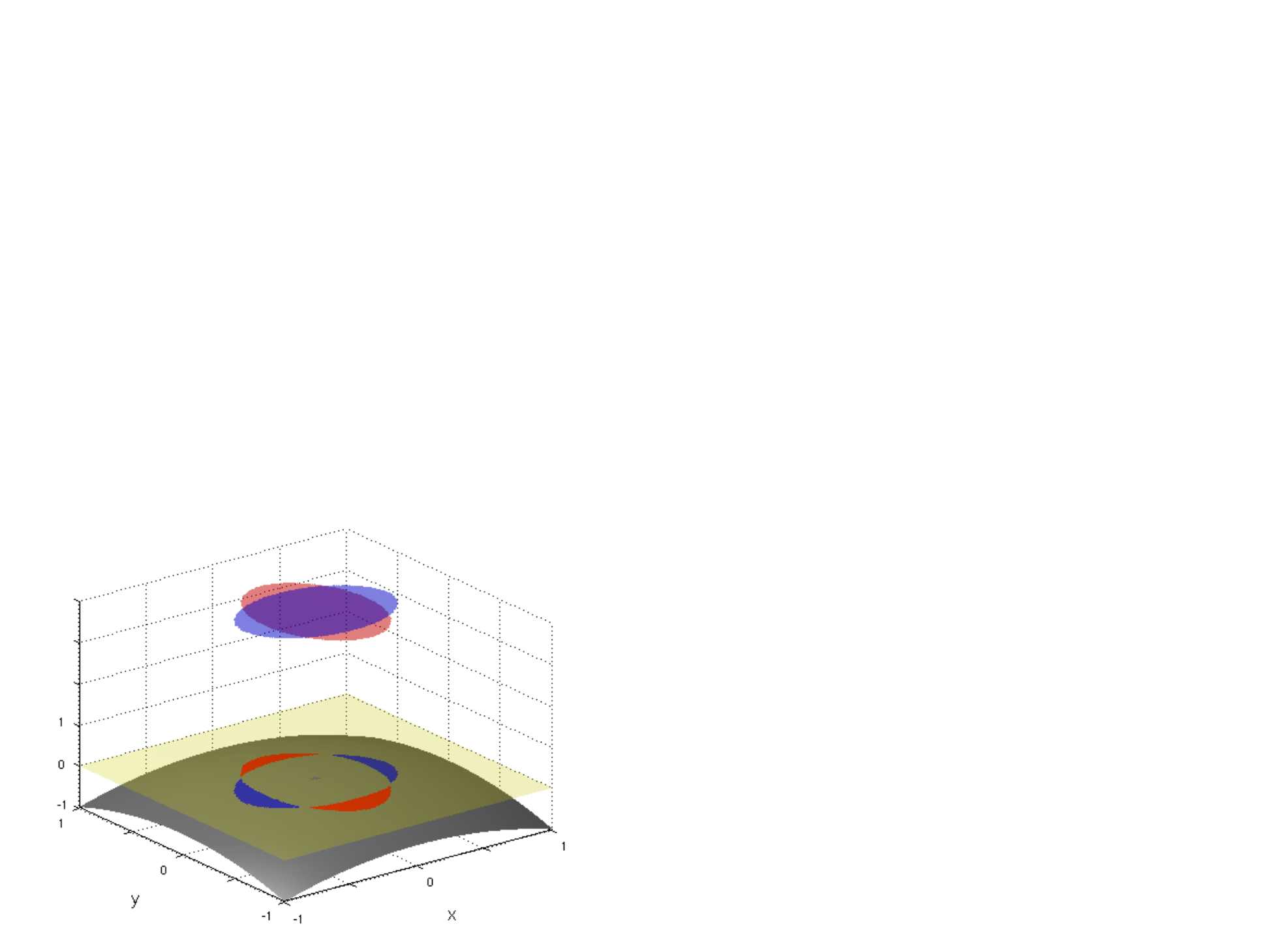} \\ (a) & (b)
      \\ \includegraphics[width=2.7in]{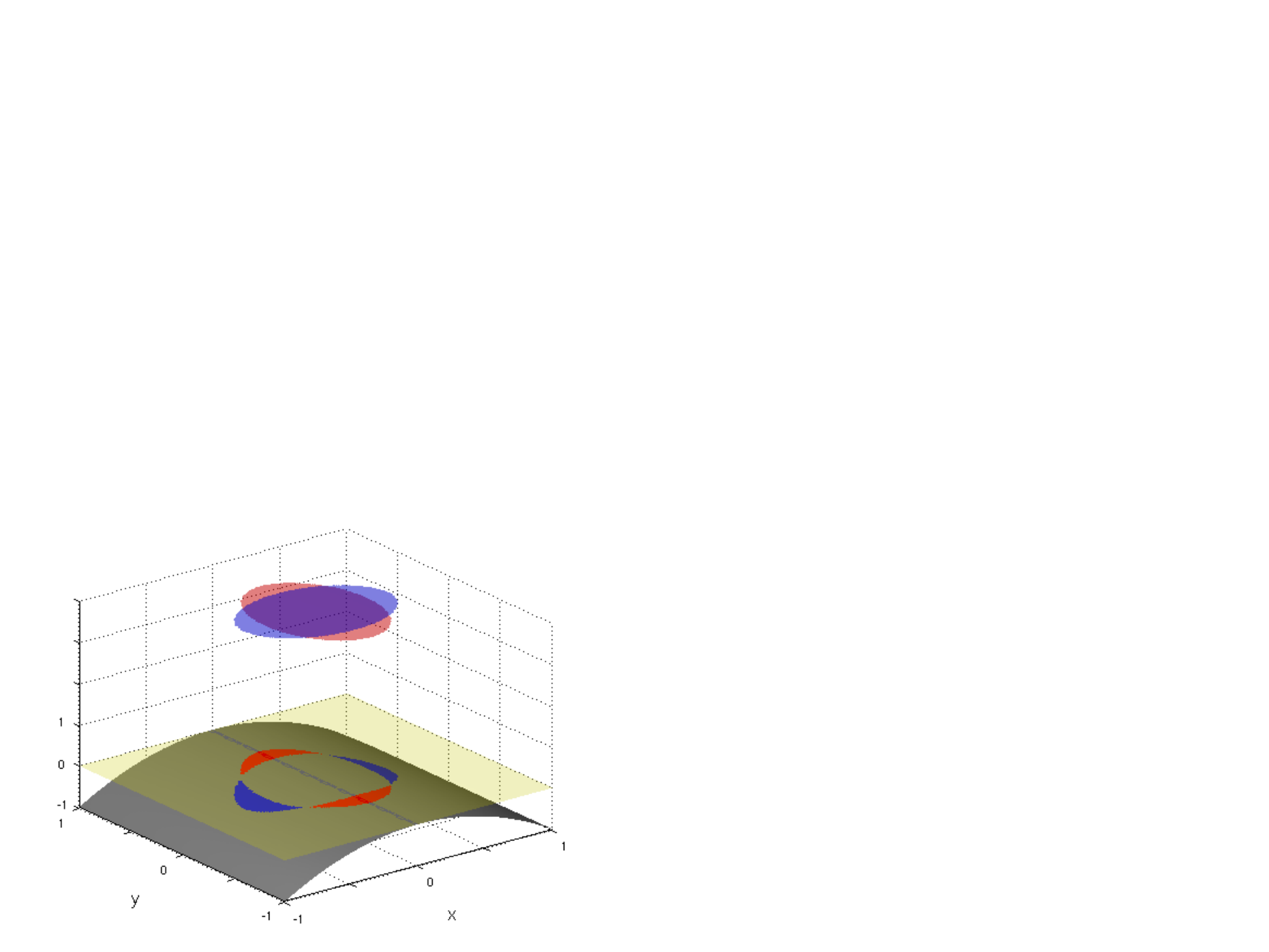} &
      \includegraphics[width=2.7in]{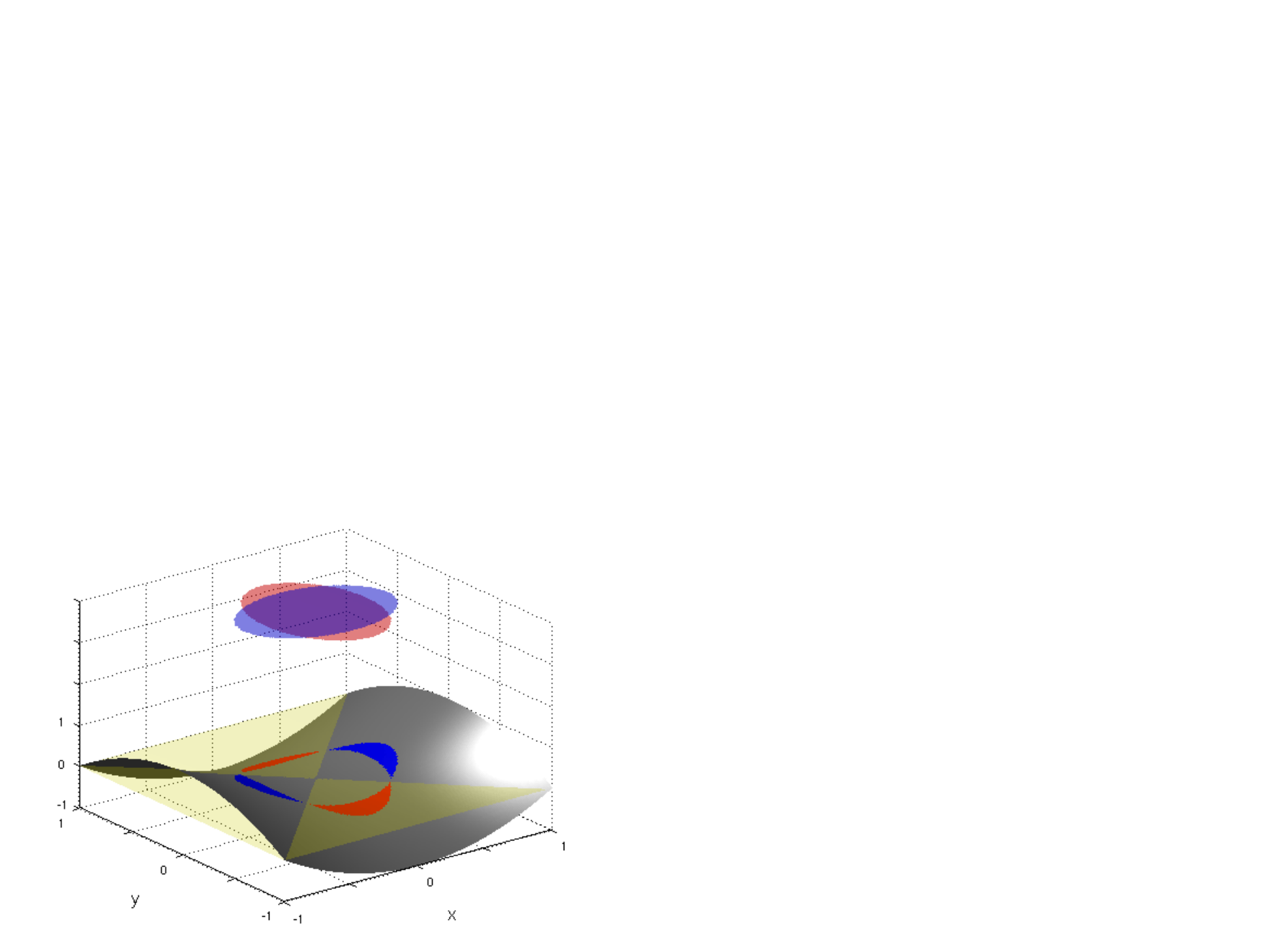} \\ (c) & (d)
      \\ \multicolumn{2}{c}{\includegraphics[width=2.7in]{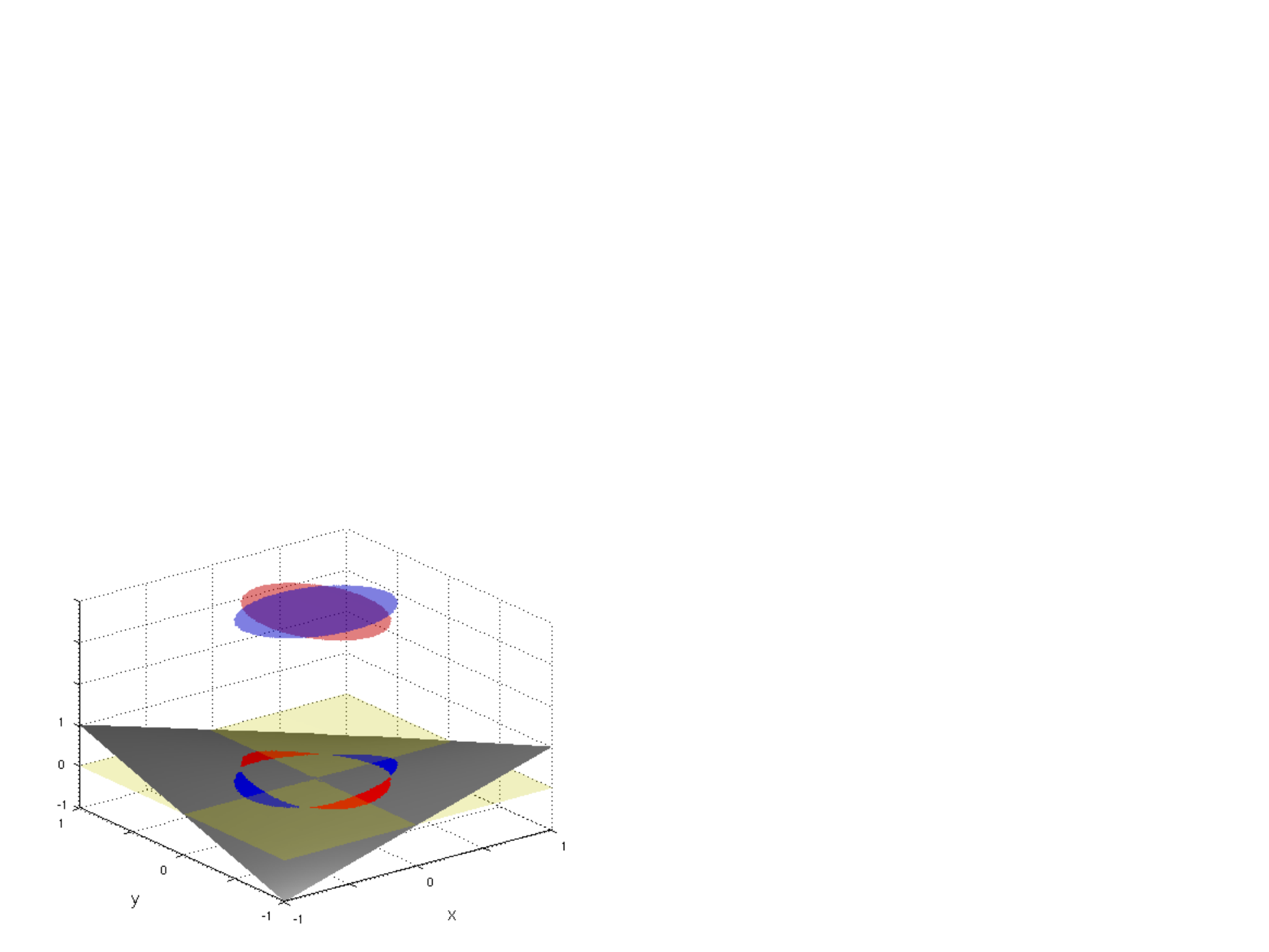}}
      \\ \multicolumn{2}{c}{(e)}
    \end{tabular}
  \end{center}
  \caption[Differential plus-ellipticity leakage]
          { \label{fig:manifoldellipp}Illustration of \ttp\ leakage resulting
            from differential plus-ellipticity. The gray plane represents a 
            $T$ sky with (a) $\nabla_xT>0$,
            $\nabla^2_xT=\nabla^2_yT=\nabla_x\nabla_yT=0$; (b)
            $\nabla_xT=\nabla_x\nabla_yT=0$,
            $\nabla^2_xT=\nabla^2_yT<0$; (c)
            $\nabla_xT=\nabla_x\nabla_yT=\nabla^2_yT=0$,
            $\nabla^2_xT<0$; (d)
            $\nabla_xT=\nabla_x\nabla_yT=0$,
            $\nabla^2_yT<0<\nabla^2_xT$; (e)
            $\nabla_xT=\nabla^2_xT=\nabla^2_yT=0$,
            $\nabla_x\nabla_y\neq0$.  Only the scenarios in (c) and (d), which
            have a non-zero difference of orthogonal second derivatives, leak
            \ttp. }
\end{figure*}

For differential plus-ellipticity it is only a \emph{difference} in the
orthogonal second derivatives of $T$ that results in \ttp\ leakage. The
coupling of differential plus-ellipticity to the second derivatives of
$T$ is illustrated in Figure~\ref{fig:manifoldellipp}. In
Figure~\ref{fig:manifoldellipp}a, neither the A nor B detector measures a signal
that is offset from the value of $T$ at the pair centroid. In
Figure~\ref{fig:manifoldellipp}b, both detectors measure an equally negatively
offset signal. In Figure~\ref{fig:manifoldellipp}c, both the A and B detectors
measure a negatively offset signal, but the B detector measures a \emph{more}
negatively offset signal, and the resulting pair-difference is non-zero. In
Figure~\ref{fig:manifoldellipp}d, the A detector measures a negatively offset
signal and the B detector measures a positively offset signal, and the resulting
pair-difference is also non-zero.  In Figure~\ref{fig:manifoldellipp}e, which
has zero first and second derivatives at the pair centroid but a non-zero cross
derivative, neither detector measures an offset signal, and the pair-difference
is zero.

A figure depicting differential cross-ellipticity analogous to
Figure~\ref{fig:manifoldellipp} is not shown. It would be identical to
Figure~\ref{fig:manifoldellipp} with the beams (or $T$) rotated by
$45\deg$. After doing this, only Figure~\ref{fig:manifoldellipp}e, which has a
non-zero cross-derivative, would leak \ttp. Thus for differential
cross-ellipticity, only a non-zero cross-derivative of $T$ produces
\ttp\ leakage.

\subsection{Crosstalk and Ghost Beams}
\label{sec:xtalkandbuddy}

Non-main-beam systematics can also be described by mismatches of elliptical
Gaussians. Crosstalk and internal reflections in the optics produce secondary
beams of smaller amplitude offset from the main beam. Again, it is only an A/B
mismatch of these secondary beams that results in \ttp\ leakage. The secondary
beams can, in general, exhibit their own mismatch, independent of the main beam
mismatch. They couple to the derivatives of $T$ not at the main beam
centroid but at their own centroid. The leakage is still entirely deterministic
and analyzable in terms of mismatched elliptical Gaussians.

\section{Mathematical Description of Deprojection}
\label{sec:deprojmath}

This section formally derives the leakage templates and fit coefficients
listed in Table~\ref{tab:deprojection}. A similar description of this formalism
is given in \citealt{aikinthesis}.

In all cases other than differential gain, we proceed by Taylor expanding an
elliptical Gaussian about the differential parameter, then expressing the
difference of perturbed beams as a linear combination of spatial derivatives of
the nominal Gaussian beam. We then express
the leaked signal as the convolution of the temperature field with the
Taylor expanded difference beam.

\subsection{Differential Gain}

We start by again noting that the \ttp\ leakage from beam mismatch
is simply $T$
convolved with a detector pair's $A-B$ difference beam, $B_{\delta}$, as
expressed in Equation~\ref{eq:diff}, and where we have dropped the explicit
dependence on $\mathbf{\hat{n}}$.

A gain mismatch between two Gaussian beams is parametrized as a difference in
the peak heights of the A and B beams,
\begin{eqnarray}
B_{\delta g}(r) & = & B_A(r)-B_B(r) \nonumber \\ & = &
\frac{g_A}{2\pi\sigma^2}\exp(-r^2/2\sigma^2) -
\frac{g_B}{2\pi\sigma^2}\exp(-r^2/2\sigma^2) \nonumber \\ & = & \delta g
\frac{\exp(-r^2/2\sigma^2)}{2\pi\sigma^2} \nonumber \\ & = & \delta g B(r)
\end{eqnarray}

\noindent where $\delta g = g_A-g_B$ and $B(r)$ without a subscript denotes the
nominal, un-differenced, circular Gaussian beam, and is a function of
1-dimensional radius $r$ instead of a 2D position vector
$\mathbf{n}$.  Note that here, the gain $g$ is defined so that the mean gain,
$(g_A+g_B)/2=1$.  The resulting leaked signal is
\begin{eqnarray}
d_{\delta g} & = & T\ast B_{\delta g}(r) \nonumber \\ & =
& \delta g \tilde{T}
\end{eqnarray}

\noindent where $\tilde{T} = T\ast B(r)$ is just $T$ convolved with the nominal,
circular Gaussian beam. 

As discussed in Section~\ref{sec:deprojalgorithm},
the method of deprojecting gain mismatch is then as follows: first, in the map
making stage, after computing the R.A./decl.\ trajectories of detector pairs, we
create a leakage template timestream, $\tilde{T}(t)$, for each detector pair by sampling a
Healpix temperature map along the pair's trajectory. (The Healpix map is
pre-smoothed to \bicep2's nominal circular Gaussian beam.) Second, we filter $\tilde{T}(t)$
exactly as is done to the real data timestreams. Lastly, we regress
the pair-difference data against $\tilde{T}(t)$ and subtract the best-fit
template.  The fit coefficient is proportional to $\delta g$. Alternately, if we
know each detector pair's 
differential gain \textit{a priori}, we can simply scale and subtract the
template.

\subsection{Differential Pointing}

To describe a Gaussian whose centroid is displaced in the focal plane
$x$-direction by a small fraction of the beamwidth, we express the nominal
circular Gaussian beam $B(r)$ in terms of $x$ and $y$ and Taylor expand about
$x$, so that
\begin{eqnarray}\label{eq:dp1}
B(x+\delta x,y) & = & B(x,y) + \left. \frac{\partial B(x',y)}{\partial x'}
\right|_x \delta x + \nonumber \\ 
& & \left.\frac{1}{2}\frac{\partial^2 B(x',y)}{\partial
  x'^2}\right|_x(\delta x)^2 + ...
\end{eqnarray}

\noindent where $\delta x$ is the displacement in $x$ and, as before,
\begin{equation}
  B(x,y)=B(r)=\exp[-(x^2+y^2)/2\sigma^2]/2\pi\sigma^2 
\end{equation}
\noindent is the unperturbed, un-differenced, circular Gaussian beam. Then,
calculating the difference beam that results from displacing the A beam by
$+\delta x /2$ and the B beam by $-\delta x /2$ (and noting that the first and
third terms in Equation~\ref{eq:dp1}, and indeed any term that is an even power
of $\delta x$, cancel),

\begin{eqnarray}
B_{\delta x}(x,y) & = & B(x+\delta x/2,y) - B(x-\delta x /2,y) \nonumber \\ 
& = & \left. \frac{\partial B(x',y)}{\partial x'} \right|_x \delta x +
\mathcal{O}[(\delta x)^3] \nonumber \\ 
& \simeq & \delta x \nabla_xB(x,y),
\end{eqnarray}
\noindent where we have defined the partial derivative with respect to focal plane
coordinate $x$
\begin{equation}
\nabla_x \equiv \frac{\partial}{\partial x}.
\end{equation}

The leaked signal resulting from differential pointing in the focal plane
$x$-direction is then
\begin{eqnarray}\label{eq:Bx2}
d_{\delta x} & = & T \ast B_{\delta x} \nonumber \\ &
\simeq & \delta x (T \ast \nabla_x B) \nonumber \\ & = & \delta x
\nabla_x (T \ast B) \nonumber \\ & = & \delta x \nabla_x
\tilde{T}.
\end{eqnarray}

\noindent Similarly, for a pointing displacement in the focal plane $y$-direction,
the leaked signal is
\begin{equation}\label{eq:Bx3}
d_{\delta y} = \delta y \nabla_y \tilde{T}.
\end{equation}

A differential pointing offset in any arbitrary direction can be expressed as
the linear combination of equations \ref{eq:Bx2} and \ref{eq:Bx3}. These
equations then tell us how to construct the leakage templates for differential
pointing. \texttt{Synfast} can produce sky maps of the first derivatives of the
circular Gaussian-smoothed temperature map expressed in the Healpix latitude/longitude
coordinate system,
$\nabla_{\theta}\tilde{T}(\mathbf{\hat{n}})$ and
$\nabla_{\phi}\tilde{T}(\mathbf{\hat{n}})$. We sample the Gaussian-smoothed
derivative maps along a detector pair's pointing trajectory as a function of time to create
two template timestreams, $\nabla_{\theta}\tilde{T}(t)$ and
$\nabla_{\phi}\tilde{T}(t)$.  Then, knowing the focal plane's orientation
on the sky at each point in the timestream, we apply the chain rule for
derivatives to transform the derivative timestreams from the Healpix coordinate
system to the focal plane $(x,y)$ coordinate system.

Once $\nabla_x \tilde{T}(t)$ and $\nabla_y \tilde{T}(t)$ have been
constructed and filtered like the real data, they are simultaneously fit to the
pair-difference data. The fit coefficients are $\delta x$ and $\delta y$.

\subsection{Differential Beamwidth}

The derivation of the templates for differential beamwidth (and differential
ellipticity) proceeds similarly to that for differential pointing. Taylor
expanding the nominal beam about the parameter to be perturbed, in this case
$\sigma$, we have
\begin{eqnarray}
B(x,y,\sigma+\delta\sigma) & = & B(x,y,\sigma) + \left.\frac{\partial
  B(x,y,\sigma')}{\partial \sigma'}\right|_\sigma 
\delta\sigma \nonumber \\  
& & + \left.\frac{1}{2}\frac{\partial^2 B(x,y,\sigma')}{\partial
  \sigma'^2}\right|_\sigma (\delta\sigma)^2 + ...
\end{eqnarray}

\noindent Again noting that even powers of $\delta\sigma$ cancel, the difference
beam is
\begin{eqnarray} 
B_{\delta\sigma}(x,y) & = & B(x,y,\sigma+\delta \sigma/2) -
B(x,y,\sigma-\delta\sigma/2) \nonumber \\ & \simeq & \left.\frac{\partial
  B(x,y,\sigma')}{\partial \sigma'}\right|_\sigma \delta\sigma +
\mathcal{O}[(\delta\sigma)^3]. \label{eq:Bsig0}
\end{eqnarray}

\noindent One can write
\begin{equation}\label{eq:Bsig1}
\frac{\partial B(x,y,\sigma)}{\partial\sigma} = \left( \frac{x^2+y^2}{\sigma^3}
- \frac{2}{\sigma}\right) B(x,y,\sigma).
\end{equation}

\noindent One can also can write
\begin{eqnarray}
\nabla^2_x B(x,y) & = & \left( \frac{x^2}{\sigma^4} - \frac{1}{\sigma^2}
\right) B(x,y) \label{eq:dxx} \\ \nabla^2_y B(x,y) & = & \left(
\frac{y^2}{\sigma^4} - \frac{1}{\sigma^2} \right) B(x,y) \label{eq:dyy}
\\ (\nabla^2_x+\nabla^2_y)B(x,y) & = & \left( \frac{x^2+y^2}{\sigma^2} - 2
\right) \frac{B(x,y)}{\sigma^2}
\end{eqnarray}

\noindent where we have defined the second partial derivatives with respect to focal
plane coordinates $x$ and $y$
\begin{equation}
\nabla^2_x \equiv \frac{\partial^2}{\partial x^2} , \nabla^2_y \equiv
\frac{\partial^2}{\partial y^2} . 
\end{equation}

\noindent Equation~\ref{eq:Bsig1} can then be written as
\begin{equation}
\frac{\partial B(x,y,\sigma)}{\partial\sigma} = \sigma ( \nabla^2_x +
\nabla^2_y ) B(x,y)
\end{equation}

\noindent and Equation~\ref{eq:Bsig0} becomes
\begin{equation}
B_{\delta\sigma} \simeq \sigma\delta\sigma ( \nabla^2_x + \nabla^2_y )
B(x,y)
\end{equation}

\noindent The \ttp\ leakage from differential beamwidth is then
\begin{eqnarray}
d_{\delta\sigma} & = & B_{\delta\sigma} \ast T \nonumber
\\ & \simeq & \sigma\delta\sigma(\nabla^2_x + \nabla^2_y)B \ast
T \nonumber \\ & \simeq &
\sigma\delta\sigma(\nabla^2_x +
\nabla^2_y)\tilde{T}.
\end{eqnarray}

Because differential beamwidth is monopole symmetric, only one template needs to
be constructed for deprojection. To construct the template, we sample 
$\nabla_{\theta\theta} \tilde{T}(\mathbf{\hat{n}})$, $\nabla_{\phi\phi} \tilde{T}(\mathbf{\hat{n}})$ and
$\nabla_{\theta\phi} \tilde{T}(\mathbf{\hat{n}})$ along each
detector pair's pointing trajectory, apply the chain rule for derivatives to construct
$\nabla^2_x \tilde{T}(t)$ and $\nabla^2_y \tilde{T}(t)$ (though the monopole symmetry
makes this step unnecessary), add them together, filter the sum like the real
data, and fit the resulting template to the pair-difference timestream. The fit
coefficient is $\sigma\delta\sigma$.

\subsection{Differential Ellipticity}

The difference beam corresponding to mismatched beam ellipticity is a quadrupole. The
orientation of the quadrupole is arbitrary, but can be approximated as the
linear combination of two orthogonal quadrupoles, chosen as the plus and cross
orientations.

An elliptical Gaussian with pure plus-ellipticity is written as
\begin{equation}
B(x,y,p) = \frac{1}{2\pi\sigma^2}\exp\left[-\frac{1}{2\sigma^2}
  \left(\frac{x^2}{1+p} + \frac{y^2}{1-p}\right)\right] .
\end{equation}

\noindent Taylor expanding the plus-ellipticity beam about $p=0$ yields
\begin{equation}
B(x,y,p) \simeq B(p=0) + \left. \frac{\partial B(x,y,p')}{\partial p'}
\right|_0p + \frac{1}{2}\left.\frac{\partial^2B(x,y,p')}{\partial p'^2}
\right|_0p^2 + ...~.
\end{equation}

\noindent Then, noting that we can write
\begin{equation}
\left.\frac{\partial B(x,y,p')}{\partial p'}\right|_0 =
\left(\frac{x^2-y^2}{2\sigma^2}\right)B(x,y) ,
\end{equation}

\noindent where again, $B(x,y)=B(x,y,p=0)$ is the nominal, circular Gaussian beam,
and using Equations~\ref{eq:dxx} and \ref{eq:dyy} to write
\begin{equation}
(\nabla^2_x-\nabla^2_y)B(x,y) = \left(\frac{x^2-y^2}{\sigma^4}\right)
  B(x,y)
\end{equation}

\noindent the difference beam is
\begin{eqnarray}
B_{\delta p}(x,y) & = & B(x,y,p_A)-B(x,y,p_B) \nonumber \\ & = & \delta p \left(
\frac{x^2-y^2}{2\sigma^2} \right) B(x,y) + \mathcal{O}(p_A^2-p_B^2) +
... \nonumber \\ & \simeq & \frac{\sigma^2}{2}\delta p
(\nabla^2_x-\nabla^2_y)B(x,y),
\end{eqnarray}

\noindent where $\delta p = p_A - p_B$. Note that the ellipticity difference
beam is accurate only to first order rather than second order like the other
Gaussian modes. Also note that the difference beam is accurate to order
$p_A^2-p_B^2$ rather than $(\delta p)^2$. Thus, large enough ellipticities can
cause a breakdown of deprojection even if the \textit{differential} ellipticity
is small. The \ttp\ leakage resulting from differential plus-ellipticity is
\begin{eqnarray}
d_{\delta p} & = & B_{\delta p}\ast T \nonumber \\ & \simeq &
\frac{\sigma^2}{2}\delta p (\nabla^2_x-\nabla^2_y) \tilde{T} .
\end{eqnarray}

Now we consider cross-ellipticity. An elliptical Gaussian with pure
cross-ellipticity can be written as
\begin{equation}
B(x,y,c) =
\frac{1}{2\pi\sigma^2}\exp\left[-\frac{1}{2(1-c^2)\sigma^2}(x^2+y^2-2cxy)\right]
,
\end{equation}

\noindent We work to linear order in $c$ and set $(1-c^2)\simeq1$. Proceeding as
above, and noting that
\begin{equation}
\nabla_x\nabla_yB(x,y) = \frac{1}{\sigma^4}xyB(x,y),
\end{equation}

\noindent where we have defined the cross-derivative with respect to focal plane
coordinates
\begin{equation}
\nabla_x\nabla_y \equiv \frac{\partial^2}{\partial x\partial y},
\end{equation}

\noindent we can write
\begin{eqnarray}
\left.\frac{\partial B(x,y,c')}{\partial c'}\right|_0 & = &
\frac{1}{\sigma^2}xyB(x,y) \nonumber \\ & = & \sigma^2\nabla_x\nabla_yB(x,y).
\end{eqnarray}

The difference beam resulting from differential cross-ellipticity is then
\begin{eqnarray}
B_{\delta c} & = & B(x,y,c_A) - B(x,y,c_B) \nonumber \\ & = & \sigma^2\delta c
\nabla_x\nabla_yB(x,y) + \mathcal{O}(c_A^2-c_B^2) + ...~.
\end{eqnarray}

\noindent where $\delta c = c_A-c_B$. The \ttp\ leakage resulting from
differential cross-ellipticity is
\begin{equation}
d_{\delta c} \simeq \sigma^2\delta c \nabla_x\nabla_y\tilde{T}.
\end{equation}

\subsection{Practical Implementation}
\label{sec:practical}

In the map making stage, we construct deprojection template timestreams for each
detector pair in the same manner as we construct simulated timestreams --- by
interpolating off a suitably pre-smoothed deprojection template map along the
pointing trajectory of a detector pair's mean centroid. The deprojection
template maps are simply a $T$ map and its first, second, and cross-derivatives,
all smoothed by \bicep2's measured mean, azimuthally averaged beam profile.

We interpolate off the deprojection $T$ map using the first and second
derivative $T$ maps to perform a second order Taylor expansion around the
nearest neighbor pixel center.  This is identical to the interpolation used in
the forward simulation pipeline. We interpolate off the first derivative maps
using the second derivative maps to perform a first order Taylor expansion
around the nearest neighbor pixel centers. We interpolate off the second
derivative maps using nearest neighbor interpolation, which we find to have
adequate accuracy for deprojection.

The interpolation scheme introduces the possibility that certain simulations of
beam mismatch might be deprojected to artificially high accuracy and indeed we
can simulate differential pointing in such a way that it deprojects to numerical
precision. However, this is not an issue in practice as (i) we perform 
simulations with an Nside=2048 input map but use an Nside=512 map for
deprojection, (ii) noise in the template map, which we account for, sets the
primary limitation on deprojection, (iii) our multiple Gaussian convolution
scheme simulates differential beamwidth and ellipticity with higher accuracy
than the leakage templates and, most importantly, (iv) the special beam map
simulations on which we ultimately rely to characterize residual contamination
from beam mismatch after deprojection do not use the same interpolation scheme
as the computation of the leakage template timestreams at all.

The template map used for deprojection of the main results is the \textit{Planck} HFI
143~GHz map, re-smoothed to \bicep2's measured mean beam profile and downgraded to
Nside=512.  We do 
not use derived data products meant to contain only CMB with no foregrounds
(i.e.\ SMICA) because \bicep2's \ttp\ leakage does, in principle, include
foreground $T$ at some very low level and we expect the \textit{Planck} 143~GHz bandpass
to most closely match \bicep2's. We smooth the \textit{Planck} map to match \bicep2's
beamwidth by computing the \textit{Planck} map's $a_{\ell m}$s using \anafast\ and
multiplying them by the ratio of \bicep2's mean azimuthally symmetric beam
window function (as 
measured from beam maps) to \textit{Planck}'s published 143~GHz beam window function,
$B_{\ell}^{B2}/B_{\ell}^{Pl}$, as presented in \citet{planckvii}. We then convert
back to map space with \synfast.

We simulate template map noise by using publicly available \textit{Planck} noise maps to
generate simulated realizations of uncorrelated, white noise temperature maps. We then
apply the same resmoothing procedure to the noise realizations as the real template map
(convert to $a_{\ell m}$s and multiply by $B_{\ell}^{B2}/B_{\ell}^{Pl}$), and use
\synfast\ to produce derivatives of the noise realizations. We add these noise realizations to
noiseless simulated maps containing lensed-\lcdm\ signal. We use the same
procedure to simulate WMAP template map noise.

In general, prior to fitting, the leakage template timestreams must be
subjected to any filtering or manipulation beyond pair differencing to which the
data timestreams themselves are subjected. For \bicep2\ this involves
third-order polynomial filtering of half scans and ground subtraction.

At this point, we could fit the deprojection templates directly to the
timestreams. However, because we co-add the data from each detector pair into
intermediate maps (the ``pair maps'' described in~\S IV.D of the Results Paper) on
approximately one hour timescales (a ``scanset'') we choose to bin the
deprojection template timestreams into map pixels on the same timescale.  When
coadding the pair maps into final $T$, $Q$ and $U$ maps, we can choose the
timescale in multiples of one hour over which to co-add the leakage templates
prior to performing deprojection.

As discussed, deprojection, like all filtering operations, removes information and produces
some amount of \etb\ mixing. In general, this mixing and mode removal becomes
worse when coadding the templates over shorter durations due to the decreased
coverage of the intermediate map, which results in more degrees of freedom being
removed from the final map. While coadding over all three years of \bicep2\ data
would produce the minimum possible mode removal, because the telescope scan
pattern repeats every 72~hr, coadding over timescales longer than this makes
no practical difference. Coadding over shorter timescales allows time variable
systematics to be filtered out.  We fully expect beam shape mismatch to remain
constant in time, but, as discussed in Section~\ref{sec:gainvar},
gain mismatch could presumably have a time variable
component.  We therefore choose to co-add the data and deprojection templates
over approximately 9~hr timescales (1 ``phase,'' equivalent to 10 scansets)
prior to performing deprojection.  Using the matrix purification power spectrum
estimator discussed in~\S VI.B of the Results Paper, we find that this does not
significantly increase the uncertainty of our final bandpowers. In any case, the
\etb\ leakage and mode removal caused by deprojection is the same as that from
any filtering operation and is fully captured by our simulation-based analysis.

\section{Beam map simulation uncertainty}
\label{sec:beammapsimappendix}

We estimate the uncertainty of the contamination predicted from beam map
simulations and shown in Figures~\ref{fig:beammapjacks} and \ref{fig:sysfigs}
using a standard resampling method known as a delete-one jackknife, not
  to be confused with the jackknife tests described in Section~\ref{sec:jackknives}
  and throughout the paper. The method is as follows:
the composite beam maps described 
in Section~\ref{sec:beamsim} are the mean
of 12 independent measurements. (For the extended composite beam, they are the
median of the measurements.) 
We remake the composite beam maps 12 times, excluding in 
turn 1 of the 12 measurements prior to taking the mean. This 
generates 12 additional sets of
simulated spectra. The standard deviation of
the 12 bandpowers in each $\ell$ bin, multiplied by $\sqrt{N-1}$, where $N=12$ is
the number of independent measurements,
 provides an estimate of the standard error in each bin. The difference of the mean
 of the 12 bandpowers and the bandpower 
from the main simulation, multiplied by $(N-1)$,  provides an estimate of the
bias in each bin.  The bias and standard error of the predicted contamination from
beam mismatch at $r<1.2\deg$ is robust because the beam maps are mean
filtered in this region and because each beam map pixel is covered by all 12
observations. (For the extended beam map an effective number of observations,
$N_{eff}<12$, must be used.)

We apply identical differential gain normalization to the composite beam
maps in each of the $12$ delete-one jackknife realizations. Uncertanties in the
differential gain measurements are therefore not accounted for. From the
standard deviation of these measurements made from different temporal subsets of
\bicep2\ data, and from simulations that include the effect of
\lcdm\ $TE$ correlation in our absolute calibration procedure, we estimate that
their uncertainty is $\sigma_{\delta g} \simeq 0.02$. We perform noiseless,
temperature-only simulations using the \textit{Planck} 143~GHz map as an input to the
standard simulation pipeline described in Section~\ref{sec:standardsim}. In each
multipole bin, we compute
the standard deviation of $50$ realizations of a random 2\% gain mismatch
($\overline{\delta g}=0$, $\langle\delta g^2\rangle^{1/2} = 0.02$) and add this to the uncertainties
predicted from the delete-one jackknife. (The sum as opposed to the quadrature
sum is appropriate for bandpowers.)

We make a correction to the simulated main beam leakage spectra by subtracting
the predicted bias, which is $\sim3-10\times$ smaller than the simulated
leakage after deprojection. The corrected leakage, shown in Figure~\ref{fig:beammapjacks},
 is statistically significant
compared to its estimated uncertainty. The extended beam maps are observed with
less redundancy and are thus noisier than the main beam maps. 
The bias corrected leakage from the extended beam simulations is consistent with
zero. Its estimated uncertainty is plotted as the
``extended beams'' line in Figure~\ref{fig:sysfigs}.

\bibliographystyle{apj}
\bibliography{syst}

\end{document}